\documentclass[preprint,12pt]{elsarticle}

\RequirePackage{amsmath}
\usepackage[hidelinks]{hyperref}
\usepackage{mathtools}
\usepackage{amssymb}
\usepackage{environ}
\usepackage{multirow}
\usepackage{longtable}
\usepackage{booktabs}
\usepackage{placeins}
\usepackage{ragged2e,array,booktabs}

\usepackage{amsmath, amssymb}
\newtheorem{example}{Example}
\usepackage{tikz} 
\usepackage{algpseudocode}
\usepackage{algorithm}
\usepackage{caption}
\usepackage{subcaption}
\usepackage{standalone} 
\usepackage{url}

\begin{document}
	
	\begin{frontmatter}
		
		\title{Longest Filled Common Subsequence for Song Identification from Degraded Audio via Construct--Merge--Solve--Adapt Optimization}
		
		\author[inst1,inst6,inst7]{Marko Djukanović\corref{cor1}}\ead{marko.djukanovic@pmf.unibl.org}

		\author[inst2]{Christian Blum}
		\author[inst3]{Aleksandar Kartelj}
		\author[inst4]{Ana Nikolikj\corref{cor1}}\ead{ana.nikolikj@ijs.si}
		
		\author[inst5]{Günther R.~Raidl}

		\cortext[cor1]{Corresponding author}
		
		\address[inst1]{Faculty of Natural Sciences and Mathematics, University of Banja Luka, Bosnia and Herzegovina \\ \texttt{marko.djukanovic@pmf.unibl.org}}
		\address[inst6]{University of Nova Gorica, Nova Gorica, Slovenia \\ \texttt{marko.dukanovic@ung.si} }
		
		\address[inst2]{Artificial Intelligence Research Institute (IIIA-CSIC), UAB, Barcelona, Spain \\ \texttt{christian.blum@iiia.csic.es}}
		\address[inst3]{Faculty of Mathematics, University of Belgrade, Serbia \\ \texttt{kartelj@math.rs}}
		\address[inst4]{Jožef Stefan Institute, Ljubljana, Slovenia \\ \texttt{ana.nikolikj@ijs.si}}
		\address[inst5]{TU Wien, Vienna, Austria \\ \texttt{raidl@ac.tuwien.ac.at}}
		\address[inst7]{Institute of Information Sciences (IZUM), Maribor, Slovenia}
		
		\begin{abstract}
			This paper addresses the Longest Filled Common Subsequence (LFCS) problem, a challenging NP-hard problem with applications in bioinformatics, including gene mutation prediction and genomic data reconstruction. Existing approaches, including exact, metaheuristic, and approximation algorithms, have primarily been evaluated on small instances, which provide limited insight into their scalability. In this work, we introduce a new benchmark dataset with significantly larger instances and demonstrate that existing datasets lack the discriminative power needed to meaningfully assess algorithm performance at scale. To solve large instances efficiently, we utilize an adaptive Construct, Merge, Solve, Adapt (CMSA) framework that iteratively generates promising subproblems via component-based construction and refines them using feedback from prior iterations. Subproblems are solved using an external black-box solver. Extensive experiments on both standard and newly introduced set of large-sized instances prove that the designed adaptive CMSA achieves state-of-the-art performance, outperforming four known approaches. Notably, among 1,510 problem instances with known optimal solutions, our approach matches the proven optimum on 1,486 instances---corresponding to 98.4\% of these instances. These results demonstrate substantially improved scalability on large problem instances. As an engineering contribution, we propose a novel application of LFCS to song identification from degraded audio excerpts, using real-world energy-profile instances derived from popular music. Finally, we conduct an empirical explainability analysis to identify critical feature combinations influencing algorithm performance, revealing the key problem features associated with the success or failure of the approaches across different instance types.
		\end{abstract}
		
		
	\end{frontmatter}
	
	\section{Introduction}\label{sec:intro}

	Strings are fundamental data structures in many programming languages, widely used to model sequences such as DNA, RNA, protein molecules, and even time series. A string $s$ is defined as a finite sequence of characters drawn from a (typically finite) alphabet~$\Sigma$. A key concept in analyzing similarities between strings is a \emph{subsequence}---a sequence derived from a string by deleting zero or more characters without changing the order of remaining characters. Detecting structural similarities among multiple strings is central to both stringology and molecular biology. One of the most widely used tools for this purpose is the identification of \emph{common subsequences}, which can help uncover relationships among sequences and reveal meaningful biological or structural patterns, such as mutations. Advances in genome sequencing have further enabled the analysis and reconstruction of genomic data using such techniques. A core optimization problem in this context is the \textit{Longest Common Subsequence} (LCS) problem~\cite{Maier78}, defined as follows: Given a set of input strings, find a longest sequence that is a subsequence of each of the input strings. The LCS problem has broad applications in computational biology~\cite{ijms160920748}, file plagiarism detection, data compression~\cite{Sto88:datacompr,Beal2016}, text editing~\cite{kruskal1983overview}, and spatial data analysis, such as identifying road intersections from GPS traces~\cite{ijgi6010001}. Practical applications include file comparison tools like the Unix \texttt{diff} command~\cite{bergroth2000survey} and revision control systems such as \textsc{Git}.
	
	To solve the LCS problem when the number of input strings $m$ is fixed, polynomial-time algorithms based on dynamic programming (DP) have been proposed~\cite{Gus97:sequence-algorithms}. However, their practical applicability remains limited due to the rapidly growing computational complexity with respect to the number of strings and the strings' lengths. To address this, several advanced metaheuristics have been developed to solve LCS variants more efficiently in practice. State-of-the-art methods are based on effectively exploring the state space graph~\cite{BluBleLop2009,DjukanovicRaidlBlum19lod,nikolic2021solving}, combined with carefully crafted, problem-specific search guidance strategies. These methods have been tested on both artificial and real benchmark datasets, revealing the strengths and drawbacks of these approaches on various instance distributions. 
	In parallel, exact approaches have advanced through the use of best-first search frameworks and their extensions to hybrid anytime algorithms~\cite{DJUKANOVIC2020106499}. Over the past two decades, practical variants of the LCS problem have gained increasing attention in the research community. Notable examples include the longest common arc subsequence problem~\cite{lin2002longest}, the repetition-free LCS problem~\cite{adi2010repetition}, and the constrained LCS problem~\cite{tsai2003constrained}, among many others.
	
	More recently, research efforts have turned toward understanding gene mutations and reconstructing genomic data, aiming to uncover the evolutionary history of molecular sequences. This goal is formalized by the \textit{Longest Filled Common Subsequence }(LFCS) problem, which is inspired by another real-world application, the scaffold filling problem~\cite{munoz2010scaffold,bulteau2015fixed} from genomic data reconstruction. The LFCS problem (LFCSP) was introduced by Castelli et al.~\cite{castelli2017longest}. Briefly, the problem involves two sequences: the original sequence and one that requires repair. Additionally, a multiset of symbols is provided, which can be used for the repair. The goal is to insert some (or all) of these symbols into the second, shorter, sequence to maximize its similarity to the original sequence. The application of the LFCSP can naturally extend to a broader range of research areas involving sequential data that may suffer from information loss, such as audio querying or corrupted text recovery~\cite{dai2014autonomous}. 

	Regarding methodological approaches for solving the LFCSP reported in the literature, in~\cite{castelli2017longest} the authors proposed two $\frac{3}{5}$-approximation algorithms and proved the APX-hardness of the LFCSP, even when the first input string contains at most two occurrences of each symbol from the alphabet~$\Sigma$. This result was established via a reduction from the maximum independent set problem on cubic graphs. The second notable work by Mincu and Popa~\cite{mincu2018heuristic} presented diverse solving approaches including an Integer Linear Programming (ILP) formulation, a randomized sampling-based method, and a local search (LS) strategy based on classical hill climbing. Their comparative computational study was conducted on a set of randomly generated instances.
	
	Several limitations of this earlier work remain unaddressed:
	\begin{itemize}
		\item \textit{Scalability}: Existing benchmark instances are limited to small and medium-sized problems, all of which can be solved efficiently by the ILP model in a matter of seconds. As a result, the practical scalability and limitations of these methods have not been fully explored. Moreover, the current literature lacks real-world problem instances, or at least instances with characteristics appearing in real-world applications. 
		
		\item\textit{Heuristic Potential}: The existing heuristic methods do not fully leverage advanced search operators or learning mechanisms commonly used in modern optimization, leaving room for significant  methodological improvements for solving the problem at hand.
		
		\item \textit{Explainability}: There is a lack of systematic analysis to explain the performance of algorithms based on instance characteristics or internal algorithmic behaviors---a growing area of interest in optimization research.
		
	\end{itemize}
	
	In this article, we address these gaps through the following contributions:
	\begin{itemize}
		\item  A state-of-the-art ILP-based metaheuristic approach is proposed for solving the LFCSP. The method is based on an adaptive variant of the \textit{Construct, Merge, Solve, Adapt} (CMSA) framework~\cite{blum2016construct,blum2024self}.
		
		\item 	Two new benchmark datasets are introduced. The first one consists of dozens of synthetic instances which are significantly larger than the instances from the existing benchmarks. This allows for a deeper investigation into the scalability and robustness of both existing and new solving approaches.  The second dataset comprises realistic LFCSP instances based on discretized energy profiles from music recordings, illustrating a way the LFCSP approach supports audio fingerprinting in challenging, low-fidelity query scenarios.


		\item An explainable AI analysis is performed using the SHAP (\textit{SHapley Additive exPlanations}) approach~\cite{holzinger2020explainable} to reveal how each instance feature influences the performance of each method. This provides novel insights into the strengths and weaknesses of different approaches across varying instance types.
	\end{itemize}
	
\begin{figure}[H]
		\centering
		\includegraphics[width=420pt,height=350pt]{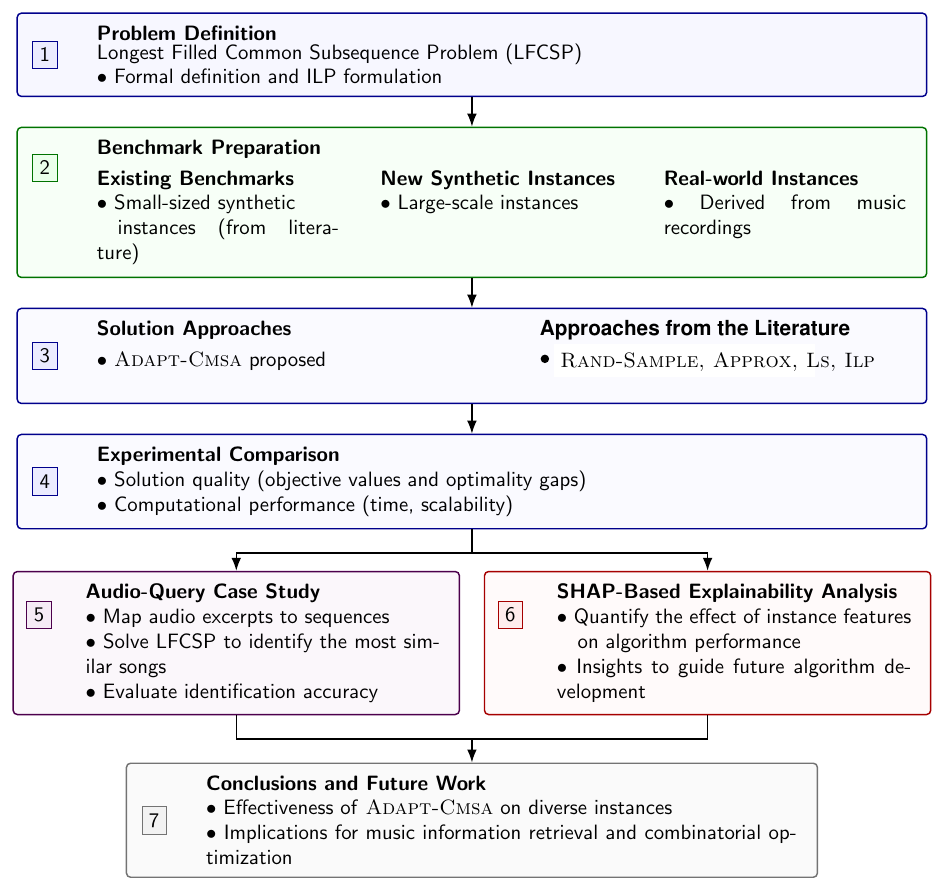}
		\caption{Overview of the research workflow. 
		} \label{fig:research-workflow}
	\end{figure}

	{Figure~\ref{fig:research-workflow} summarizes the overall workflow of this research. Starting from the formal definition of the LFCSP, the study combines benchmark preparation, the development of the proposed \textsc{Adapt-Cmsa} approach, comparison with existing solution methods, validation through an audio-query case study, and a SHAP-based analysis of the influence of instance features on algorithmic performance.}
	
	The remainder of the paper is organized as follows. In Section~\ref{sec:problem_def}, a formal problem definition is given and some notations are introduced. Brief descriptions of the existing approaches from the literature are given in Section~\ref{sec:existing-algs}. Section~\ref{sec:adapt_cmsa} provides the details on the newly developed approach to solve the LFCSP. A thorough experimental analysis is provided in Section~\ref{sec:experimental_section}. Section~\ref{sec:lfcsp-case-study} presents a case study in the context of audio querying and fingerprinting. Section~\ref{sec:explainable_AI} is devoted to analyzing the performance of all approaches through the lens of instance features and explainable artificial intelligence. In Section~\ref{sec:conclusions}, we conclude the paper and outline promising future research questions.

	\section{Problem Definition and Notations}\label{sec:problem_def}
	
	Given two input strings $A$ and $B$ over an alphabet $\Sigma$, and a multiset $\mathcal{M}$ in which symbols from $\Sigma$ may appear arbitrarily many times, we call a sequence $B^*$ a \emph{filled subsequence} of $B$ w.r.t.\ $\mathcal{M}$ if it can be obtained by inserting (some) symbols from $\mathcal{M}$ into the string $B$. The LFCSP seeks to find a filled sequence $B^*$ such that the length of the LCS among $A$ and $B^*$ is maximized across all possible filled versions of the initial string $B$.
	
	This LFCSP can alternatively be formulated in terms of deleting symbols from input string $A$, allowing a possibly more intuitive interpretation and more constrained search space for solving the LFCSP:
	Consider looking for symbols from $\mathcal{M}$ that match symbols in $A$ and
	deleting these symbols from $A$, resulting in a subsequence $A'$. Let $r$ be the number of these removed symbols, i.e., $r = |A| - |A'|$.
	We therefore seek a solution that maximizes this number plus the length of the longest common subsequence of $A'$ and $B$, i.e.,
	\begin{equation}\label{eq:alternative_lfcsp_definition}
		\text{obj}(A') = r + \text{LCS}(A', B).
	\end{equation}
	
	\begin{example} \label{ex:ecample_lfcsp}
		Given the two strings $A=\texttt{EGHGBCBEGECEEHDA}$ and $B=\texttt{EGGHHD}$ and the multiset $\mathcal{M}= \{ \texttt{E,D,B,C,B,E,G,E,E,A,G} \}$. A LFCS of this instance is $\texttt{EGHGBCBEGEEEHDA}$ and its length is 15. 
		Considering the second interpretation of the LFCSP based on symbol deletion, this solution is obtained by removing from $A$ the symbols $\texttt{HGBCBEGEEDA}$ at positions \{4, 5, 6, 7, 8, 9, 10, 12, 13, 15, 16\}, obtaining $A'=\texttt{EGHCH}$. 
		Note that the removed symbols are indeed taken from $\mathcal{M}$ and their number is $r=11$. As $\text{LCS}(A', B) = \texttt{EGHH}$ and its length is four, we obtain as solution value $11+4=15$.
		This solution is also visualized in Figure~\ref{fig:example_lcfsp}. 
		Following the original LFCSP definition, the same solution can be obtained by inserting symbols from $\mathcal{M}$ into string $B$: 
		Inserting \texttt{GBCBEGEEE} after position four in string $B$, i.e., between the two symbols \texttt{H} and appending \texttt{DA} at the end results in the filled sequence $B^*=\texttt{EGGHGBCBEGEEEHDDA}$, and $\text{LCS}(A, B^*) = \texttt{EGHGBCBEGEEEHDA}$. 
	\end{example}
	
	\begin{figure}[!ht]
		\scalebox{0.7}{
			
			\begin{tikzpicture}[every node/.style={minimum size=1cm, font=\sffamily\bfseries}]
				
				\node[draw] (Ax) at (-2, 0) {$\textbf{A}$:};
				\node[circle, draw] (A1) at (0, 0) {E};
				\node[circle, draw] (A2) at (1, 0) {G};
				\node[circle, draw] (A3) at (2, 0) {H};
				\node[rectangle, draw, red, rounded corners=5pt] (A4) at (3, 0) {G};
				\node[rectangle, draw, red, rounded corners=5pt] (A5) at (4, 0) {B};
				\node[rectangle, draw, red, rounded corners=5pt] (A6) at (5, 0) {C};
				\node[rectangle, draw, red] (A7) at (6, 0) {B};
				\node[rectangle, draw, red, rounded corners=5pt] (A8) at (7, 0) {E};
				\node[rectangle, draw, red,rounded corners=5pt] (A9) at (8, 0) {G};
				\node[rectangle, draw, red,rounded corners=5pt] (A10) at (9, 0) {E};
				\node[circle, draw] (A11) at (10, 0) {C};
				\node[rectangle, draw, red,rounded corners=5pt] (A12) at (11, 0) {E};
				\node[rectangle, draw, red,rounded corners=5pt] (A13) at (12, 0) {E};
				\node[circle, draw] (A14) at (13, 0) {H};
				\node[rectangle, draw, red,rounded corners=5pt] (A15) at (14, 0) {D};
				\node[rectangle, draw, red,rounded corners=5pt] (A16) at (15, 0) {A};
				
				\node[draw] (Axx) at (-2, -2) {$\textbf{B}$:};
				\node[circle, draw] (B1) at (0, -2) {E};
				\node[circle, draw] (B2) at (1, -2) {G};
				\node[circle, draw] (B3) at (2, -2) {G};
				\node[circle, draw] (B4) at (3, -2) {H};
				\node[circle, draw] (B5) at (4, -2) {H};
				\node[circle, draw] (B6) at (5, -2) {D};
				
				\node[draw] (Axx) at (-2, -4) {$\mathcal{M}$:};
				\node[circle, draw, fill=lightgray] (C1) at (0, -4) {E};
				\node[circle, draw, fill=lightgray] (C2) at (1, -4) {D};
				\node[circle, draw, fill=lightgray] (C3) at (2, -4) {B};
				\node[circle, draw, fill=lightgray] (C4) at (3, -4) {C};
				\node[circle, draw, fill=lightgray] (C5) at (4, -4) {B};
				\node[circle, draw, fill=lightgray] (C6) at (5, -4) {E};
				\node[circle, draw, fill=lightgray] (C7) at (6, -4) {G};
				\node[circle, draw, fill=lightgray] (C8) at (7, -4) {E};
				\node[circle, draw, fill=lightgray] (C9) at (8, -4) {E};
				\node[circle, draw, fill=lightgray] (C8) at (9, -4) {A};
				\node[circle, draw, fill=lightgray] (C9) at (10, -4) {G};
				
				\draw[-, thick, blue] (A1) -- (B1);
				\draw[-, thick, blue] (A2) -- (B2);
				\draw[-, thick, blue] (A3) -- (B4);
				\draw[-, thick, blue] (A14) to[out=-120, in=50]  (B5);
		\end{tikzpicture}}
		
		\caption{Visualization of an LFCSP solution. }
		\label{fig:example_lcfsp}
	\end{figure}

	\subsection{Further Notations}\label{sec:notation}
	
	We now introduce some basic notation used throughout this work. 
	
	By $I=(A, B, \mathcal{M})$, we denote an LFCSP instance with input strings $A$ and $B$ and the multiset of symbols $\mathcal{M}$. Let $n=|A|$ and $m=|B|$ be the lengths of these two strings and $k=|\mathcal{M}|$ the cardinality of $\mathcal{M}$. 
	By $s[i]$ we denote the $i$-th character of some string $s$, with $i=1,\ldots,|s|$. 
	Again, $A'$ is a subsequence of string $A$ if $A'$ can be obtained by deleting zero or more characters from $A$. 
	By $A[i, j]$ we denote the substring of $A$ that starts with the character at index $i$ and ends with the character at index $j$; specifically, $A[i, i] =A[i]$ and if $i>j$ then $A[i, j]=\varepsilon$, where $\varepsilon$ denotes the empty string. By $|\mathcal{M}_{\sigma}|$, $\sigma \in \Sigma$ we denote the number of occurrences of symbol $\sigma$ in multiset $\mathcal{M}$; {similarly, for a sequence $s$ and a symbol $\sigma\in \Sigma$, the number of occurrences of $\sigma$ in $s$ is denoted by $|s|_{\sigma}$.}

	\section{Approaches from the Literature}\label{sec:existing-algs}
	
	In the following, we briefly explain the existing approaches for solving the LFCSP, which we will consider in our experimental comparisons. As corresponding implementations were not freely available, nor were they provided to us by the authors, we re-implemented all these approaches from scratch.
	
	\subsection{Approximation Algorithms}\label{sec:approx-lcfsp}
	
	Castelli et al.~\cite{castelli2017longest} described two approximation algorithms for the LFCSP. As both provide the same approximation factor of 3/5, and as, during experimentation, we observed no statistically significant difference between the results of the two approaches, we limit ourselves to providing the details of one of them here.
	
	The basic workflow of the algorithm, henceforth called \textsc{Approx}, is as follows. 
	\begin{itemize}
		\item Determine an LCS of $A$ and $B$. Let $R_{1,\mathrm{a}}$ denote the set of positions in $A$ matched by the alignment of the LCS.
		\item Remove all positions of $R_{1, \mathrm{a}}$ from $A$, and denote the resulting string by $A'$.
		\item By applying the procedure \textsc{GreedyMatch} as described below, compute a set $R_{1, \mathrm{i}}$ of positions of $A'$ of maximal size that matches $\mathcal{M}$ by insertion.
		\item Let $R_1 = R_{1,\mathrm{a}} \cup R_{1,\mathrm{i}}$ be the set of resulting positions returned by the algorithm, representing the LFCSP solution.
	\end{itemize}

	Procedure \textsc{GreedyMatch}$(A', \mathcal{M})$ works as follows:
	\begin{itemize}
		\item Set $\tilde A =\varepsilon$,  $R_{1, \mathrm{i}}=\emptyset$ 
		\item Proceed letter-by-letter from left to right over string $A'$ and check if the current symbol $A'[i]=\alpha$ is free/available to be matched, i.e., appended to $\tilde A$.   
		\item If yes ($\alpha \in \mathcal{M}$ is confirmed), append $\alpha$ to $\tilde A$ ($\tilde A\leftarrow \tilde A \cdot \alpha$), remove $\alpha$ from $\mathcal{M}$ ($\mathcal{M}\leftarrow \mathcal{M} \setminus \{\alpha \}$), and add $i$ to $R_{1, \mathrm{i}}$. Otherwise, skip the letter and move to the next letter of $A'$;
		\item When all letters of $A'$ were considered, the algorithm terminates by returning $\tilde A$ and the set of positions $R_{1, \mathrm{i}}$. 
	\end{itemize}

	\subsection{A Randomized Greedy Construction Method} \label{sec:uniform-sampling}
	
	The following randomized greedy solution construction was proposed by Mincu and Popa~\cite{mincu2018heuristic}. 
	
	\begin{itemize}
		\item Select a subset of symbols from \( A \) to match as many symbols as possible from \( \mathcal{M} \). Let the positions of these matched symbols in \( A \) be denoted by \( R_{1, \mathrm{a}} \). For instance, if \( A \) contains three $\texttt{a}\in\Sigma$ symbols and \( \mathcal{M} \) contains four, all three \texttt{a} symbols in \( A \) (and their positions) are subsequently matched. Conversely, if \( A \) has four \texttt{a} symbols and \( \mathcal{M} \) has three, then three out of the four \texttt{a} positions in \( A \) are randomly selected to match between symbols. 
		
		\item Obtain string $A'$ by removing all selected positions in $R_{1, \mathrm{a}}$. 
		\item Determine an LCS for $A'$ and $B$; the objective value of the obtained solution equals LCS($A', B$)$+|A|-|A'|$.
	\end{itemize}
	
	Mincu and Popa apply this randomized construction $n_\mathrm{rand}=10000$ times and finally return a best obtained solution, and so do we in our experimental comparison.

	\subsection{Local Search Methods}\label{sec:ls-based-approach}
	
	The following local search method has also been proposed by Mincu and Popa~\cite{mincu2018heuristic}.
	\begin{itemize}
		\item For a {fixed $l$}, we consider all consecutive substrings of size {$l$} of string $A$, i.e., {$A[1, l]$, $A[2, l+1]$}, etc. obtained by windowing sequence $A$.
		\item For each such substring, we take all possible combinations of letters ({$2^l$} of them) and match them with symbols from $\mathcal{M}$. For the remaining (non-matched) parts of sequences $A$ and $B$, the LCS is determined, and the objective value of the resulting filled common subsequence (i.e., solution) is reported. Following a first-improvement strategy, any new best solution identified is accepted as new incumbent.
		\item In each major iteration, up to {$2^l \cdot(n-l)$} candidate solutions are therefore considered and evaluated.
	\end{itemize} 
	The authors refer to the algorithm as {S$l$}, and the two settings $l=2$ and $l=4$ are considered in the evaluation. We will do the same in our experimental comparisons.
	
	\subsection{Integer Linear Programming Model}\label{sec:ilp_lcfsp}
	
	Mincu and Popa~\cite{mincu2018heuristic} further modeled the LFCSP by an ILP.
	Let $\mathcal{R}$ denote the set of all matchings between input strings $A$ and $B$, i.e., $(i. j) \in \mathcal{R}$ iff $A[i]=B[j]$ for $i=1,\ldots,n$ and $j=1,\ldots,m$. Two sets of binary decision variables are used:
	\begin{itemize}
		\item $x_{i, j}$ for $(i, j) \in \mathcal{R}$: $x_{i,j} = 1$ means that the letter $A[i]=B[j]$ is included in the LFCS, otherwise it is not. 
		\item $y_i \in \{0, 1\}$, for $i=1, \ldots, n$: $y_i=1$ means that $A[i]$ is included in the solution as a symbol that is filled in, or in other words, $A[i]$ is deleted from $A$ for obtaining $A'$ in the deletion-based interpretation of the LFCSP. 
		
	\end{itemize}
	
	The ILP can now be stated as:
	\begin{align}
		\max\ &\sum_{i=1}^n y_i + \sum_{(i, j) \in \mathcal{R}} x_{i, j} \label{ilp:obj}\\
		\mathrm{s.t.\ }	& x_{i, j} + x_{k, l} \leq 1 & \forall (i, j), (k, l) \in \mathcal{R}\ \text{if in conflict} \label{ilp:conflicting-constraints} \\
		& \sum_{i=1}^n x_{i, j} \leq 1 & \forall j=1, \ldots, m \label{ilp:match-of-a-to-b}  \\
		&  y_i + \sum_{j=1}^m x_{i, j} \leq 1 & \forall i=1, \ldots, n  \label{ilp:connection_xij_yj}  \\
		&  \sum_{i=1,\ldots,m \mid A[i]=\sigma} y_i \leq |\mathcal{M}_\sigma| &  \forall \sigma \in \mathcal{M}  \label{ilp:budget_per_symbols} \\
		& x_{i, j} \in  \{0, 1\} & (i, j) \in \mathcal{R}\\
		& y_i \in \{0, 1\} & i=1, \ldots, n
	\end{align}
	
	We say that two matchings $(i, j), (k, l)\in \mathcal{R}$ are \emph{in conflict} iff they are different and $(i \leq k\ \wedge \ j \geq l) \vee (i\ \geq\ k \wedge\ j\ \leq l)$ holds, i.e., they cannot both appear in a common subsequence.
	
	Inequalities~(\ref{ilp:conflicting-constraints}) ensure that the selected matchings are conflict-free and thus form a proper subsequence of both input strings. 
	Constraints~(\ref{ilp:match-of-a-to-b}) enforce that each symbol in $B$ is matched with at most one symbol in $A$. 
	On the other hand, constraints~(\ref{ilp:connection_xij_yj}) ensure that each symbol from $A$, if it is part of the solution, is either matched with a symbol in $\mathcal{M}$ or with at most one symbol from $B$. Constraints~(\ref{ilp:budget_per_symbols}) permit symbols $\sigma \in \mathcal{M}$ to be matched to symbols in $A$ at most $\mathcal{M}_{\sigma}$ times.

	\textbf{Remark}. Before presenting the core section on the design of our algorithm, which explicitly uses solution components, note that each solution of the filled common subsequence problem instance is defined by a set of positions of symbols in $A$ that match appropriate symbols from $\mathcal{M}$. In Example~\ref{ex:ecample_lfcsp}, the selected positions $\{4, 5, 6, 7, 8, 9, 10, 12, 13, 15, 16\}$ are the components of solution $A'$, which is obtained by omitting the symbols from $A$ that appear at these positions. The LCS between $A'$ and $B$ is then determined, yielding a set of non-conflicting matchings. In Example~\ref{ex:ecample_lfcsp}, the set of non-conflicting pairs is given by the set $\{ (1, 1), (2, 2), (3, 4), (14, 5) \}$. Thus, an LFCSP solution $A'$ induces the solution components $\{ \{4, 5, 6, 7, 8, 9, 10, 12, 13, 15, 16\}, \{ (1, 1), (2, 2), (3, 4), (14, 5) \} \}$ forming the resulting common subsequence with the symbols taken from the positions $\{1,\ldots,16\}\setminus\{11, 14\}$ of $A$.

	\section{A New Hybrid ILP Approach for the LFCSP}\label{sec:adapt_cmsa}
	
	The ILP from the last section performs well on small- to medium-sized instances, roughly up to $n = 80$, see~\cite{mincu2018heuristic}. We build on this and set out to develop an ILP-based metaheuristic capable of efficiently solving significantly larger instances. A promising framework for this purpose is the Construct, Merge, Solve, and Adapt (CMSA) approach originally proposed by Blum et al.~\cite{blum2016construct}. Since its introduction, CMSA has achieved state-of-the-art results in a wide range of problem domains, including bioinformatics~\cite{blum2016constructMcsp}, software engineering and software product lines~\cite{ferrer2021cmsa}, planning~\cite{dupin2021matheuristics}, packing and routing~\cite{akbay2024cmsa}, among others. For more information on CMSA and its variants, the reader is referred to the recent survey paper~\cite{blum2025hybrid}. The core idea of CMSA is to iteratively generate promising subproblems that are solved using an external solver, which---in our case---is an ILP solver applied to a model based on the previously presented one. Over successive iterations, the components that form these subproblems are adapted based on their historical contribution to high-quality solutions. This is achieved through an aging mechanism, where promising components are retained for longer in the search process, while less useful components are discarded more quickly. This dynamic adjustment enables the approach to continuously refine the search space and improve solution quality over time.

	In each major iteration, CMSA performs four main steps, as already indicated by the title of the algorithm. The \textit{construction} phase rapidly generates a set of diverse solutions of reasonable quality. 
	The \textit{merge} phase defines a restricted-size sub-instance that considers only the solution components in an initially empty pool together with the components of the newly generated solutions. 
	The \textit{solving} phase solves this sub-instance subject to a time limit. 
	Finally, the \textit{adapt} phase rewards or penalizes solution components and prepares the pool for the next iteration. The commonly used model is the \textit{aging} model: each component that appears in the new incumbent solution is rewarded by resetting its age to zero, while the age of every other component is incremented. Components whose age exceeds a threshold are removed from the pool. 
	
	To improve the robustness of CMSA---particularly its sensitivity to parameter tuning across different instance sizes---several variants have been proposed in the literature. Among them, the \textit{self-adaptive} CMSA (or \textsc{Adapt-Cmsa}) has demonstrated enhanced performance on various problems, including influence maximization in social networks~\cite{akbay2022self} and graph partitioning~\cite{djukanovic2023self}. For a comprehensive overview of CMSA and its variants, we refer to~\cite{blum2024self}. 
	In contrast to the basic CMSA, the self-adaptive variant eliminates the need for a manually parameterized adapt phase.
	Instead, the components of a sub-instance are retained only during the iteration in which they are generated. 
	The algorithm dynamically adjusts key parameters such as the number of generated solutions ($n_\mathrm{a}$) and a similarity threshold ($\alpha_\mathrm{bsf}$) that govern the balance between intensification and diversification. This parameter adaptation is driven by the solution obtained during the solve phase of each iteration. At the end of each iteration, components from the best-so-far solution are passed to the subsequent iteration and, together with newly generated solutions, are used to create a new subproblem. 
	The size of the sub-instance in the next iteration is controlled by the value $\alpha_\mathrm{bsf}$---the larger $\alpha_\mathrm{bsf}$ is, the greater the diversity among the components of the newly generated solutions. This mechanism allows the algorithm to continuously refine its focus based on accumulated search experience, leading to improved scalability and robustness across diverse problem instances.

	\subsection{The Self-Adaptive CMSA}
	
	This section provides details of the self-adaptive CMSA designed to solve the LFCSP. Before presenting these details, recall from Section~\ref{sec:ilp_lcfsp} that a possible solution $s$ (being a subsequence of $A$) to a problem instance $\mathcal{I}=(A, B, \mathcal{M})$ is obtained by determining a set of positions in $A$ matched with $\mathcal{M}$, deleting these positions from $A$ to generate $s$, and evaluating $s$ by identifying matching positions between $s$ and $B$ (obtained by determining the LCS($s$, $B$)). Note that a solution $s$ was previously denoted by $A'$.
	
	\textit{The merge procedure and subproblems}. A core step in the design of a CMSA is to define what the solution components are. Suppose $S$ is a set of diverse solutions to an LFCSP instance $I=(A, B, \mathcal{M})$. For each $s\in S$, we use the following notation:
	\begin{itemize}
		
		\item $\texttt{Match}_{A\mathcal{M}}^{s}=\{ j_1, \ldots, j_r \}$, a set of positions of $A$ whose corresponding symbols are found in $\mathcal{M}$. These positions (resp.~symbols) are those that, when removed from $A$, generate the solution string $s$. 
		\item $\texttt{Match}_{B}^{s}=\{ (p_1^s, q_1^s), \ldots, (p_{u}^s, q_{u}^s)\}$, $0 \leq u \leq \min\{|s|, |B|\}$, a set of mutually non-conflicting matching positions (i.e., pairs) between string $s$ and $B$. This set results from solving the LCS($s$, $B$). Again, two (different) components $(i, j)$ and ($k, l$) are in conflict iff either $i \leq k\  \wedge\  j \geq l$ or $i \geq k\  \wedge\  j \leq l$, following Section~\ref{sec:ilp_lcfsp}. 
		
		
		
	\end{itemize}
	
	In other words, a solution $s$ is completely determined by these two sets of components, i.e., actions that correspond to decision variables \textbf{x} and \textbf{y} following their meaning in the ILP model. \\
	
	A subproblem can be derived from the solution set $S$ by merging the respective sets of matchings:
	$\mathcal{C}(S)=(\texttt{Match}^{S}_{B}, \texttt{Match}_{A\mathcal{M}}^{S})$, where $\texttt{Match}^{S}_{B}:=\bigcup_{s\in S} \texttt{Match}_{B}^{s}$ and $\texttt{Match}_{A\mathcal{M}}^{S}:=\bigcup_{s\in S} \texttt{Match}_{A\mathcal{M}}^{s}$. 
	Let $\mathcal{C}_i = \{ j \mid (i, j) \in \texttt{Match}^{S}_{B} \}$ represents
	all positions in $B$ that are matched (in at least one of the solutions
	from $S$) to the position $i$ of $A$, $i=1,\ldots,n$, and let $\mathcal{C}^j = \{ i \mid (i, j) \in \texttt{Match}^{S}_{B} \}$ represent all positions in $A$ that are matched (in at least one of the solutions from $S$) to the position $j$ of $B$, $j=1,\ldots, m$. 
	
	We can now formulate the ILP model for the subproblem to find a LFCSP solution with the restriction that only solution components from $\mathcal{C}(S)$ are allowed (this can be seen as a task of finding an optimal recombination between the given solution components). Boolean decision variables $x_{i,j}$ are now only defined for $(i, j) \in \texttt{Match}^{S}_{B}$, and variables $y_i$ only for $i \in \texttt{Match}_{A\mathcal{M}}^{S}$. We refer to this subproblem formulation as ILP($\mathcal{C}(S)$), which is given by

	\begin{align}
		\max\ & \sum_{i\in \texttt{{Match}}_{A\mathcal{M}}^{S}} y_i + \sum_{(i, j) \in  \texttt{Match}_{B}^{S}} x_{i, j} \label{constraint:begin} \\
		\mathrm{s.t.\ }& x_{i,j} + x_{k,l} \leq 1 & \forall (i,j), (k,l) \in \texttt{Match}_{B}^{S} \text{ if in conflict}\\  	
		&\sum_{ i \in \mathcal{C}^j  } x_{i, j} \leq 1 & \forall j = 1,\ldots,m \mid \mathcal{C}^j \neq \emptyset \\   
		& y_i + \sum_{j \in \mathcal{C}_i} x_{i, j} \leq 1 &  \forall i \in \texttt{Match}_{A\mathcal{M}}^S \\
		&\sum_{i \in \texttt{Match}_{A\mathcal{M}}^{S} \mid  A[i]=\sigma} y_i \leq |\mathcal{M}_\sigma| & \forall \sigma \in \mathcal{M} \label{constraint:end}\\
		& x_{i,j} \in \{0, 1\} & (i,j) \in \texttt{Match}_{B}^{S} \\
		& y_i \in \{0, 1\} & i \in \texttt{Match}_{A\mathcal{M}}^{S} \label{constraint:y_i_vars_end}
	\end{align}

	\textit{Objective function}. Given solution $s$ and its accompanying component sets $\texttt{Match}_{A\mathcal{M}}^{s}$ and $\texttt{Match}_{B}^{s}$, by  
	\begin{equation}\label{eq:obj_solution_components}
		\mathrm{obj}(s):=|\texttt{Match}^{s}_{A\mathcal{M}}| +  \texttt{Match}^{s}_{B}|
	\end{equation}
	we denote the objective value of $s$, which is in line with Eq.~(\ref{eq:alternative_lfcsp_definition}) and the definition of the problem itself, 
	because $\mathrm{obj}(s)= |\texttt{Match}^s_{A\mathcal{M}}| + |\textrm{LCS}(s, B)|$. 
	

	\textit{The solution construction procedure}.  We basically apply the randomized greedy approach from~\cite{mincu2018heuristic} in the CMSA's construction phase. However, this approach was changed for a stronger randomization and thus diversification that is controlled by a bias parameter.  
	The resulting method is shown in the form of pseudo-code in Algorithm~\ref{alg:alg-generation-adapt-cmsa} and works as follows. 
	
	First, the solution $s_\mathrm{bsf}$---represented by the corresponding solution components---is generated using the randomized greedy approach from Section~\ref{sec:uniform-sampling} (with a setting of $n_{\textrm{rand}}=1$). These components are further copied to $\texttt{Match}_{A\mathcal{M}}^{s_\mathrm{mutate}}$ and $\texttt{Match}_{B}^{s_\mathrm{mutate}}$. In the main loop, the algorithm iterates through the elements (positions) of $\texttt{Match}_{A\mathcal{M}}^{s_\mathrm{bsf}}$, attempting to replace each element with a new random position in $A$ that matches the same symbol not already included in $\texttt{Match}_{A\mathcal{M}}^{s_\mathrm{mutate}}$. If there is no such position, the procedure skips the current iteration, and the algorithm moves to the next one. Otherwise, the current position is replaced with the new one, affecting the set $\texttt{Match}_{A\mathcal{M}}^{s_\mathrm{mutate}}$ (see Line~\ref{alg2:line-mutation}). This replacement is performed with probability of ${\alpha_\mathrm{bsf}}$. 
	In other words, each component from $\texttt{Match}_{A\mathcal{M}}^{s_\mathrm{bsf}}$ faces a replacement under the controlled probability. The replacement candidates are positions in $A$ not yet matched with $\mathcal{M}$. In this way, the number of symbols matched between $A$ and $\mathcal{M}$ remains unchanged. Therefore, {this modification never decreases the solution quality at this construction stage}, while it may introduce substantial changes in the selected components. After generating the set $\texttt{Match}_{A\mathcal{M}}^{s_\mathrm{mutate}}$ as described above, it is left to derive the corresponding components in $\texttt{Match}_{\mathcal{B}}^{s_\mathrm{mutate}}$. This is done by calculating an LCS between $s_{\textrm{mutate}} $ (generated from $\texttt{Match}_{\mathcal{M}}^{s_\mathrm{mutate}}$ by omitting appropriate letters from $A$) and $B$. In this way, both component sets are determined and returned. Let us denote this procedure by \texttt{ModifyComponents}().

	\begin{algorithm}[t!]
		
		\caption{\texttt{ModifyComponents()}}
		\begin{algorithmic}[1]
			\State \textbf{Input:} Solution $s_\mathrm{bsf},$ threshold of mutation $\alpha_\mathrm{bsf}$, an LFCSP instance $I=(A, B, \mathcal{M})$
			\State \textbf{Output:} An mutated solution $s_\mathrm{mutate}$ represented by solution component sets $\texttt{Match}_\mathcal{B}^{s_\mathrm{mutate}}$ and $\texttt{Match}_{A\mathcal{M}}^{s_\mathrm{mutate}}$
			\State $\texttt{Match}_{A\mathcal{M}}^{s_\mathrm{mutate}} \gets \texttt{Match}_{A\mathcal{M}}^{s_\mathrm{bsf}}$
			\For{ $i \in$  $\texttt{Match}_{A\mathcal{M}}^{s_\mathrm{bsf}}$}
			\State $c \gets A[i]$
			\If{  $ \emph{prob} \ni \mathcal{U}([0,1]) {<} \alpha_\mathrm{bsf}$}
			\State $i' \gets$ pick a position $pos$ uniformly so that  $A[pos] = c$ and ${pos} \not  \in \texttt{Match}_{A\mathcal{M}}^{s_\mathrm{mutate}}$
			\If{no such position $i'$}
			\State \textbf{continue} 
			\EndIf
			\State $\texttt{Match}_{A\mathcal{M}}^{s_\mathrm{mutate}} \gets \texttt{Match}_{A\mathcal{M}}^{s_\mathrm{mutate}} \setminus \{ i \} \cup \{i'\}$ // swap positions but keep quality of the  solution w.r.t.\ the number of matches between $A$ and  $\mathcal{M}$  \label{alg2:line-mutation}
			\EndIf
			\EndFor
			\State $A'_{s_\mathrm{mutate}} \gets$ remainder of $A$ after deleting positions $\texttt{Match}_{A\mathcal{M}}^{s_\mathrm{mutate}}$
			\State $\texttt{Match}^{s_\mathrm{mutate}}_{B} \gets $ matchings of  LCS between $A'_{s_\mathrm{mutate}}$ and $B$ w.r.t.~initial positional numbers in $A$ 
			\State \textbf{return} ($\texttt{Match}_{A \mathcal{M}}^{s_\mathrm{mutate}},\texttt{Match}^{s_\mathrm{mutate}}_{B}$)
		\end{algorithmic}	\label{alg:alg-generation-adapt-cmsa}
	\end{algorithm}

	Finally, the complete pseudocode of the self-adaptive \textsc{Cmsa} (labeled \textsc{Adapt-Cmsa}) for the LFCSP is given in Algorithm~\ref{alg:adapt-cmsa-lcfsp}. First, the algorithm initializes a random solution $s_\mathrm{bsf}$ by running the randomized algorithm described in Section~\ref{sec:uniform-sampling} (with $n_\mathrm{rand}=1$). Initially, variables $\alpha_\mathrm{bsf}$ and $n_\mathrm{a}$ are set to $\alpha_\mathrm{lb}$ (an algorithm parameter that requires tuning) and 1, respectively. 
	At each iteration, the following steps are executed until the specified time limit is reached. First, $n_\mathrm{a}$ randomized solutions are generated by executing the \texttt{ModifyComponents}($s_{\mathrm{bsf}}, \alpha_{\mathrm{bsf}}$) procedure. All these solutions are added to the set $S$, which---at the beginning of each iteration---is initialized to contain $s_{\textrm{bsf}}$. If, at some point, a generated solution $s$ is better than $s_\mathrm{bsf}$ ($\mathrm{obj}(s) > \mathrm{obj}(s_\mathrm{bsf})$), it becomes the best solution so far ($s_{\mathrm{bsf}}=s$). After the solutions are generated, the sub-instance $\mathcal{C}(S) = (\texttt{Match}^{S}_{B}, \texttt{Match}_{A\mathcal{M}}^{S})$ based on the solutions in $S$ (as explained above) is constructed, inducing the restricted ILP model~(\ref{constraint:begin})--(\ref{constraint:y_i_vars_end}), labeled $\text{ILP}(C(S))$.

	\begin{algorithm}[H]
		\caption{ The \textsc{Adapt-Cmsa} algorithm for the LFCSP}
		\label{alg:adapt-cmsa-lcfsp}
		\begin{algorithmic}[1]
			\State \textbf{Input:} An instance  $I=(A, B, \mathcal{M})$ of the LFCSP, alphabet $\Sigma$,  $t_\mathrm{prop}, t_\mathrm{ILP}, \alpha_\mathrm{lb}, \alpha_\mathrm{ub}, \alpha_\mathrm{red}$ 
			\State \textbf{Output:} A feasible solution $s_\mathrm{bsf}$
			\State $s_\mathrm{bsf} \gets \texttt{UniformSampling}(I)$ // via the randomized algorithm ($n_\mathrm{rand}=1$) from Section~\ref{sec:uniform-sampling}
			\State $\alpha_\mathrm{bsf} \gets \alpha_\mathrm{lb}, n_\mathrm{a} \gets 1$
			\While{\emph{time limit is not exceeded}}
			
			\State $S \gets \{s_{\textrm{bsf}}\}$ 
			\For{$l=1,\ldots,n_a$}
			\State $s \gets$ \texttt{ModifyComponents}($s_\mathrm{bsf}, \alpha_\mathrm{bsf}$)
			\State $S \gets S \cup \{s\}$
			\State \textbf{if} $\mathrm{obj}(s) > \mathrm{obj}(s_{\textrm{bsf}})$ \textbf{then} $s_{\textrm{bsf}} \gets s$ \textbf{end if}
			\EndFor
			\State Generate sub-instance $\mathcal{C}(S) = (\texttt{Match}^{S}_{B}, \texttt{Match}_{A\mathcal{M}}^{S}) $   
			
			\State $(s', t_\mathrm{solve}) \gets \texttt{ExactSolver}(\text{ILP}(C(S)), t_\mathrm{ILP})$ // solve the ILP (\ref{constraint:begin})--(\ref{constraint:y_i_vars_end}) 
			\If{$t_\mathrm{solve} < t_{prop} \cdot t_\mathrm{ILP} \wedge \alpha_\mathrm{bsf} > \alpha_\mathrm{lb} $}
			\State $\alpha_\mathrm{bsf} \gets \alpha_\mathrm{bsf} - \alpha_\mathrm{red}$
			\EndIf
			\If{$\mathrm{obj}(s') > \mathrm{obj}(s_{\textrm{bsf}})$}
			\State $s_\mathrm{bsf} \gets s'$ 
			\State $n_\mathrm{a} \gets 1$
			\State $\alpha_\mathrm{bsf} \gets \alpha_\mathrm{ub}$ \label{line:adapt_cmsa_line20}
			
			\ElsIf{$\mathrm{obj}(s') < \mathrm{obj}(s_{\textrm{bsf}})$} // weaker solution, move closer to $s_\mathrm{bsf}$
			
			\If{$n_\mathrm{a}=1$}
			\State $\alpha_\mathrm{bsf} \gets \min(\alpha_\mathrm{bsf} + \frac{\alpha_\mathrm{red}}{10}, \alpha_\mathrm{ub})$
			\Else
			\State $n_\mathrm{a} \gets 1$
			\State $\alpha_\mathrm{bsf} \gets \alpha_\mathrm{ub} $ \label{line:adapt_cmsa_line26}
			\EndIf
			\Else // equally good solutions
			\State $n_\mathrm{a} \gets n_\mathrm{a} + 1$
			\EndIf 
			\EndWhile
			\State \textbf{return} $s_{\textrm{bsf}}$ 
		\end{algorithmic}
	\end{algorithm}
	
	After this, the solving phase takes place by applying \texttt{ExactSolver}() to the model $\text{ILP}(C(S))$. The time limit assigned to the solving phase is parameterized by the parameter $t_\mathrm{ILP}$. The procedure returns a pair $(s', t_\mathrm{solve})$ where $s'$ is the solution produced---as induced by the values assigned to the binary decision variables of the model---and $t_\mathrm{solve}$ represents the time used for solving. Note that an additional termination criterion is employed in the solving phase: We noticed that the time spent by the solver proving optimality was generally much longer than the time required to find high-quality solutions. Consequently, we immediately stop the solving process whenever the ILP solver finds a solution better than $s_{\mathrm{bsf}}$. If $t_\mathrm{solve}$ is relatively short, this generally indicates that the sub-instance $\text{ILP}(C(S))$ was easy to solve, which allows more diverse solutions to be generated, usually resulting in larger and potentially more promising sub-instances. This is achieved by reducing $\alpha_\mathrm{bsf}$ linearly by a value of $\alpha_\mathrm{red}$. Subsequently, the solution $s'$ produced by the ILP solver is compared to $s_\mathrm{bsf}$. If $s'$ is a new incumbent, $s_\mathrm{bsf}$ is updated to $s'$, $n_\mathrm{a}$ is reset to one, and $\alpha_\mathrm{bsf}$ to $\alpha_\mathrm{ub}$. Otherwise, if the objective value of $s'$ is worse than $s_\mathrm{bsf}$ and $n_\mathrm{a}=1$, then $\alpha_\mathrm{bsf}$ is slightly increased, provided that it does not exceed $\alpha_\mathrm{ub}$. This results in the intensification of the search around the solution $s_\mathrm{bsf}$. If $n_\mathrm{a}>1$, $n_\mathrm{a}$ is reset to one, and $\alpha_\mathrm{bsf}$ is set to the maximum allowed value $\alpha_\mathrm{ub}$. If the objective values of $s'$ and $s_\mathrm{bsf}$ are equal, the value of $n_\mathrm{a}$ is incremented by one to increase the diversification of the search process. Once the time limit is exceeded, the algorithm returns the best found solution $s_\mathrm{bsf}$. 
	
	{Compared with the original \textsc{Adapt-Cmsa} algorithm proposed in~\cite{akbay2022self}, our variant modifies the adaptation mechanism in Algorithm~\ref{alg:adapt-cmsa-lcfsp}. In Lines~\ref{line:adapt_cmsa_line20} and~\ref{line:adapt_cmsa_line26}, the diversification parameter is reset according to $\alpha_{\mathrm{bsf}} \gets \alpha_{\mathrm{ub}}$. This reset deliberately increases diversification after relevant search events and proved beneficial for the final solution quality. Beyond this modification, the method introduces non-trivial LFCS-specific design choices, including the construction and modification of solutions together with the adaptive generation of promising ILP subproblems. We also tested the basic \textsc{Cmsa} variant, but its performance was not sufficiently robust for the LFCSP because it was highly sensitive to the parameter configuration, consistent with the observations reported in~\cite{akbay2022self}.}
	
	\section{Experimental Evaluation}\label{sec:experimental_section}
	
	In this section, a thorough experimental comparison of four literature approaches for solving the LFCSP and our proposed approach is conducted. More precisely, our \textsc{Adapt-Cmsa} is compared against the following approaches:
	\begin{itemize}
		\item An approximation algorithm, described in Section~\ref{sec:approx-lcfsp}, which is labeled \textsc{Approx} here;
		\item The randomized greedy approach, described in Section~\ref{sec:uniform-sampling}, labelled \textsc{RandSample};
		\item The local search method, described {in Section~\ref{sec:ls-based-approach}}, labelled \textsc{Ls}. Two settings of this algorithm are executed depending on the value of parameter $l$:
		\begin{itemize}
			\item the setting with {$l=2$}, therefore labelled \textsc{Ls}2;
			\item the setting with {$l=4$}, therefore labelled \textsc{Ls}4;
		\end{itemize}
		\item ILP method, described in Section~\ref{sec:ilp_lcfsp}, labelled \textsc{Ilp}. 
	\end{itemize}

	All algorithms are implemented in $C$++ and compiled under GCC 11.4.0. All experiments are executed in single-threaded mode on a server equipped with an Intel Xeon E5-2640 v4 processor with 2.40 GHz, allocating 32 GB of memory per run. {For all literature-based approaches, the algorithms are run to completion, except for \textsc{Ilp} and our \textsc{Adapt-Cmsa}, both of which are subject to the same time limit of 600 CPU seconds in all experiments, including the synthetic benchmarks in Section~\ref{sec:experimental_section} and the real-world audio case study.} The \textsc{Ilp} model is solved using the general-purpose solver \textsc{Cplex}, under version 12.8. For \textsc{Adapt-Cmsa}, a custom \texttt{IncumbentCallbackI} is implemented to enforce an additional termination condition. {Within each \textsc{Adapt-Cmsa} subproblem, the \textsc{Cplex} run is stopped as soon as it finds an incumbent solution that improves the current best solution. This criterion is used only inside the iterative \textsc{Adapt-Cmsa} framework: it avoids spending a disproportionate amount of time proving subproblem optimality after an improving solution has already been obtained, thereby allowing the algorithm to frequently construct and explore additional subproblems. 
	}
	
	\subsection{Problem Instances}
	
	Two sets of synthetic instances are generated. The first set, denoted as \textsc{Small}, consists of the instances generated by the procedure as introduced in~\cite{mincu2018heuristic}. For each combination of sequence length $n \in \{16, 32, 48, 64, 80 \}$ and alphabet size $|\Sigma| \in \left\{\frac{n}{8}, \frac{n}{4}, \frac{n}{2}\right\}$, 100 instances are generated using the instance generator as described in the next paragraph. In contrast to the small- and medium-sized instances, a second benchmark set, \textsc{Large}, consists of significantly larger instances---approximately an order of magnitude larger in size. Using the same instance generator, 10 instances are generated for each combination of $n \in \{200, 400, 500, 600, 800, 1000\}$ and $|\Sigma| \in \{4, 20\}$. These alphabet sizes are chosen to reflect more realistic scenarios: RNA sequences are those with an alphabet of size 4, while protein sequences are modelled with an alphabet of size 20. In total, 1,620 problem instances are generated, spanning 27 distinct instance groups. \\
	\textit{Instance Generator}. The instance generator, originally proposed in~\cite{mincu2018heuristic}, operates as follows. Each instance $I=(A,B,\mathcal{M})$ is created by first generating a random string $A$ over alphabet $\Sigma$. The sequence $B$ is then derived from $A$ by iterating through its symbols and applying a modification operation with a probability of 50\%. The modifications are selected uniformly between three possible operations: duplication, deletion, or substitution with another symbol. After applying these operations, sequence $B$ is randomly partitioned into segments of maximum length $n$/8, and more than 30\% of these segments are randomly discarded. The remaining segments are concatenated to form the final string $B$, while the discarded segments are stored in $\mathcal{M}$. {This established generator has been adopted to preserve comparability with previous studies. We used it to introduce a new set of large-scale synthetic problem instances. Although the generator is not intended to reproduce every structural property of real biological or audio data, it produces substantial variation in sequence length, alphabet size, and the number of feasible matchings, which directly affect the size and difficulty of the resulting ILP models. The complementary audio-based instances in Section~\ref{sec:lfcsp-case-study} provide an additional evaluation on non-synthetic data and therefore a structurally different dataset.}
	
	Note that the synthetic problem instances and source codes can be downloaded from GitHub \url{https://github.com/markodjukanovic90/LFCSP_public/tree/main}.

	\subsection{Parameters Tuning}
	
	Note that the algorithms \textsc{Approx}, \textsc{Ls}2, and \textsc{Ls}4 are all parameter-free. For \textsc{Ilp}, the \textsc{Cplex} solver is run with default settings. For the algorithm \textsc{Rand-Simple}, $n_{\text{rand}}$ is set to 10,000, which means that 10,000 random solutions are generated; this setting is chosen in accordance with~\cite{mincu2018heuristic}.
	
	To tune the parameters of \textsc{Adapt-Cmsa}, the \texttt{irace} tool~\cite{lopez2016irace} was applied separately to each of the two benchmark sets. For the \textsc{Small} benchmark set, five instances--indexed with $\{0, 1, 2, 3, 4\}$---were selected for each of the 15 groups, resulting in a total of 75 instances selected for parameter tuning. For the \textsc{Large} benchmark set, one instance was selected from each of the 12 groups, yielding a total of 12 instances for \texttt{irace} tuning process. 
	
	The parameters and their respective tuning domains are listed in Table~\ref{tab:irace-params}.

	\begin{table}[ht]
		\centering
		\begin{tabular}{|l|l|}
			\hline
			\textbf{Parameter} & \textbf{Domain}  \\
			\hline \hline
			${\alpha_\mathrm{lb}}$     & \{0.1, 0.2, 0.25, 0.3, 0.5, 0.7\} \\
			${\alpha_\mathrm{ub}}$     & \{0.9, 0.95, 1.0\} \\
			${t_\mathrm{ILP}}$ & \{5, 10, 20, 30, 60\} \\
			${t_{prop}}$      & \{0.1, 0.2, 0.5, 0.7\} \\
			${\alpha_\mathrm{red}}$   & \{0.05, 0.1, 0.2, 0.3\} \\
			\hline
		\end{tabular}
		\caption{Irace tuning parameters and their candidate values}
		\label{tab:irace-params}
	\end{table}
	
	The budget for \texttt{irace} at each tuning scenario is set to 5000 runs, while each run is assigned a time limit of 600 CPU seconds.
	
	{We performed separate tuning runs for the \textsc{Small} and \textsc{Large} benchmark sets because a preliminary unified tuning run on the combined training set led to a substantial deterioration in \textsc{Adapt-Cmsa}'s performance. The two sets differ by approximately one order of magnitude in instance size and induce markedly different ILP subproblem sizes and search dynamics as shown in Section~\ref{sec:ilp_problm_size_study}. For this reason, a single configuration lacks the balance between intensification and diversification across both instance types.}
	
	\textsc{Irace} identified the following configuration as the best for the \textsc{Small} benchmark set:
	$$\alpha_\mathrm{lb}=0.2, \alpha_\mathrm{ub}=1.0, t_\mathrm{ILP}=10, t_{prop}=0.7, \alpha_\mathrm{red}= 0.05$$  
	Irace has produced the following configuration of the \textsc{Adapt-Cmsa} as the best on benchmark set \textsc{Large}:
	
	$$\alpha_\mathrm{lb}=0.25, \alpha_\mathrm{ub}=0.95, t_\mathrm{ILP}=30, t_{prop}=0.2, \alpha_\mathrm{red}= 0.1$$
	
	
	
	\subsection{Numerical Results}\label{sec:numerical-results}

	This section presents numerical results of all algorithms evaluated on the two benchmark sets: \textsc{Small} and \textsc{Large}.
	Table~\ref{tab:small-numerical-results} shows the results for the benchmark set \textsc{Small}. The table is organized as follows. The first block provides a description of each instance (group), defined by two parameters---the instance size $n$ and the alphabet size $|\Sigma|$. The remaining columns correspond to six different algorithms, each grouped into its own block. For each algorithm, two metrics are reported: the average solution quality ($\overline{obj}$) and the average runtime ($\overline{t}[s]$, in seconds), computed over the 100 problem instances in each group. For the \textsc{Adapt-Cmsa} algorithm, instead of reporting the average total runtime, we provide $\overline{t}_\mathrm{best}[s]$, indicating the average time required to reach the best solution across 100 instances per given group. (Note that \textsc{Adapt-Cmsa} is executed with a fixed runtime limit of 600 CPU seconds per instance.) Additionally, the block corresponding to the \textsc{Ilp} results, includes an extra column $\overline{gap}$[\%], representing the average relative optimality gap reported by \textsc{Cplex} upon termination.
	
	The following conclusions are drawn from these results:
	
	\begin{itemize}
		\item The \textsc{Ilp} approach successfully solves all problem instances to optimality, which is in line with the findings already reported in the existing literature~\cite{mincu2018heuristic}.
		
		\item For the smallest instances with $n=16$, the heuristic methods \textsc{Rand-Sample}, \textsc{Ilp}, and \textsc{Adapt-Cmsa} consistently achieved optimal solutions across all 300 problem instances. In contrast, \textsc{Approx}, \textsc{Ls}2, and \textsc{LS}4, while producing near-optimal results, did not reach the optimum in every case, even for these small instances.
		
		\item For slightly larger instances with $n=32$, \textsc{Adapt-Cmsa} was the only heuristic approach capable of consistently finding optimal solutions. The second-best performer for this instance group was \textsc{Rand-Sample}, though it fell short of optimality on several occasions.
		
		\item  On the instances with $n=48$, \textsc{Adapt-Cmsa} produced average solutions that closely matched the average optimal values. Notably, it found proven optimal solutions for 298 out of the 300 instances—substantially outperforming the other heuristic methods, which struggled to reach such consistency.
		
		\item For medium-sized instances with $n \in \{64, 80\}$, \textsc{Adapt-Cmsa} achieved solution quality corresponding to 99.9\% of the known optimal values. The remaining heuristic methods demonstrate considerably weaker performance in comparison.
		
		\item The average runtimes of all approaches (and the time to reach best solution in the case of \textsc{Adapt-Cmsa}) were relatively short, indicating that the existing literature instances are not computationally challenging. This highlights a strong need for introducing significantly larger instances to better assess the performance boundaries and limitations of the proposed algorithms.
		
	\end{itemize}

\begin{table}[H]
\caption{Results on the benchmark set \textsc{Small} (the instances from literature). } \label{tab:small-numerical-results}
\centering
\scalebox{0.7}{
\begin{tabular}{|c|c|l l|l l|l l|l l|l l  l|l l|}
\hline \hline
\multicolumn{2}{|c|}{Inst.} & \multicolumn{2}{c|}{\textsc{Rand-Sample}}  &
\multicolumn{2}{|c|}{\textsc{Approx}} & \multicolumn{2}{c|}{\textsc{Ls}2} & 
\multicolumn{2}{|c|}{\textsc{Ls}4} & \multicolumn{3}{c|}{\textsc{Ilp}} &  \multicolumn{2}{c|}{\textsc{Adapt-cmsa}} \\ 
\cmidrule(lr){1-2} \cmidrule(lr){3-4}
\cmidrule(lr){5-6} \cmidrule(lr){7-8}
\cmidrule(lr){9-10} \cmidrule(llr){11-13} \cmidrule(lr){14-15}
$n$ & $|\Sigma|$ & $\overline{obj}$ & $\overline{t}[s]$ &  $\overline{obj}$ &  $\overline{t}[s]$ & $\overline{obj}$ &  $\overline{t}[s]$ & $\overline{obj}$ &  $\overline{t}[s]$ & $\overline{obj}$ & $\overline{gap}[\%]$ & $\overline{t}[s]$ & $\overline{obj}$  & $\overline{t}_{best}[s]$ \\
\hline \hline
16 & 2 & \textbf{14.35} & 0.01 & 14.30 & 1.39 & 14.32 & 0.00 & 14.27 & 0.00 & \textbf{14.35} & 0.00 & 0.04 & \textbf{14.35} & 0.02 \\
 & 4 & \textbf{13.67} & 0.01 & 13.59 & 1.31 & 13.61 & 0.00 & 13.46 & 0.00 & \textbf{13.67} & 0.00 & 0.01 & \textbf{13.67} & 0.01 \\
 & 8 & \textbf{12.66} & 0.01 & 12.59 & 1.31 & 12.58 & 0.00 & 12.55 & 0.00 & \textbf{12.66} & 0.00 & 0.01 & \textbf{12.66} & 0.01 \\ \hline

32 & 16 & \textbf{24.57} & 0.02 & 24.45 & 2.52 & 24.47 & 0.00 & 24.36 & 0.01 & \textbf{24.57} & 0.00 & 0.06 & \textbf{24.57} & 0.02 \\
 & 4 & 28.00 & 0.02 & 27.97 & 2.41 & 27.97 & 0.00 & 27.70 & 0.01 & \textbf{28.20} & 0.00 & 0.12 & \textbf{28.20} & 1.31 \\
 & 8 & 26.13 & 0.02 & 25.87 & 2.42 & 25.88 & 0.00 & 25.68 & 0.01 & \textbf{26.15} & 0.00 & 0.07 &  \textbf{26.15} & 0.06 \\ \hline

48 & 12 & 38.33 & 0.03 & 38.34 & 3.57 & 38.39 & 0.01 & 38.20 & 0.02 & \textbf{38.62} & 0.00 & 0.46 & 38.60 & 0.22 \\
 & 24 & 36.36 & 0.03 & 36.11 & 3.58 & 36.18 & 0.01 & 36.10 & 0.02 & \textbf{36.39} & 0.00 & 0.33 &  \textbf{36.39} & 0.04 \\
 & 6 & 41.13 & 0.03 & 41.40 & 3.50 & 41.29 & 0.01 & 40.94 & 0.03 & \textbf{41.78} & 0.00 & 0.63 & \textbf{41.78} & 1.85 \\ \hline

64 & 16 & 51.21 & 0.04 & 51.35 & 4.97 & 51.47 & 0.03 & 51.12 & 0.06 & \textbf{51.80} & 0.00 & 1.25 & 51.74 & 0.22\\
 & 32 & 48.27 & 0.04 & 48.22 & 5.08 & 48.28 & 0.02 & 48.09 & 0.04 & \textbf{48.45} & 0.00 & 1.18 & 48.42 & 0.07 \\ 
 & 8 & 53.19 & 0.04 & 54.14 & 4.80 & 53.94 & 0.04 & 53.80 & 0.08 & \textbf{54.74} & 0.00 & 1.77 & 54.73 & 1.87 \\ \hline

80 & 10 & 65.62 & 0.05 & 67.11 & 6.35 & 66.99 & 0.07 & 66.54 & 0.16 & \textbf{67.87} & 0.00 & 4.63 & 67.84 & 2.85 \\
 & 20 & 62.46 & 0.05 & 63.30 & 6.45 & 63.31 & 0.06 & 62.89 & 0.13 & \textbf{63.77} & 0.00 & 3.56 & 63.69 & 0.59 \\
 & 40 & 59.97 & 0.05 & 59.98 & 6.52 & 60.02 & 0.05 & 59.84 & 0.09 & \textbf{60.32} & 0.00 & 3.17 & 60.28 & 0.06 \\
\hline \hline\end{tabular}}
\end{table}

	Numerical results of the six approaches on benchmark set \textsc{Large} are reported in Table~\ref{tab:large-comparison}. This table is organized as follows. The first block refers to instance characteristics, provided by two columns ($n$ and $|\Sigma|$). The next six columns report the average results of the six methods, \textsc{Rand-Simple}, \textsc{Approx}, \textsc{Ls}2, \textsc{Ls}4, \textsc{Ilp}, and \textsc{Adapt-Cmsa}, in this order. The average is obtained over the 10 instances in each instance group (rows).
	
	\begin{table}[H]
  \caption{Results on the benchmark set \textsc{Large}. }
   \label{tab:large-comparison}
   \centering
  \scalebox{0.8}{
\begin{tabular}{|c l|l|l|l|l|l|l|}
\toprule
$n$ & $|\Sigma|$ & \textsc{Rand-Sample} & \textsc{Approx} & \textsc{Ls}2 & \textsc{Lp}4 & \textsc{Ilp} & \textsc{Adapt-cmsa} \\
\hline \hline
200 & 20 & 157.10 & 98.40 & 165.00 & 164.70 & \textbf{166.60}$^*$ & \textbf{166.60} \\
 & 4 & 169.50 & 170.00 & 181.50 & 178.60 & 119.80 & \textbf{184.60} \\ \hline
400 & 20 & 305.80 & 229.10 & 339.20 & 337.00 & 273.60 & \textbf{341.50} \\
 & 4 & 341.60 & 343.40 & 370.80 & 362.60 & 0.00 & \textbf{379.10} \\ \hline
500 & 20 & 380.90 & 312.40 & 426.70 & 422.40 & 151.50 & \textbf{430.00} \\
 & 4 & 424.50 & 441.20 & 461.10 & 452.10 & 0.00 & \textbf{471.10} \\ \hline
600 & 20 & 471.30 & 365.00 & 521.30 & 516.70 & 150.90 & \textbf{526.50} \\
 & 4 & 511.50 & 546.30 & 557.50 & 544.70 & 0.00 &\textbf{565.00} \\ \hline
800 & 20 & 641.70 & 519.70 & 705.40 & 700.80 & 0.00 & \textbf{709.30} \\
 & 4 & 663.80 & 705.30 & 732.80 & 714.70 & 0.00 & \textbf{739.70} \\ \hline
1000 & 20 & 784.00 & 720.80 & 885.50 & 882.00 & 0.00 & \textbf{889.80} \\
 & 4 & 828.70 & 898.90 & 918.10 & 894.00 & 0.00 & \textbf{925.00} \\
\hline \hline
\end{tabular}}
\end{table}
	
	

\begin{figure}[H]
		\centering
		\includegraphics[width=340pt,height=200pt]{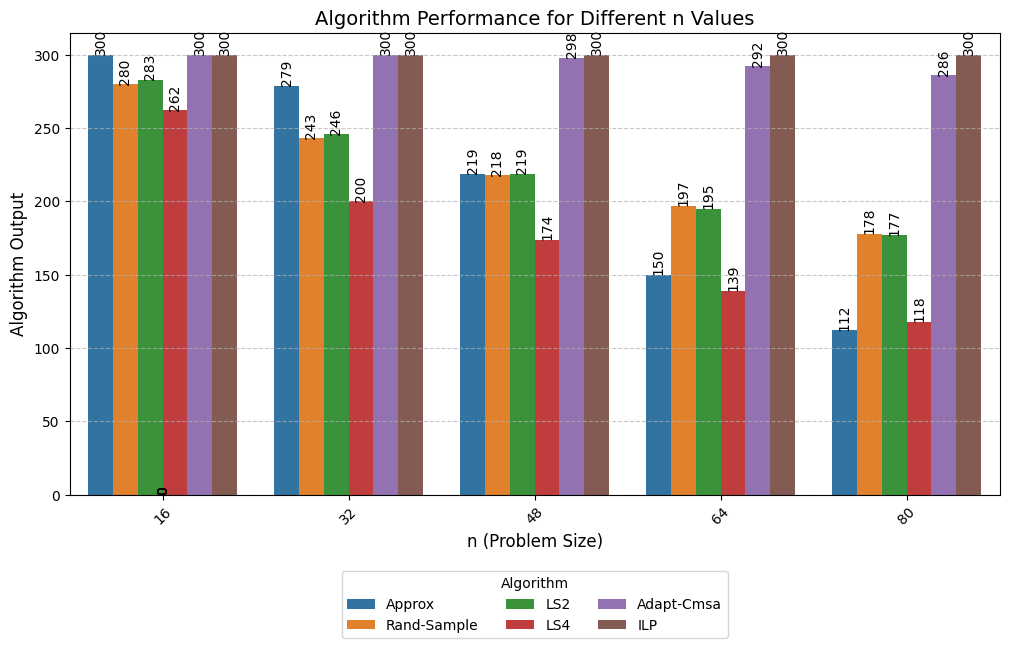}
		\caption{{Number of instances in the \textsc{Small} dataset for which each approach reached a known optimal solution.}} \label{fig:num-matched-optima}
	\end{figure}

	The following conclusions are drawn from the results on the benchmark set \textsc{Large}:
	
	\begin{itemize}
		\item The exact approach \textsc{Ilp} fails to deliver solutions of reasonable quality, except for the smallest case with $n = 200$ and $|\Sigma| = 20$. For the instances with $|\Sigma| = 4$ and $n > 200$, it was unable to produce any non-trivial solution, returning only the (trivial) empty solution after exceeding the time limit.  These results indicate that the performance of the \textsc{Ilp} approach degrades rapidly with increasing instance size, indicating its limited practical applicability for large-scale problems.
		
		\item Among the heuristic methods, the most effective in terms of final solution quality is \textsc{Adapt-Cmsa},  achieving the best results across all 12 instance groups.
		
		\item The average solution quality achieved by \textsc{Adapt-Cmsa} across all 120 instances is 527.38—significantly higher than the second-best method, \textsc{Ls}2, which achieved an average of 522.07. This is followed by \textsc{Ls}4 with an average solution quality of 514.19. All other approaches perform noticeably worse in terms of overall average solution quality.
		
		\item Figure~\ref{fig:num-matched-optima} shows the number of instances solved optimally by each algorithm on the dataset \textsc{Small}. Instances are grouped by sequence length $n$ (300 instances per group) and plotted along the x-axis. For $n \in \{16, 32\}$, both \textsc{Ilp} and \textsc{Adapt-Cmsa} are able {to find optimal solutions.} For $n = 48$, \textsc{Adapt-Cmsa} failed to find the optimal solution in only two cases. For the medium-sized instances with $n \in \{64, 80\}$, \textsc{Adapt-Cmsa} successfully finds  (known) optimal solutions in 292 and 286 cases (out of 300), respectively. In comparison, the next-best approach, \textsc{Rand-Sample}, returns {solutions that match the (known) optimal solutions} in just 197 and 178 cases (out of 300), respectively.
	\end{itemize}

	\subsection{Evolution of the Restricted Subproblem Size}\label{sec:ilp_problm_size_study}
	\label{subsec:subproblem-size-evolution}
\begin{figure}[!h]
		\centering
		\begin{subfigure}[t]{0.485\textwidth}
			\centering
			\includegraphics[width=\linewidth]{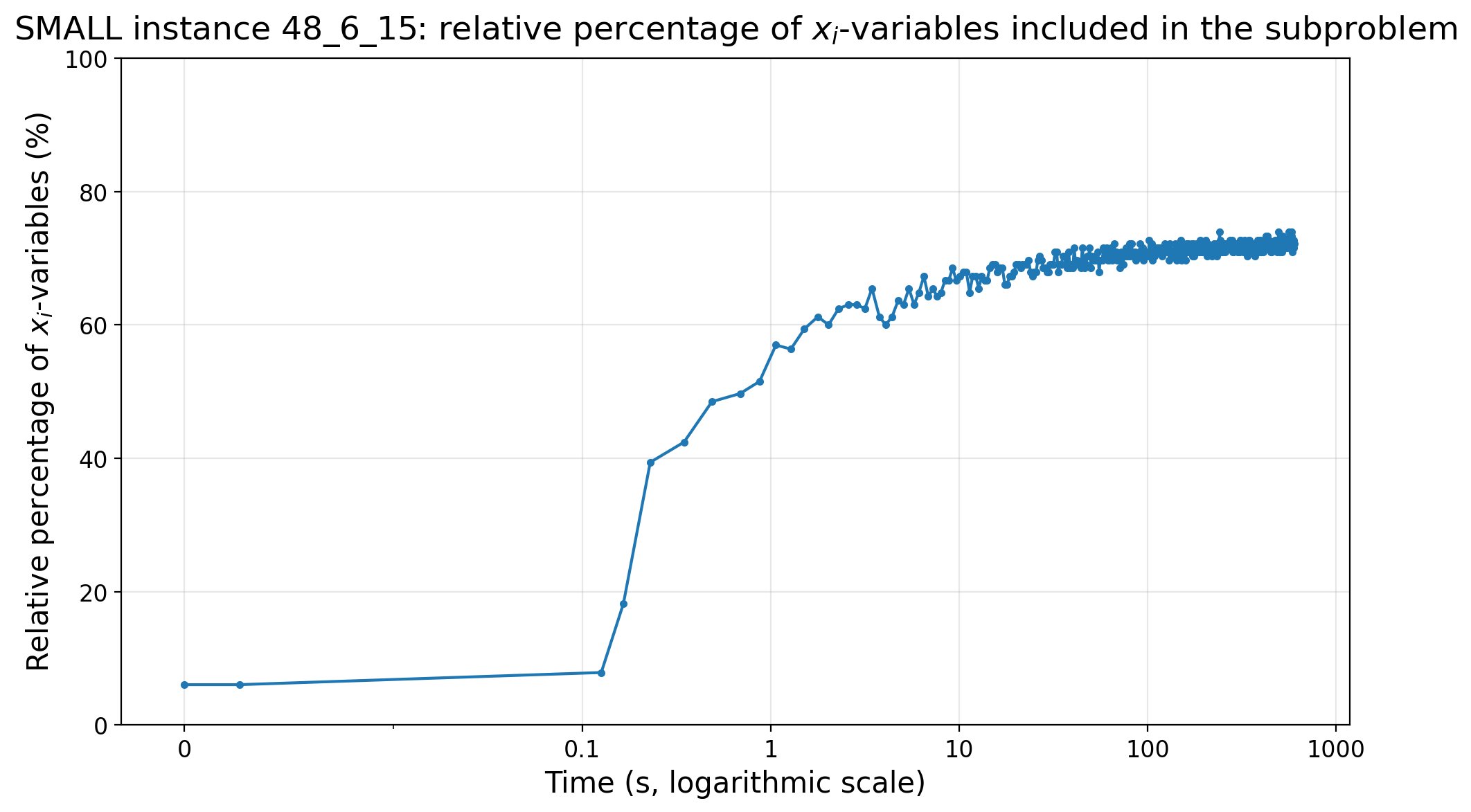}
			\caption{Relative pct. of $\textbf{x}$-variables for the \textsc{Small} instance $n=48, |\Sigma|=6, \textit{index}=15$.  The number of $\textbf{x}$-variables in the complete ILP model equals 165.}
			\label{fig:small-x-vars}
		\end{subfigure}
		\hfill
		\begin{subfigure}[t]{0.485\textwidth}
			\centering
			\includegraphics[width=\linewidth]{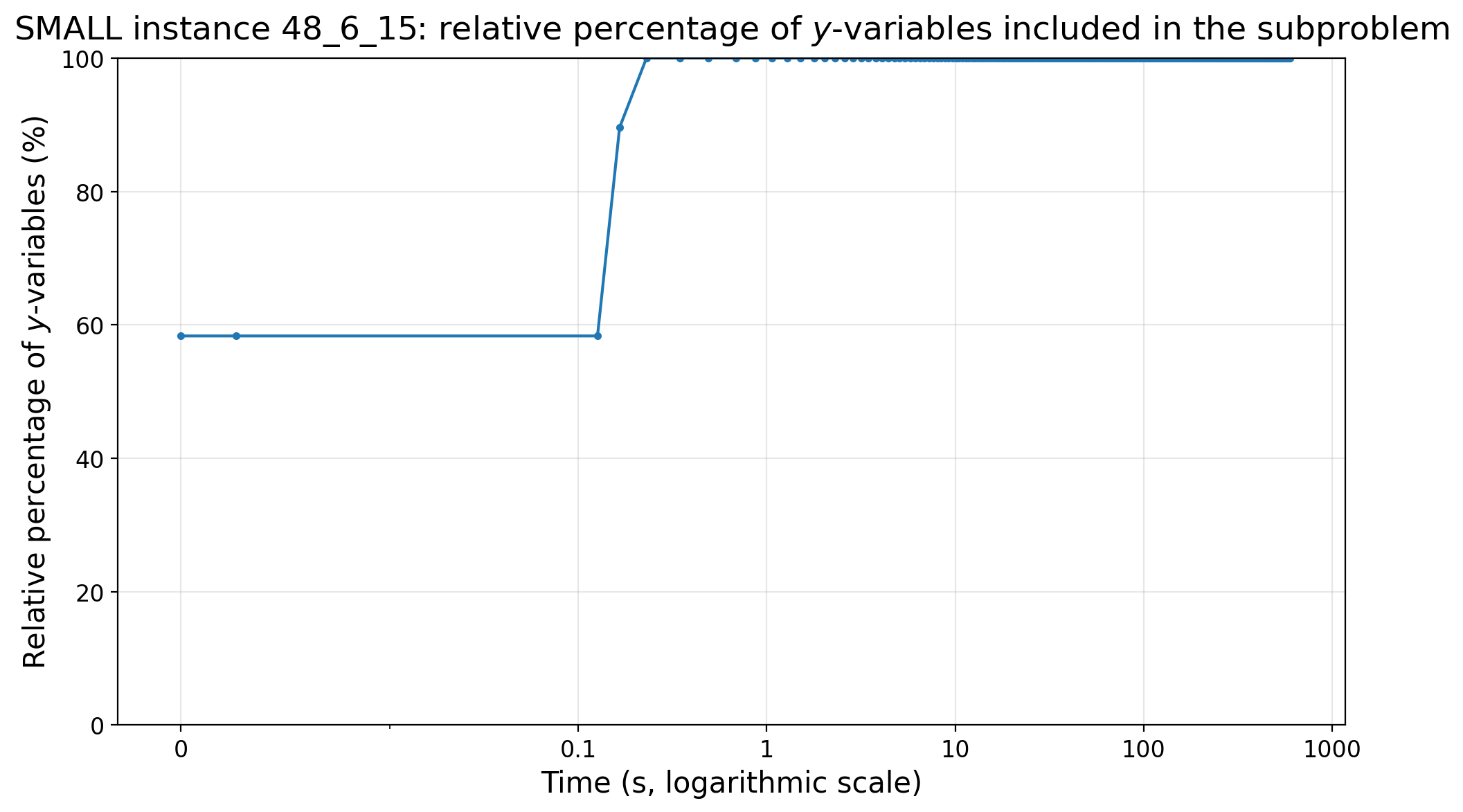}
			\caption{Relative pct. of $\textbf{y}$-variables for the \textsc{Small} instance \(n=48, |\Sigma|=6, \textit{index}=15\). The number of $\textbf{y}$-variables in the complete ILP model equals 48.}
			\label{fig:small-y-vars}
		\end{subfigure}
		
		\vspace{0.2em}
		
		\begin{subfigure}[t]{0.485\textwidth}
			\centering
			\includegraphics[width=\linewidth]{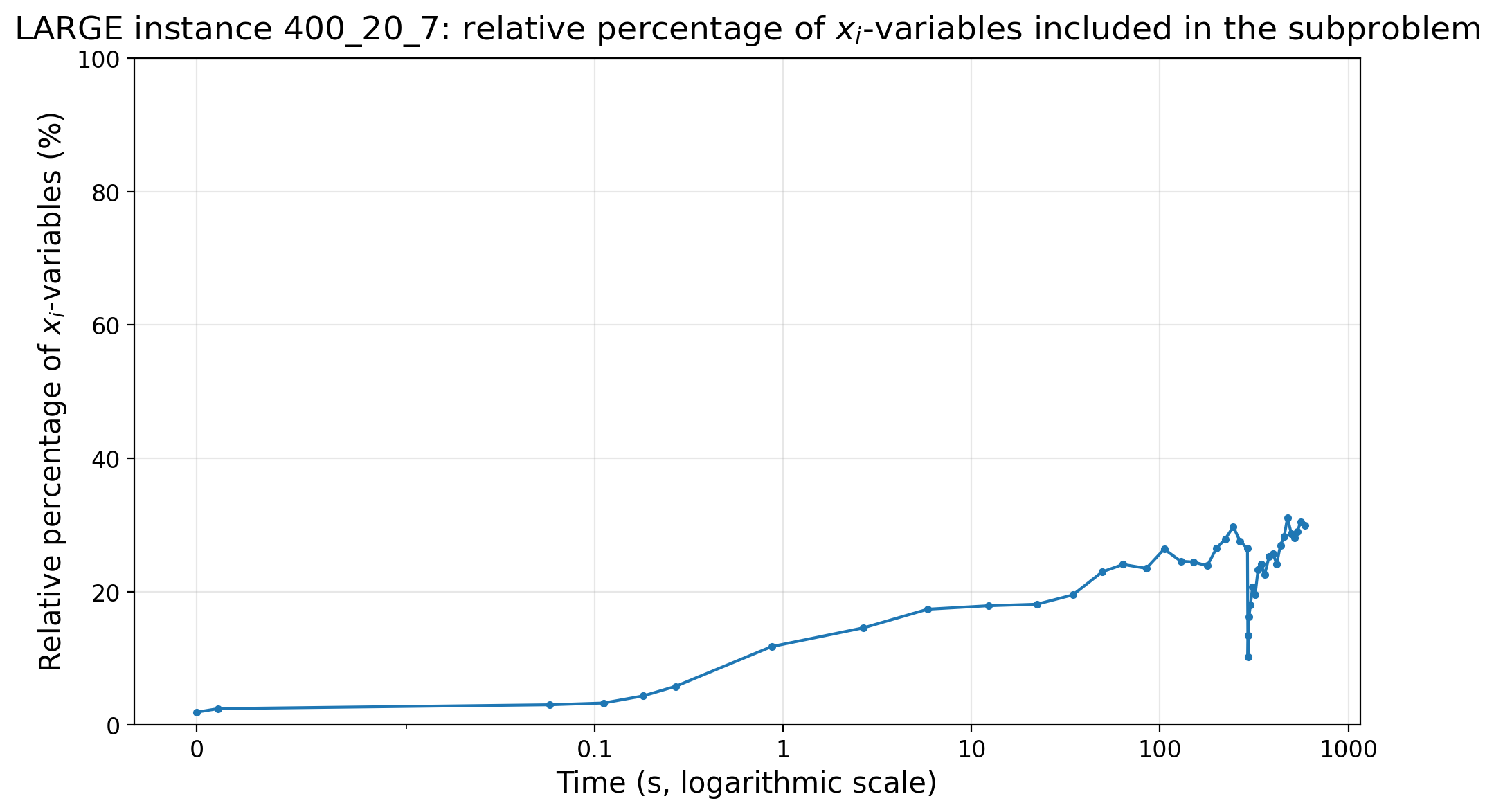}
			\caption{Number of $\textbf{x}$-variables for the \textsc{Large} instance \(n=400, |\Sigma|=20, \textit{index}=7\).  The number of $\textbf{x}$-variables in the complete ILP model equals 3086. }
			\label{fig:large-x-vars}
		\end{subfigure}
		\hfill
		\begin{subfigure}[t]{0.485\textwidth}
			\centering
			\includegraphics[width=\linewidth]{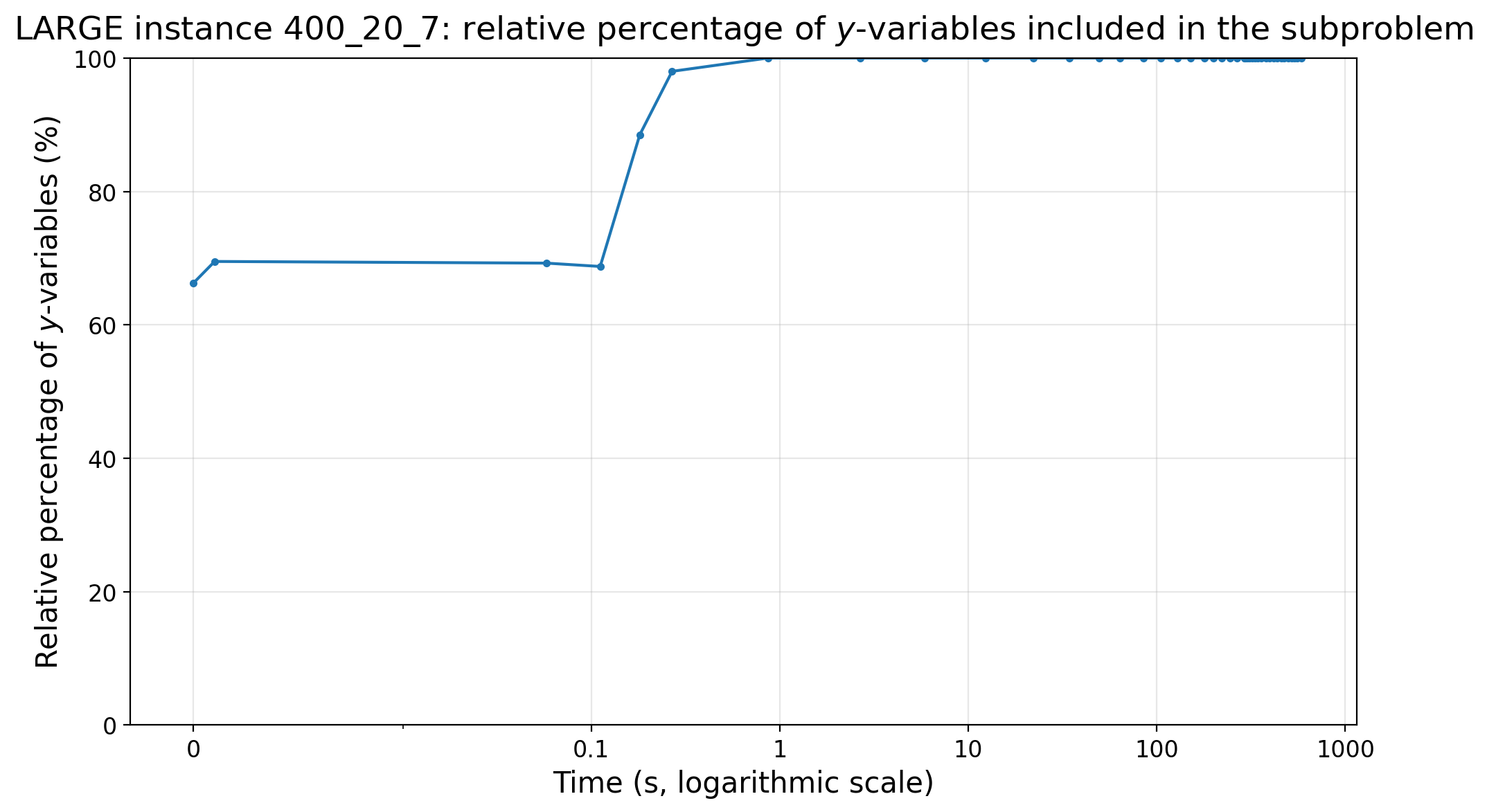}
			\caption{Relative pct. of $\textbf{y}$-variables for the \textsc{Large} instance \(n=400, |\Sigma|=20, \textit{index}=7\). The number of $\textbf{y}$-variables in the complete ILP model equals 400.}
			\label{fig:large-y-vars}
		\end{subfigure}
		
		\caption{Evolution of the restricted ILP subproblem size during the execution of
			\textsc{Adapt-Cmsa}.}
		\label{fig:subproblem-size-evolution}
	\end{figure}

	{
		To provide further insight into the behavior of \textsc{Adapt-Cmsa}, we analyze how the size of the restricted ILP subproblem evolves during the execution. In particular, we record the number of $\textbf{y}$-variables and $\textbf{x}$-variables included in each restricted subproblem, along with the elapsed computational time required to construct the corresponding subproblem. Remember that the $\textbf{y}$-variables represent the sequence positions retained in the restricted model, whereas the $\textbf{x}$-variables represent the compatible pairwise relations induced by the selected positions of input sequences $A$ and $B$. Consequently, the number of $\textbf{x}$-variables depends not only on the number of retained $\textbf{y}$-variables, but also on the compatibility structure of the considered instance.\\ 
		Figure~\ref{fig:subproblem-size-evolution} reports this evolution for  the instance $n=48, |\Sigma|=6, \textit{index}=15$  from benchmark set \textsc{Small} and the instance $n=400, |\Sigma|=20, \textit{index}=7$ from benchmark set \textsc{Large}. The plots on the left show the relative percentage of \textbf{x}-variables included in the restricted subproblem w.r.t.\ the number of all possible $\textbf{x}$-variables (equals to $|\mathcal{R}|$), while those on the right show the relative percentage of $\textbf{y}$-variables w.r.t.\ the number of all  possible \textbf{y}-variables (equals to $n$). In all plots, the horizontal axis represents elapsed computational time in seconds.}

{
	For the instance from \textsc{Small}, the number of $\textbf{y}$-variables increases rapidly from its initial value and reaches the complete set of 48 variables within the first few seconds. It then remains constant for the rest of the execution. In contrast, the number of $\textbf{x}$-variables grows more gradually. After an initial rapid increase, it stabilizes at approximately 75\% of the total number of available variables, with relatively small fluctuations. This indicates that the construction mechanism quickly incorporates all position variables for this comparatively small instance, while the set of compatible pairwise relations continues to evolve toward a relatively large proportion of the complete model.\\
	For the instance from \textsc{Large} ($n=400$, $|\Sigma|=20$, $\textit{index}=7$), the number of $\textbf{y}$-variables likewise reaches its maximum value of 400 near the beginning of the run and remains constant thereafter. The evolution of the number of $\textbf{x}$-variables is considerably less stable, as expected for a large-scale problem instance. Their number increases from approximately 2\% (about 60 variables) to more than 30\% (about 900 variables) of the total available set, but a pronounced reduction occurs around the middle of the execution. Following this reduction, the number of $\textbf{x}$-variables increases again. This behavior is consistent with the adaptive reconstruction of model components guided by the value of the self-adaptive parameter $\alpha_{\textrm{bsf}}$.\\
	Overall, the results show that the two variable types exhibit different dynamics. The $\textbf{y}$-variables tend to reach a high proportion of the complete set relatively early in the search, particularly for small-sized instances. In contrast, the set of $\textbf{x}$-variables remains dynamic throughout the execution. Since the number of pairwise variables is determined by the compatibility relations among the retained sequence positions, relatively small changes in the constructed solution components may lead to more pronounced variations in the number of $\textbf{x}$-variables.\\
	The results also illustrate the role of the adaptive model restriction mechanism. Rather than solving a single static ILP formulation throughout its execution, \textsc{Adapt-Cmsa} repeatedly modifies the set of model components according to the generated solutions, with this process being influenced by $\alpha_{\textrm{bsf}}$. The restricted model may therefore expand when new components are introduced, for instance when the value of $\alpha_{\textrm{bsf}}$ increases. This behavior is especially pronounced for the instance from \textsc{Large}, where controlling the number of pairwise variables is particularly important because the complete ILP formulation may contain a substantially larger number of such variables.\\
	The corresponding evolution of the restricted ILP subproblem sizes for instances from the \textsc{Song} benchmark set is reported in~\ref{app:evolution_subproblem_songs}.}

	\subsection{Statistical Evaluation}\label{sec:statistical-analysis-results}
	
	In this section, we perform a statistical analysis on the results reported in Section~\ref{sec:numerical-results}. To this end, we apply a post-hoc analysis based on the non-parametric one-sided Wilcoxon signed-rank test~\cite{woolson2005wilcoxon}. For each pair among the six algorithms, a separate test is conducted to evaluate the null hypothesis ($H_0$): the two algorithms perform equally well, against the alternative hypothesis ($H_1$): the first algorithm performs significantly better than the second algorithm.
	
	\begin{figure}[H]
		\centering
		\includegraphics*[width=320pt,height=200pt]{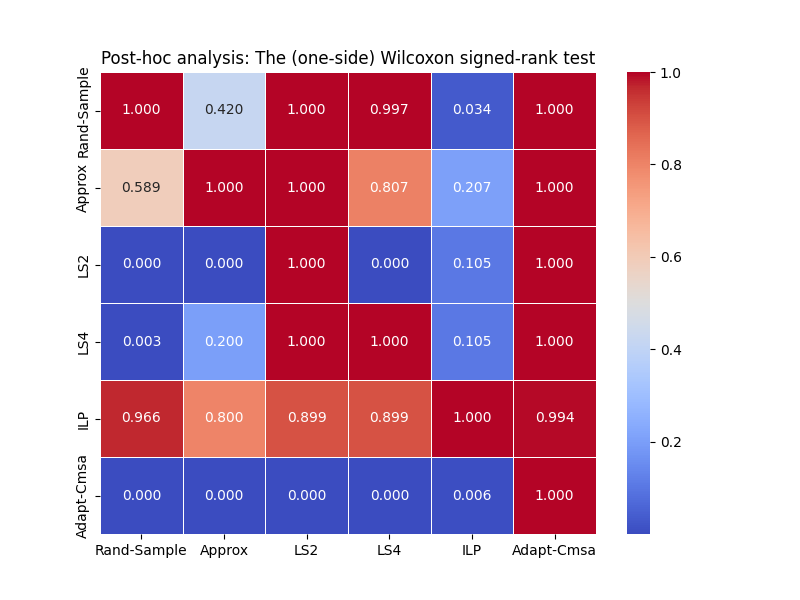}
		\caption{Pairwise post-hoc statistical analysis: the results of two benchmark sets combined.}
		\label{fig:stats_results}
	\end{figure}  
	
	This post-hoc statistical analysis, based on the combined results from both benchmark sets, is illustrated in Figure~\ref{fig:stats_results}. Each algorithm is positioned along both the horizontal (x) and vertical $y$ axes. The square at the intersection of two algorithms is filled with the corresponding $p$-value obtained from testing the null hypothesis ($H_0$): the algorithm on the y-axis performs equally well as the algorithm on the x-axis, against the alternative hypothesis ($H_1$): the algorithm on the y-axis performs significantly better (i.e., yields larger results) than the one on the x-axis. For example, the bottom-left square shows the test result whether \textsc{Adapt-Cmsa} performs equally well as \textsc{Rand-Sample}. The test yields a $p$-value of $0.000 < 0.05$, indicating that $H_0$ can be rejected at the significance level $\alpha = 0.05$ in favor of the alternative hypothesis. Thus, it can be concluded that \textsc{Adapt-Cmsa} performs significantly better than \textsc{Rand-Sample}. 
	
	The following conclusions are drawn from this analysis:
	\begin{itemize}
		\item The results of \textsc{Adapt-Cmsa} are statistically significantly better than all other competitor approaches (see the lowest row).
		\item The \textsc{Ls}2 approach is significantly better from \textsc{Random-Sample}, \textsc{Approx}, and \textsc{Ls}4; however, \textsc{Ls}4 better than \textsc{Rand-Sample} but there is insufficient evidence that it performs better than the \textsc{Approx} approach.   
	\end{itemize}

	\section{A Case Study: Audio Fingerprinting}
	\label{sec:lfcsp-case-study}
	Audio fingerprinting is a widely used technique for identifying songs based on short, potentially noisy audio excerpts. A canonical system is Shazam~\cite{wang2003shazam}, which extracts prominent spectral peaks from a song and hashes pairs of them into compact fingerprints. These are stored in a large-scale database and enable fast, noise-tolerant matching even for low-quality recordings.
	Other robust alternatives include Panako~\cite{joren2014panako} and Audfprint~\cite{ellis20142014}, among others. These methods generally rely on exact or approximate matching of robust audio features and assume at least partial preservation of time-local structure.
	
	When audio is recorded using low-end microphones or under challenging conditions such as high noise levels, lossy compression, or variable latency, much of the structure typically relied upon by fingerprinting systems may be lost or severely degraded. While established systems like Shazam and its variants are highly effective in many practical settings, they are not designed to operate with extremely fragmented or partially observed data. In such scenarios, alternative formulations like the LFCSP approach offer a different perspective. Rather than competing directly with conventional fingerprinting, LFCSP represents one of several directions that could be further explored to address highly degraded or incomplete audio inputs.

	To compare songs meaningfully, a variety of signal representations have been proposed:
	\begin{itemize}
		\item Spectrograms: A basic but powerful representation computed using the Short-Time Fourier Transform (STFT), showing how amplitudes are distributed across frequencies over time.
		
		\item Mel-Spectrograms: Spectrograms warped onto the Mel scale to better reflect human perception of pitch and loudness.
		
		\item MFCCs: Mel-Frequency Cepstral Coefficients summarize the spectral envelope of a sound and are popular in speech and speaker recognition.
		
		\item Learned Embeddings: Models such as OpenL3~\cite{cramer2019look}, VGGish~\cite{hershey2017cnn} produce high-level embeddings from audio using deep learning trained on large datasets. These embeddings capture abstract musical or semantic properties.
		
		\item Energy Profiles: A simplified yet informative representation used in this work is the energy profile, where consecutive parts of audio (seconds in our study) are converted into discretized energy levels (e.g., one of 10 bins from $\{0,.., 9\}$). This results in a one-dimensional symbolic sequence that captures loudness and dynamics, and is robust to time stretching and pitch shifts.
	\end{itemize}
	
	LFCSP in the context of audio fingerprinting is defined over three components:
	\begin{enumerate}
		\item Sequence $A$ (candidate song),
		\item Sequence $B$ (noisy audio query),
		\item Multiset $\mathcal{M}$ of additional symbols derived from $B$ but without known temporal positioning.
	\end{enumerate}
	
	In this application, the LFCSP solution is used to identify a song from a known database of high-quality recordings, i.e., a set of candidate sequences $A_s = \{A_1, A_2, \ldots \}$ based on a noisy or incomplete energy profile $B$ (evidence), supplemented with additional unordered energy information $\mathcal{M}$, which acts as a histogram of symbols with unknown temporal positions.
	
	We construct audio fingerprinting LFCSP instances\footnote{Instances and generator are available at: \url{https://github.com/markodjukanovic90/LFCSP_public/tree/main}} based on energy profiles of 8 well-known songs:
	\begin{itemize}
		\item $A_1$: Adele – Rolling in the Deep
		\item $A_2$: Daft Punk – One More Time
		\item $A_3$: Fugees – Killing Me Softly (\textbf{reference song})
		\item $A_4$: Madonna – Like a Prayer
		\item $A_5$: Queen – Bohemian Rhapsody
		\item $A_6$: Radiohead – Creep
		\item $A_7$: Rihanna \& Calvin Harris – We Found Love
		\item $A_8$: Roberta Flack – Killing Me Softly
	\end{itemize}
	
	These songs simulate a reference database (which, in a real-world scenario, would be much larger), with each represented as a complete sequence $A_i$, obtained by discretizing per-second energy levels into 10 energy-level bins (referring to an alphabet $\Sigma = \{0, \dots, 9\}$).
	In a practical system, identification would typically proceed in two phases. The first phase would quickly eliminate the vast majority of songs from a large-scale database (potentially millions of entries) by comparing coarse global features, such as energy histograms. Only a small set of candidates---those whose histograms closely resemble that of the query---would proceed to the second, more precise phase in which the LFCSP algorithm is applied to this refined set of similar candidates.
	
	Accordingly, the eight selected songs were chosen from a larger pool as those with the most similar energy histograms to the reference song $A_3$. To simulate a query, we constructed a sequence $B$ by first computing the common histogram $\mathcal{M}$ across all sequences $A_i$, resulting in the following 70-symbol multiset $\mathcal{M}$:
	
	\begin{verbatim}
		0000000000000000111111111111112222222222333333333344444444555555555666
	\end{verbatim}
	
	We then removed $\mathcal{M}$ from the reference song $A_3$, producing the initial query sequence $B$ for the simplest experimental condition, in which the parameter $rem$ was set to 0.
	
	To simulate more challenging conditions, we further reduced the length of $B$ by randomly removing additional symbols. This removal was governed by the parameter $rem$, which was varied across values $\{0.2, 0.4, 0.6, 0.8\}$, modeling increasing levels of information loss due to poor recording quality or environmental noise.
	
	It is important to note that $\mathcal{M}$ represents content inferred from $B$, but without known temporal alignment. As such, it serves as a statistically reliable fragment of the original song. Since typical background noise does not systematically distort energy levels in one direction, $\mathcal{M}$ remains a meaningful prior in the presence of random degradation.
	
	The LFCSP algorithm is applied to each candidate song $A_i$, $i \in \{1,\ldots,8\}$, together with the noisy query $B$, with the multiset $\mathcal{M}$ used to extend $B$ in the matching process.
	The candidate $A_i$ with the highest LFCSP score is selected as the predicted match. Note that the results of our \textsc{Adapt-Cmsa} for each problem instance are reported as the average over 10 independent runs of the algorithm on the {given} instance. {We use the same algorithm settings reported in Table~\ref{tab:irace-params} also in this case study.}
	
	The results for $rem=0$ in Table~\ref{tab:small-songs-rem0.0} show that all algorithms except \textsc{Ilp} correctly identified \emph{Fugees} as the target song, in most cases with a substantial margin. \textsc{Ilp}, however, consistently reached the time limit across all instances, making its results difficult to interpret. It is likely that, with more time, \textsc{Ilp} would produce more meaningful differences among the candidate solutions.

\begin{table}[H]
	\caption{Results on the benchmark set \textsc{Songs}, $rem=0.0$. } \label{tab:small-songs-rem0.0}
	\centering
	\scalebox{0.7}{
		\begin{tabular}{|l|c|l l|l l|l l|l l|l l  |l l|}
			\hline \hline
			\multicolumn{2}{|c|}{Inst.} & \multicolumn{2}{c|}{\textsc{Rand-Sample}}  &
			\multicolumn{2}{|c|}{\textsc{Approx}} & \multicolumn{2}{c|}{\textsc{Ls}2} & 
			\multicolumn{2}{|c|}{\textsc{Ls}4} & \multicolumn{2}{c|}{\textsc{Ilp}} &  \multicolumn{2}{c|}{\textsc{Adapt-cmsa}} \\ 	
			\cmidrule(lr){1-2} \cmidrule(lr){3-4}
			\cmidrule(lr){5-6} \cmidrule(lr){7-8}
			\cmidrule(lr){9-10} \cmidrule(lr){11-12} \cmidrule(lr){13-14}
			name & rem & $\overline{obj}$ & $\overline{t}[s]$ &  $\overline{obj}$ &  $\overline{t}[s]$ & $\overline{obj}$ &  $\overline{t}[s]$ & $\overline{obj}$ &  $\overline{t}[s]$ & $\overline{obj}$ &   $\overline{t}[s]$ & $\overline{obj}$  & $\overline{t}_{best}[s]$ \\
			\hline \hline
			
Adele & 0.00 & 136 & 2.10 & 93 & 0.00 & 137 & 1.92 & 137 & 4.11 & 108 & 602.66 & 137 & 0.23 \\
DaftPunk & 0.00 & 179 & 2.51 & 130 & 0.00 & 185 & 2.21 & 184 & 5.21 & \textbf{127} & 610.42 & 185 & 15.87 \\
\underline{Fugees} & 0.00 & \textbf{240} & 2.28 & \textbf{254} & 0.00 & \textbf{248} & 1.82 & \textbf{247} & 3.83 & 124 & 607.42 & \textbf{254} & 0.54 \\
Madonna & 0.00 & 158 & 2.57 & 127 & 0.00 & 157 & 1.67 & 157 & 3.28 & 0 & 619.74 & 159 & 3.93 \\
Queen & 0.00 & 162 & 2.67 & 126 & 0.00 & 162 & 2.48 & 162 & 5.08 & 0 & 778.99 & 162 & 0.04 \\
Radiohead & 0.00 & 134 & 2.02 & 90 & 0.00 & 134 & 0.99 & 134 & 2.10 & 119 & 606.42 & 134 & 0.13 \\
Rihanna & 0.00 & 146 & 2.18 & 98 & 0.00 & 146 & 2.88 & 146 & 7.29 & 118 & 606.53 & 146 & 0.05 \\
Roberta Flack & 0.00 & 164 & 2.20 & 130 & 0.00 & 162 & 1.39 & 161 & 3.07 & 125 & 609.66 & 165 & 0.16 \\
\hline \hline\end{tabular}}
\end{table}

	A similar pattern is observed for $rem=0.2$ in Table~\ref{tab:small-songs-rem0.2}, although the difference in favor of \emph{Fugees} becomes smaller. This behavior is expected, as the amount of information retained in the sequence $B$ is reduced.

\begin{table}[H]
	\caption{Results on the benchmark set \textsc{Songs}, $rem=0.2$. } \label{tab:small-songs-rem0.2}
	\centering
	\scalebox{0.7}{
		\begin{tabular}{|l|c|l l|l l|l l|l l|l l  |l l|}
			\hline \hline
			\multicolumn{2}{|c|}{Inst.} & \multicolumn{2}{c|}{\textsc{Rand-Sample}}  &
			\multicolumn{2}{|c|}{\textsc{Approx}} & \multicolumn{2}{c|}{\textsc{Ls}2} & 
			\multicolumn{2}{|c|}{\textsc{Ls}4} & \multicolumn{2}{c|}{\textsc{Ilp}} &  \multicolumn{2}{c|}{\textsc{Adapt-cmsa}} \\ 	
			\cmidrule(lr){1-2} \cmidrule(lr){3-4}
			\cmidrule(lr){5-6} \cmidrule(lr){7-8}
			\cmidrule(lr){9-10} \cmidrule(lr){11-12} \cmidrule(lr){13-14}
			name & rem & $\overline{obj}$ & $\overline{t}[s]$ &  $\overline{obj}$ &  $\overline{t}[s]$ & $\overline{obj}$ &  $\overline{t}[s]$ & $\overline{obj}$ &  $\overline{t}[s]$ & $\overline{obj}$ &   $\overline{t}[s]$ & $\overline{obj}$  & $\overline{t}_{best}[s]$ \\
			\hline \hline

Adele & 0.20 & 131 & 1.79 & 90 & 0.00 & 132 & 1.58 & 132 & 3.84 & 130 & 601.64 & 132 & 5.86 \\
DaftPunk & 0.20 & 165 & 2.21 & 113 & 0.00 & 169 & 1.93 & 169 & 4.43 & \textbf{138} & 606.60 & 168 & 476.92 \\
\underline{Fugees} & 0.20 & \textbf{211} & 1.99 & \textbf{190} & 0.00 & \textbf{217} & 1.38 & \textbf{214} & 2.91 & 133 & 605.02 & \textbf{218} & 0.42 \\
Madonna & 0.20 & 148 & 2.22 & 115 & 0.00 & 147 & 1.32 & 147 & 3.21 & 114 & 608.23 & 148 & 1.15 \\
Queen & 0.20 & 151 & 2.34 & 109 & 0.00 & 151 & 2.21 & 150 & 4.76 & 0 & 610.99 & 151 & 0.85 \\
Radiohead & 0.20 & 125 & 1.68 & 78 & 0.00 & 126 & 0.87 & 125 & 1.76 & 111 & 604.33 & 126 & 0.12 \\
Rihanna & 0.20 & 135 & 2.07 & 87 & 0.00 & 135 & 2.33 & 135 & 5.77 & 116 & 604.98 & 135 & 2.88 \\
Roberta Flack & 0.20 & 153 & 1.90 & 119 & 0.00 & 151 & 1.15 & 149 & 2.37 & 134 & 606.59 & 153 & 0.08 \\
\hline \hline
\end{tabular}}
\end{table}

	A continued reduction in margin is evident in Table~\ref{tab:small-songs-rem0.4} for $rem=0.4$. Notably, the \textsc{Ilp} solver was able to find the proven optimal solution for the \emph{Fugees} candidate, resulting in a correct identification of the song. As shown in the table, only \textsc{Adapt-Cmsa}, in addition to \textsc{Ilp}, succeeded in finding the optimal solution in this setting.

\begin{table}[H]
	\caption{Results on the benchmark set \textsc{Songs}, $rem=0.4$. } \label{tab:small-songs-rem0.4}
	\centering
	\scalebox{0.7}{
		\begin{tabular}{|l|c|l l|l l|l l|l l|l l  |l l|}
			\hline \hline
			\multicolumn{2}{|c|}{Inst.} & \multicolumn{2}{c|}{\textsc{Rand-Sample}}  &
			\multicolumn{2}{|c|}{\textsc{Approx}} & \multicolumn{2}{c|}{\textsc{Ls}2} & 
			\multicolumn{2}{|c|}{\textsc{Ls}4} & \multicolumn{2}{c|}{\textsc{Ilp}} &  \multicolumn{2}{c|}{\textsc{Adapt-cmsa}} \\ 	
			\cmidrule(lr){1-2} \cmidrule(lr){3-4}
			\cmidrule(lr){5-6} \cmidrule(lr){7-8}
			\cmidrule(lr){9-10} \cmidrule(lr){11-12} \cmidrule(lr){13-14}
			name & rem & $\overline{obj}$ & $\overline{t}[s]$ &  $\overline{obj}$ &  $\overline{t}[s]$ & $\overline{obj}$ &  $\overline{t}[s]$ & $\overline{obj}$ &  $\overline{t}[s]$ & $\overline{obj}$ &   $\overline{t}[s]$ & $\overline{obj}$  & $\overline{t}_{best}[s]$ \\
			\hline \hline
Adele & 0.40 & 122 & 1.44 & 82 & 0.00 & 123 & 1.35 & 122 & 3.08 & 123 & 56.80 & 123 & 7.54 \\
DaftPunk & 0.40 & 153 & 1.91 & 101 & 0.00 & 155 & 1.65 & 155 & 3.86 & 122 & 603.80 & 155 & 0.24 \\
\underline{Fugees} & 0.40 & \textbf{180} & 1.69 &  \textbf{155} & 0.00 &  \textbf{180} & 1.13 &  \textbf{180} & 2.35 &  \textbf{181} & 161.55 & \textbf{181} & 0.41 \\
Madonna & 0.40 & 133 & 2.04 & 101 & 0.00 & 133 & 1.10 & 133 & 3.17 & 119 & 604.75 & 133 & 0.21 \\
Queen & 0.40 & 140 & 2.19 & 100 & 0.00 & 139 & 1.88 & 140 & 3.92 & 114 & 606.74 & 142 & 0.49 \\
Radiohead & 0.40 & 114 & 1.33 & 67 & 0.00 & 114 & 0.70 & 113 & 1.55 & 114 & 410.90 & 114 & 0.06 \\
Rihanna & 0.40 & 123 & 1.61 & 73 & 0.00 & 123 & 1.85 & 123 & 4.38 & 100 & 602.64 & 123 & 0.04 \\
Roberta Flack & 0.40 & 141 & 1.72 & 109 & 0.00 & 140 & 1.08 & 138 & 1.81 & 112 & 604.24 & 142 & 0.12 \\
\hline \hline
\end{tabular}}
\end{table}

	As shown in Table~\ref{tab:small-songs-rem0.6}, the trend of decreasing margin persists for $rem=0.6$. At this level of removal, the problem instance becomes notably easier, so the reduced length of $B$ enables four out of six algorithms to find the optimal solution.

\begin{table}[H]
	\caption{Results on the benchmark set \textsc{Songs}, $rem=0.6$. } \label{tab:small-songs-rem0.6}
	\centering
	\scalebox{0.7}{
		\begin{tabular}{|l|c|l l|l l|l l|l l|l l  |l l|}
			\hline \hline
			\multicolumn{2}{|c|}{Inst.} & \multicolumn{2}{c|}{\textsc{Rand-Sample}}  &
			\multicolumn{2}{|c|}{\textsc{Approx}} & \multicolumn{2}{c|}{\textsc{Ls}2} & 
			\multicolumn{2}{|c|}{\textsc{Ls}4} & \multicolumn{2}{c|}{\textsc{Ilp}} &  \multicolumn{2}{c|}{\textsc{Adapt-cmsa}} \\ 	
			\cmidrule(lr){1-2} \cmidrule(lr){3-4}
			\cmidrule(lr){5-6} \cmidrule(lr){7-8}
			\cmidrule(lr){9-10} \cmidrule(lr){11-12} \cmidrule(lr){13-14}
			name & rem & $\overline{obj}$ & $\overline{t}[s]$ &  $\overline{obj}$ &  $\overline{t}[s]$ & $\overline{obj}$ &  $\overline{t}[s]$ & $\overline{obj}$ &  $\overline{t}[s]$ & $\overline{obj}$ &   $\overline{t}[s]$ & $\overline{obj}$  & $\overline{t}_{best}[s]$ \\
			\hline \hline
			
Adele & 0.60 & 113 & 1.22 & 73 & 0.00 & 113 & 0.99 & 113 & 2.44 & 113 & 13.43 & 113 & 5.95 \\
DaftPunk & 0.60 & 134 & 1.62 & 82 & 0.00 & 134 & 1.26 & 134 & 3.22 & 134 & 132.78 & 134 & 0.31 \\
\underline{Fugees} & 0.60 & \textbf{144} & 1.33 & \textbf{98} & 0.00 & \textbf{143} & 0.73 & \textbf{144} & 1.61 & \textbf{144} & 47.76 & \textbf{144} & 0.05 \\
Madonna & 0.60 & 117 & 1.57 & 83 & 0.00 & 117 & 0.90 & 116 & 1.94 & 117 & 63.60 & 117 & 0.12 \\
Queen & 0.60 & 122 & 1.72 & 80 & 0.00 & 120 & 1.40 & 121 & 3.30 & 122 & 198.48 & 122 & 0.08 \\
Radiohead & 0.60 & 103 & 1.12 & 60 & 0.00 & 102 & 0.56 & 103 & 1.23 & 103 & 19.76 & 103 & 0.06 \\
Rihanna & 0.60 & 108 & 1.37 & 56 & 0.00 & 108 & 1.52 & 108 & 3.59 & 108 & 106.15 & 108 & 0.02 \\
Roberta Flack & 0.60 & 124 & 1.36 & 92 & 0.00 & 123 & 0.76 & 120 & 1.33 & 124 & 174.27 & 124 & 3.49 \\
\hline \hline
\end{tabular}}
\end{table}

	Finally, for the highest reduction level, Table~\ref{tab:small-songs-rem0.8} shows that the instance becomes trivial to solve, where all algorithms except \textsc{Approx} successfully identified the optimal solution.

\begin{table}[H]
	\caption{Results on the benchmark set \textsc{Songs}, $rem=0.8$. } \label{tab:small-songs-rem0.8}
	\centering
	\scalebox{0.7}{
		\begin{tabular}{|l|c|l l|l l|l l|l l|l l  |l l|}
			\hline \hline
			\multicolumn{2}{|c|}{Inst.} & \multicolumn{2}{c|}{\textsc{Rand-Sample}}  &
			\multicolumn{2}{|c|}{\textsc{Approx}} & \multicolumn{2}{c|}{\textsc{Ls}2} & 
			\multicolumn{2}{|c|}{\textsc{Ls}4} & \multicolumn{2}{c|}{\textsc{Ilp}} &  \multicolumn{2}{c|}{\textsc{Adapt-cmsa}} \\ 	
			\cmidrule(lr){1-2} \cmidrule(lr){3-4}
			\cmidrule(lr){5-6} \cmidrule(lr){7-8}
			\cmidrule(lr){9-10} \cmidrule(lr){11-12} \cmidrule(lr){13-14}
			name & rem & $\overline{obj}$ & $\overline{t}[s]$ &  $\overline{obj}$ &  $\overline{t}[s]$ & $\overline{obj}$ &  $\overline{t}[s]$ & $\overline{obj}$ &  $\overline{t}[s]$ & $\overline{obj}$ &   $\overline{t}[s]$ & $\overline{obj}$  & $\overline{t}_{best}[s]$ \\
			\hline \hline

Adele & 0.80 & 95 & 1.03 & 42 & 0.00 & 95 & 0.72 & 95 & 1.70 & 95 & 2.12 & 95 & 0.00 \\
DaftPunk & 0.80 & 103 & 1.28 & 55 & 0.00 & 103 & 1.02 & 103 & 2.35 & 103 & 9.21 & 103 & 0.02 \\
\underline{Fugees} & 0.80 & \textbf{107} & 1.09 & \textbf{60} & 0.00 & \textbf{107} & 0.55 & \textbf{107} & 1.15 & \textbf{107} & 7.33 & \textbf{107} & 0.01 \\
Madonna & 0.80 & 99 & 1.31 & 59 & 0.00 & 99 & 0.68 & 99 & 1.46 & 99 & 6.86 & 99 & 0.03 \\
Queen & 0.80 & 98 & 1.35 & \textbf{60} & 0.00 & 98 & 1.18 & 97 & 2.65 & 98 & 14.46 & 98 & 0.00 \\
Radiohead & 0.80 & 88 & 0.91 & 42 & 0.00 & 88 & 0.38 & 88 & 0.90 & 88 & 3.01 & 88 & 0.01 \\
Rihanna & 0.80 & 91 & 1.11 & 38 & 0.00 & 91 & 1.19 & 91 & 2.71 & 91 & 7.05 & 91 & 0.02 \\
Roberta Flack & 0.80 & 102 & 1.12 & 57 & 0.00 & 102 & 0.52 & 99 & 0.94 & 102 & 6.64 & 102 & 0.23 \\
\hline \hline
\end{tabular}}
\end{table}

	Overall, the results indicate that \textsc{Adapt-Cmsa} consistently achieved the best performance across all reduction levels. However, most other methods also succeeded in correctly identifying the song in nearly all scenarios except for \textsc{Ilp} in the more challenging (high-information) setting, and \textsc{Approx} in the easiest (low-information) scenario.
	
	Notably, the experimental evidence suggests that the LFCSP framework is well aligned with the needs of audio fingerprinting in a specific context: the secondary song detection phase. In this phase, majority of irrelevant candidates are already filtered out, and the challenge lies in selecting the correct match from a small set of highly similar candidates, typically tens or possibly hundreds. In such settings, the LFCSP formulation proves to be a viable and effective approach.

	{Beyond this case study on eight manually curated songs, we further validate the underlying idea at a considerably larger and less curated scale, using 1000 real, full-length tracks (500 rock songs and 500 classical pieces) from the MTG-Jamendo dataset~\cite{bogdanov2019mtg} and the Audfprint fingerprinting system. The results, reported in \ref{app:jamendo}, suggest a hybrid usage pattern in which LFCSP re-ranks the shortlist returned by an existing fingerprinting system, improving retrieval rank on heavily degraded queries where the underlying system's own confidence is low; a preliminary comparison indicated a prefiltering effect similar to the one of Panako, though we focused the full analysis on Audfprint for its more direct integration with our existing Python pipeline.}

	\section{Model Explainability Analysis}\label{sec:explainable_AI}
	
	To provide a more comprehensive comparison of the algorithms, we apply explainable algorithm performance prediction. This analysis is conducted on both the \textsc{Small} and \textsc{Large} benchmark sets. Each dataset is partitioned into training and test set scenarios, with 25\% of the problem instances randomly selected for testing. For each algorithm, a separate single-target regression (STR) model is trained using four instance features to predict its final solution quality. To evaluate the contribution of these features, we employ the \textit{SHapley Additive exPlanations} (SHAP) method~\cite{nohara2022explanation}. SHAP provides both global feature importance—indicating the overall impact of each feature on the model’s predictions across the entire dataset—and local feature importance—reflecting the influence of each feature on individual instance predictions. By computing feature importance for each algorithm-specific model, we enable a comparative analysis of how instance characteristics affect predicted performance across the different algorithms.

	\subsection{Algorithm Performance Prediction}
	
	We selected the \texttt{XGBoost} (XGB) model~\cite{avanijaa2021prediction} to predict algorithm performance on both benchmark sets. The model was evaluated using its default hyperparameters and trained to minimize the \emph{Mean Squared Error} (MSE). For comparison, a baseline model was also evaluated and produced substantially worse results, as demonstrated below. Several alternative regression models, including Random Forests and Decision Trees, were additionally explored, but none outperformed XGB in terms of predictive accuracy.
	
	Table~\ref{tab:small_train} presents the predictive performance of the considered models using 5-fold cross-validation on the training samples of the benchmark set \textsc{Small}, while Table~\ref{tab:large_train} reports their performance on the corresponding test samples.
	
	{
		The prediction analysis is based on eleven instance features. The first four features describe the basic dimensions of an instance: the sequence length $n=|A|$, the alphabet cardinality $|\Sigma|$, the length of sequence $B$, denoted by $|B|$, and the multiset cardinality $k=|\mathcal{M}|$. To provide a more detailed characterization of the internal instance structure, seven additional features are considered.
	}
	{
		The relative multiset overlap between $\mathcal{M}$ and $A$ is defined as
		$\rho_{\mathcal{M},A}
		=
		\frac{
			\sum_{\sigma\in\Sigma}
			\min\{|\mathcal{M}|_\sigma,|A|_\sigma\}
		}{
			|A|
		}$,
		This feature measures the proportion of symbols in $A$, including multiplicities, that are available in the multiset $\mathcal{M}$.\\
		The normalized density of potential matches between $A$ and $B$ is defined as
		$\rho_{A,B}^{\mathrm{match}}
		=
		\frac{
			\sum_{\sigma\in\Sigma}
			|A|_\sigma |B|_\sigma
		}{
			|A||B|
		}$.
		It measures the proportion of position pairs $(i,j)\in A\times B$ for which the corresponding symbols are equal.\\
		The symbol-frequency distributions of the two ordered sequences are described by their mean normalized Shannon entropy~\cite{ali2023shannon},
		$\overline{H}_{A,B}
		=
		\frac{  
			H_{\mathrm{norm}}(A)+H_{\mathrm{norm}}(B)
		}{2}$, 
		where
		$H_{\mathrm{norm}}(s) =
		\frac{
			-\sum_{\sigma\in\Sigma}
			p_s(\sigma)\log_2 p_s(\sigma)
		}{
			\log_2|\Sigma|
		}$
		and
		$p_s(\sigma)=|s|_{\sigma}/|s|$.
		The normalized entropy of the multiset, denoted by
		$H_{\mathcal{M}}=H_{\mathrm{norm}}(\mathcal{M})$,
		is included as a separate feature because $\mathcal{M}$ represents symbol availability and has a different role from the ordered sequences $A$ and $B$.\\
		The concentration of the symbol-frequency distributions in $A$ and $B$ is measured by
		$\overline{f}_{\max,A,B}
		=
		\frac{1}{2}
		\left(
		\frac{\max_{\sigma \in \Sigma}|A|_\sigma}{|A|}
		+
		\frac{\max_{\sigma \in \Sigma} |B|_\sigma}{|B|}
		\right)$.  
		Finally, two features describe local repetition patterns in the ordered sequences. The mean adjacent-repetition ratio is defined as
		$\overline{r}_{1,A,B}
		=
		\frac{r_1(A)+r_1(B)}{2}$,
		where
		$r_1(s)
		=
		\frac{|\{i \mid s[i]=s[i+1]\}|}{|s|-1}$.
		The mean repeated-bigram ratio is defined as
		$\overline{r}_{2,A,B}
		=
		\frac{r_2(A)+r_2(B)}{2}$,
		where
		$r_2(s)
		=
		1-
		\frac{
			|\operatorname{distinct}\{s[i] s[i+1] \mid 1\leq i < |s|\}|
		}{|s|-1}$.
		Since $\mathcal{M}$ is a multiset without an intrinsic ordering, the adjacent-repetition and bigram-based features are computed only for $A$ and $B$. \\ Consequently, the complete feature vector used to explain algorithm-performance prediction is given by
		$$(n,|\Sigma|,|B|,k,\rho_{\mathcal{M},A},
		\rho_{A,B}^{\mathrm{match}},\overline{H}_{A,B},
		H_{\mathcal{M}},\overline{f}_{\max,A,B}, \overline{r}_{1,A,B},\overline{r}_{2, A,B}) \;\;\;.$$} 
	
	Tables~\ref{tab:small_train} and~\ref{tab:large_test} report the predictive performance on the benchmark set \textsc{Large}, using the same eleven features, regression models, and evaluation procedure.

	\begin{table}[htbp]
		\centering
		\caption{{Error metrics of cross-validation (train) of the XGBoost model when predicting
				the performance of competitors on the benchmark set \textsc{Small}}.}
		\label{tab:small_train}
		\begin{tabular}{ll rr}
			\toprule
			model & algorithm & RMSE & $R^2$ \\
			\midrule
			\multirow{6}{*}{Baseline}
			& \textsc{Adapt-Cmsa}  & 18.2327 & -0.0021 \\
			& \textsc{Approx}                & 17.7651 & -0.0020 \\
			& \textsc{Rand-Sample}           & 18.0762 & -0.0021 \\
			& \textsc{Ls2}                   & 18.0780 & -0.0020 \\
			& \textsc{Ls4}                    & 17.9937 & -0.0020 \\
			& \textsc{Ilp}                  & 18.2499 & -0.0021 \\
			\midrule
			\multirow{6}{*}{XGB default}
			& \textsc{Adapt-Cmsa}  & 2.1786 & 0.9857 \\
			& \textsc{Approx}                & 2.0357 & 0.9868 \\
			& \textsc{Rand-Sample}           & 2.2670 & 0.9842 \\
			& \textsc{Ls2}                   & 2.2211 & 0.9849 \\
			& \textsc{Ls4}                    & 2.2591 & 0.9842 \\
			& \textsc{Ilp}                  & 2.1937 & 0.9855 \\
			\bottomrule
		\end{tabular}
	\end{table}

	\begin{table}[htbp]
		\centering
		\caption{{Error metrics (test) of the XGBoost model when predicting the performance of
				competitors on the benchmark set \textsc{Small}.}}
		\label{tab:small_test}
		\begin{tabular}{ll rr}
			\toprule
			model & algorithm & RMSE & $R^2$ \\
			\midrule
			\multirow{6}{*}{Baseline}
			& \textsc{Adapt-Cmsa}  & 18.1725 & -0.0018 \\
			& \textsc{Approx}                & 17.5854 & -0.0015 \\
			& \textsc{Rand-Sample}           & 18.0345 & -0.0018 \\
			& \textsc{Ls2}                   & 17.9385 & -0.0016 \\
			& \textsc{Ls4}                    & 17.8490 & -0.0017 \\
			& \textsc{Ilp}                  & 18.1932 & -0.0018 \\
			\midrule
			\multirow{6}{*}{XGB default}
			& \textsc{Adapt-Cmsa}  & 2.0911 & 0.9867 \\
			& \textsc{Approx}                & 1.9422 & 0.9878 \\
			& \textsc{Rand-Sample}           & 2.2105 & 0.9849 \\
			& \textsc{Ls2}                   & 2.1462 & 0.9857 \\
			& \textsc{Ls4}                    & 2.1065 & 0.9860 \\
			& \textsc{Ilp}                  & 2.1439 & 0.9861 \\
			\bottomrule
		\end{tabular}
	\end{table}

	\begin{table}[htbp]
		\centering
		\caption{{Error metrics of cross-validation (train) of the XGBoost model when predicting
				the performance of competitors on the benchmark set \textsc{Large}}.}
		\label{tab:large_train}
		\begin{tabular}{ll rr}
			\toprule
			model & algorithm & RMSE & $R^2$ \\
			\midrule
			\multirow{6}{*}{Baseline}
			& \textsc{Adapt-Cmsa}  & 251.2170 & -0.0429 \\
			& \textsc{Approx}                & 244.0035 & -0.0321 \\
			& \textsc{Rand-Sample}           & 222.5714 & -0.0409 \\
			& \textsc{Ls2}                   & 250.3454 & -0.0429 \\
			& \textsc{Ls4}                    & 245.9836 & -0.0433 \\
			& \textsc{Ilp}                  & 129.8468 & -0.0081 \\
			\midrule
			\multirow{6}{*}{XGB default}
			& \textsc{Adapt-Cmsa}  & 19.8007 & 0.9935 \\
			& \textsc{Approx}                & 64.4326 & 0.9280 \\
			& \textsc{Rand-Sample}           & 25.2501 & 0.9866 \\
			& \textsc{Ls2}                   & 18.7154 & 0.9942 \\
			& \textsc{Ls4}                    & 16.2624 & 0.9954 \\
			& \textsc{Ilp}                  & 60.1825 & 0.7834 \\
			\bottomrule
		\end{tabular}
	\end{table}

	\begin{table}[htbp]
		\centering
		\caption{{Error metrics (test) of the XGBoost model when predicting the performance of
				competitors on the benchmark set \textsc{Large}}.}
		\label{tab:large_test}
		\begin{tabular}{ll rr}
			\toprule
			model & algorithm & RMSE & $R^2$ \\
			\midrule
			\multirow{6}{*}{Baseline}
			& \textsc{Adapt-Cmsa}  & 225.8739 & -0.2071 \\
			& \textsc{Approx}                & 221.4666 & -0.2762 \\
			& \textsc{Rand-Sample}           & 202.3635 & -0.1823 \\
			& \textsc{Ls2}                   & 224.1258 & -0.2077 \\
			& \textsc{Ls4}                    & 220.1046 & -0.1955 \\
			& \textsc{Ilp}                  & 122.8566 & -0.0124 \\
			\midrule
			\multirow{6}{*}{XGB default}
			& \textsc{Adapt-Cmsa}  & 17.5627 & 0.9927 \\
			& \textsc{Approx}                & 40.2522 & 0.9578 \\
			& \textsc{Rand-Sample}           & 20.3870 & 0.9880 \\
			& \textsc{Ls2}                   & 18.6943 & 0.9916 \\
			& \textsc{Ls4}                    & 14.3857 & 0.9949 \\
			& \textsc{Ilp}                  & 133.5828 & -0.1969 \\
			\bottomrule
		\end{tabular}
	\end{table}

	The predictive performance of the models is assessed using two metrics: Root Mean Squared Error (RMSE) and the coefficient of determination ($R^2$ score). Lower RMSE values and higher $R^2$ scores indicate better predictive accuracy. The results show that all XGB regression models perform well on the \textsc{Small} benchmark set, achieving $R^2$ scores of {0.98} or higher. Notably, the XGB model with default hyperparameters demonstrates excellent performance, highlighting its strong predictive capabilities. On the \textsc{Large} benchmark set, XGB also performs well for all algorithms, except in the case of the \textsc{Ilp} model on the test data. This exception can be attributed to the limited effectiveness of the \textsc{Ilp} approach: for the vast majority of large instances, it produces only trivial solutions (i.e., empty or near-zero), with meaningful solutions obtained only for a few smaller instances. Consequently, the model struggles to achieve high RMSE and $R^2$ scores, as most target values are constant (equal to zero), and only a few deviate significantly. 
	
	\subsection{Model Explainability Analysis}
	
	\begin{figure}[H]
		\centering
		\begin{subfigure}[b]{0.48\textwidth}
			\centering        \includegraphics[width=\textwidth]{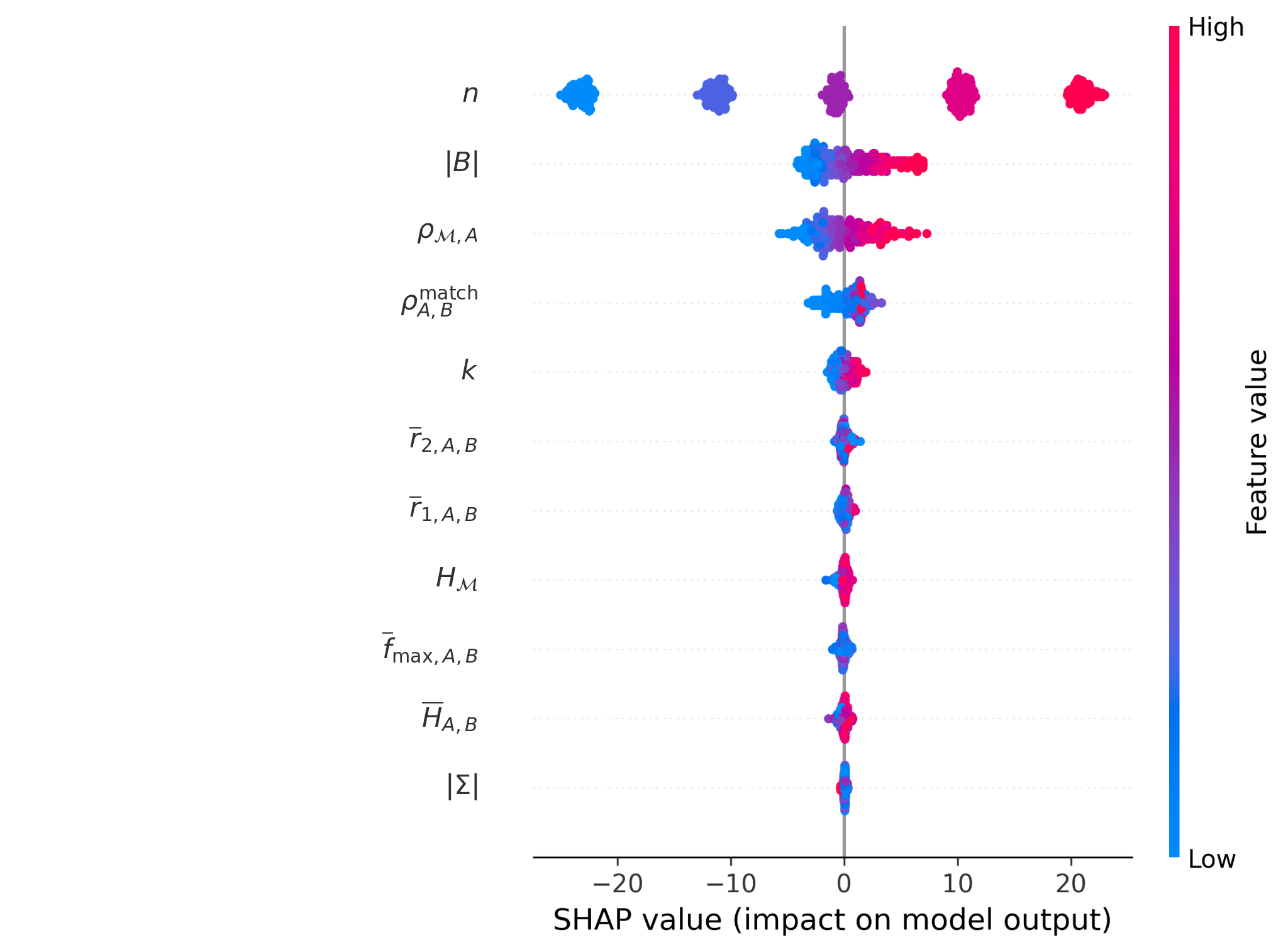}
			\caption{SHAP summary plot for \textsc{Adapt-Cmsa}}
		\end{subfigure}
		\hfill
		\begin{subfigure}[b]{0.48\textwidth}
			\centering
			\includegraphics[width=\textwidth]{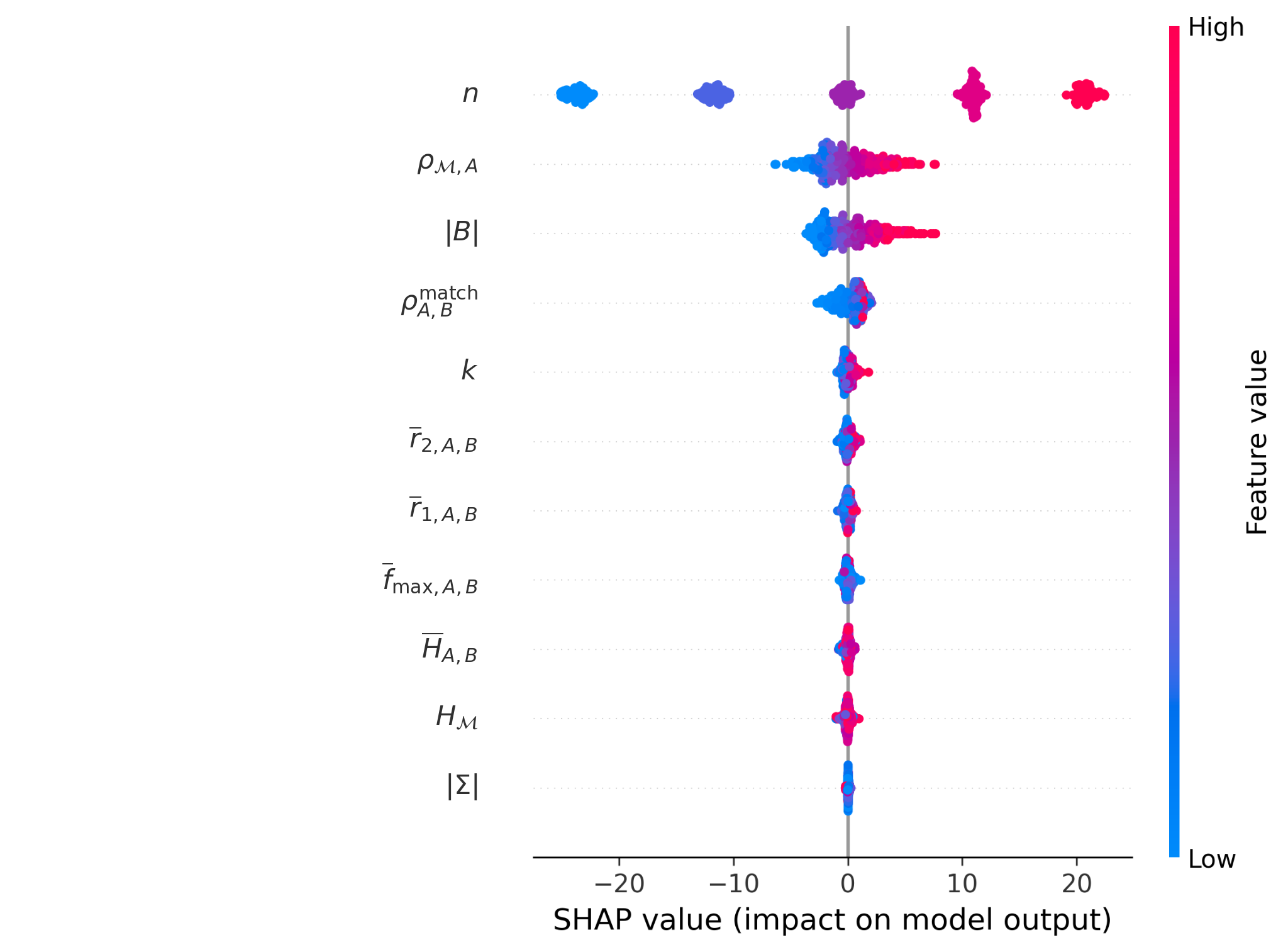}
			\caption{SHAP summary plot for \textsc{Approx}}    
		\end{subfigure}
		\hfill
		\begin{subfigure}[b]{0.48\textwidth}
			\centering
			\includegraphics[width=\textwidth]{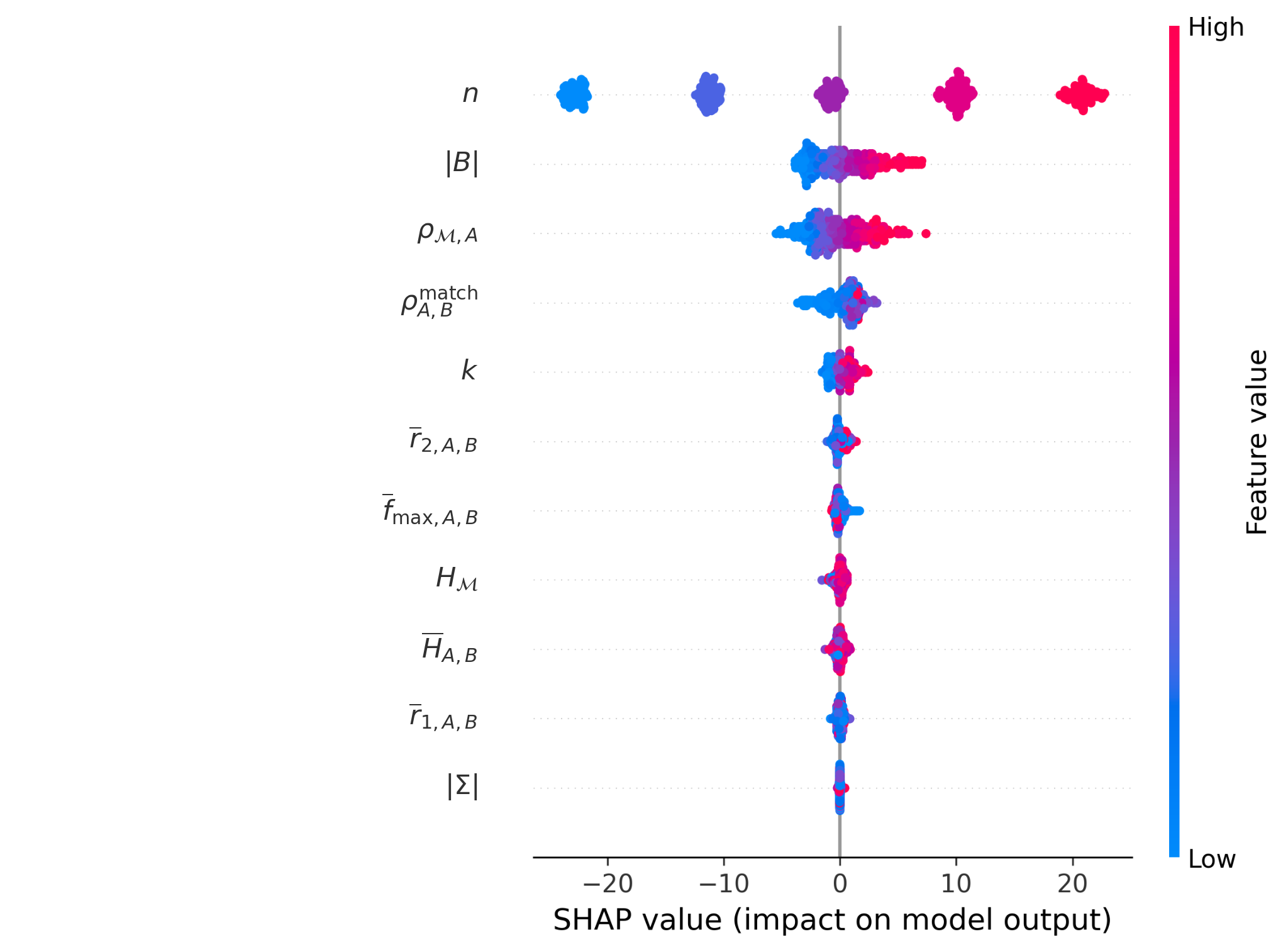}
			\caption{SHAP summary plot for \textsc{Rand-Sample}}    
		\end{subfigure}
		\hfill
		\begin{subfigure}[b]{0.48\textwidth}
			\centering
			\includegraphics[width=\textwidth]{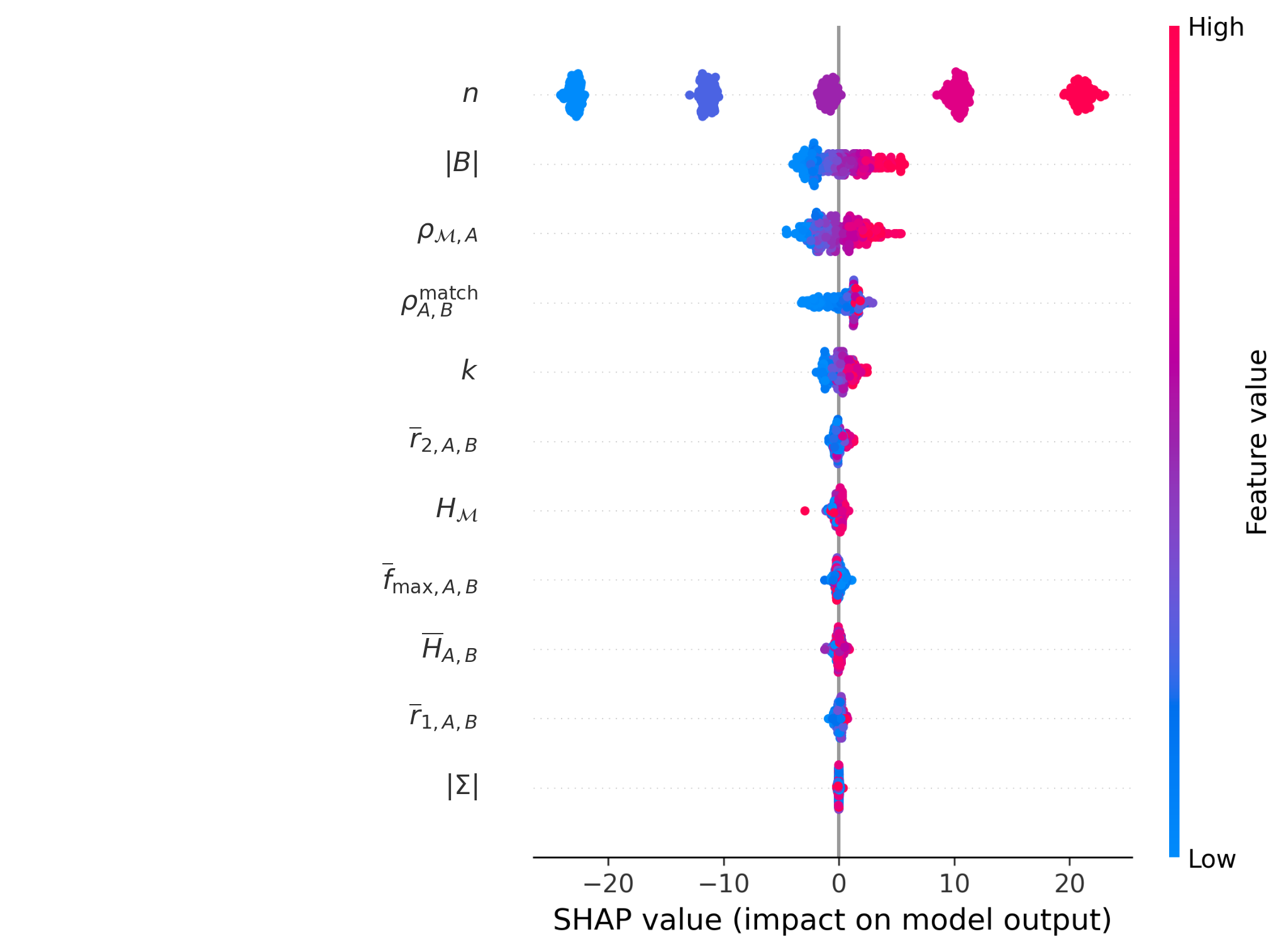}
			\caption{SHAP summary plot for \textsc{LS}2}    
		\end{subfigure}
		
		\begin{subfigure}[b]{0.48\textwidth}
			\centering
			\includegraphics[width=\textwidth]{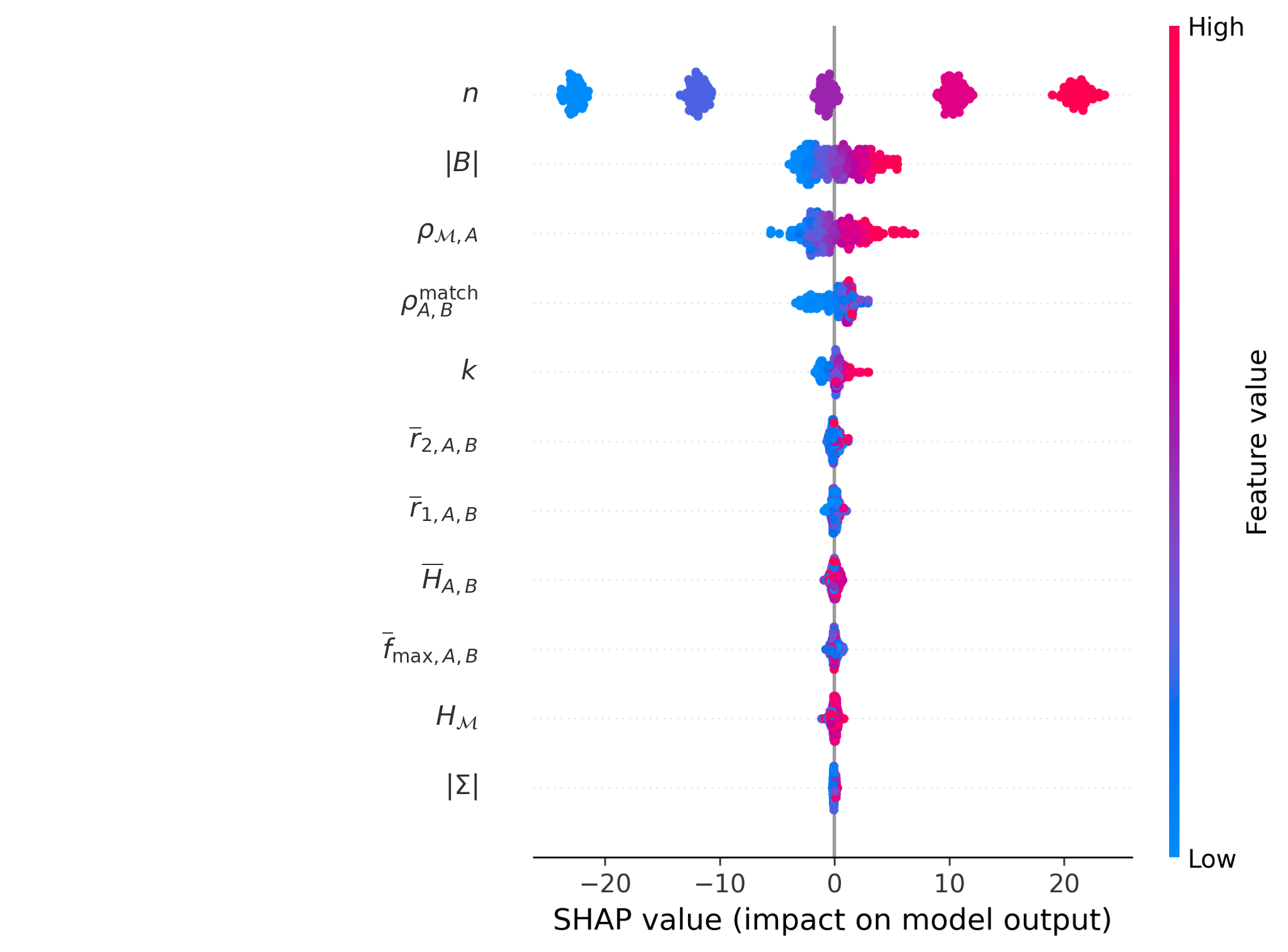}
			\caption{SHAP summary plot for \textsc{LS}4}
		\end{subfigure}
		\hfill
		\begin{subfigure}[b]{0.48\textwidth}
			\centering
			\includegraphics[width=\textwidth]{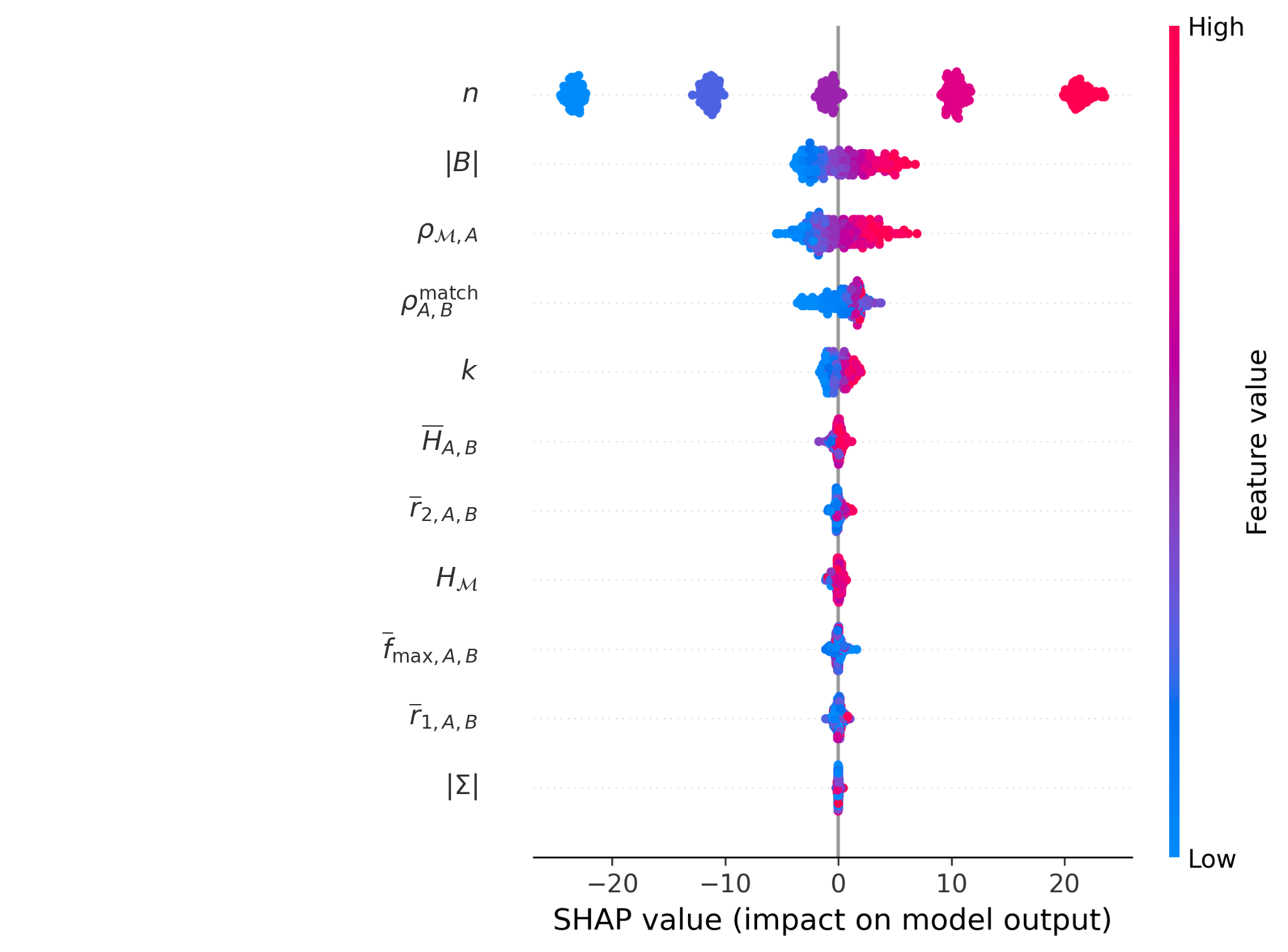}
			\caption{SHAP summary plot for \textsc{Ilp}}    
		\end{subfigure}    
		
		\caption{{SHAP summary plots showing the global feature importance and the direction of influence of different feature values (i.e., low in blue to high in red)  on the  benchmark set~\textsc{Small} (the test data). }}
		\label{fig:summary-globalplots-lcfsp}
	\end{figure}
	
	{
		The following conclusions are drawn from the SHAP plots on global feature importance across all instances of the \textsc{Small} benchmark set. \\
		Figure~\ref{fig:summary-globalplots-lcfsp} shows the SHAP summary plots for all six algorithms using the complete eleven-feature instance description. Across all algorithms, a consistent feature-importance ordering emerges. The sequence length $n$ dominates by a wide margin. Its SHAP values form five almost non-overlapping colour clusters, ranging approximately from $-25$ for the shortest sequences to $+22$ for the longest ones. This behaviour reflects the fact that $n$ represents a hard upper bound on the achievable objective value, since
		$\operatorname{obj}(s) \leq n$
		for every method, independently of the way in which the corresponding algorithm explores the solution space.\\
		For five of the six algorithms, the next most influential features are $|B|$ and $\rho_{\mathcal{M},A}$, followed by $\rho_{A,B}^{\mathrm{match}}$ and $k=|\mathcal{M}|$. All these features exhibit the expected influence direction: larger values of $|B|$, $\rho_{\mathcal{M},A}$, $\rho_{A,B}^{\mathrm{match}}$, and $k$ tend to increase the predicted objective value, as indicated by red points being predominantly located on the positive side of the SHAP axis. This is consistent with the fact that larger values of these features increase the number of potentially exploitable matches among $A$, $B$, and $\mathcal{M}$.\\
		The remaining structural and statistical features,
		$\overline{r}_{2,A,B}$,
		$\overline{r}_{1,A,B}$,
		$\overline{f}_{\max,A,B}$,
		$\overline{H}_{A,B}$,
		$H_{\mathcal{M}}$,
		and $|\Sigma|$, have only a marginal contribution. Their SHAP values are tightly concentrated around zero and exhibit little colour separation. This indicates that fine-grained characteristics of the symbol distributions have limited influence on the predicted solution quality once $n$, $|B|$, $\rho_{\mathcal{M},A}$, $\rho_{A,B}^{\mathrm{match}}$, and $k$ are already taken into account.}
	
	Algorithm-by-algorithm comparison reveals the following conclusions.
	
	\begin{itemize}
		
		\item {\textsc{Approx}. The
			\textsc{Approx} approach is the only clear outlier in terms of feature ordering. For this method, $\rho_{\mathcal{M},A}$ is ranked above $|B|$ and therefore occupies the second position in the global feature-importance ranking. This observation is consistent with the way in which \textsc{Approx} constructs a solution. After fixing the LCS-based matches, its \textsc{GreedyMatch} subroutine can add only those symbols that are explicitly available in $\mathcal{M}$, matching them greedily from left to right without backtracking. Consequently, its success is more strongly related to the fraction of symbols of $A$, including multiplicities, that are present in $\mathcal{M}$ than to the length of $B$. Unlike the other methods, \textsc{Approx} does not explicitly explore the matching space induced by $B$.}
		
		\item {\textsc{Rand-Sample},  \textsc{LS2}, \textsc{LS4}, and \textsc{Adapt-Cmsa}. 
			These four methods exhibit almost identical top-five feature orderings:
			$n$, $|B|$, $\rho_{\mathcal{M},A}$, $\rho_{A,B}^{\mathrm{match}}$, and $k$.
			This similarity can be explained by their shared algorithmic structure but also by their stochastic nature, which increases robustness. All four approaches construct a solution by first selecting a subset of positions of $A$ to match with the multiset $\mathcal{M}$ and then computing an LCS-style alignment of the remaining sequence with $B$.\\
			In particular, the construction phases of \textsc{Rand-Sample} and \textsc{Adapt-Cmsa} rely on the same randomized-greedy mechanism, while \textsc{LS2} and \textsc{LS4} differ mainly in the size of the local windows used for re-matching. Since all four methods are constrained by the same two principal factors, namely how much of $A$ can be matched with $\mathcal{M}$ and how effectively the remaining part can be aligned with $B$, their SHAP feature-sensitivity profiles naturally converge.\\ 
			The corresponding plots differ mainly in the low-importance tail, which consists of the bigram, repetition, and entropy-related features. These differences are more likely to reflect statistical variation than fundamental differences in the underlying search mechanisms.}
		
		\item {\textsc{ILP}. 
			The SHAP profile of \textsc{ILP} appears superficially similar to those of the heuristic methods, since the same five features dominate its importance ranking. However, this similarity should not be interpreted as evidence that the underlying \textsc{Ilp} search process resembles the heuristic search mechanisms. Instead, it is largely a consequence of the characteristics of the \textsc{Small} benchmark set.\\
			Since \textsc{ILP} solves these instances to optimality within the prescribed time limit, its predicted performance surface closely follows the dependence of the true optimum on the principal instance-size characteristics. On these comparatively small instances, the exact and near-exact methods therefore produce similar objective values, which results in similar SHAP importance profiles.\\
			In comparison with the heuristic methods, $\overline{H}_{A,B}$ occupies a noticeably higher position for \textsc{ILP}, namely the sixth position. This may indicate that, even at this instance scale, the branch-and-cut behaviour of \textsc{Cplex} is more sensitive to the shape of the symbol-frequency distributions in $A$ and $B$ than the constructive heuristics are. One possible explanation is that sequence entropy influences the density and structure of the conflict graph induced by the conflict constraints created between pairs of matchings in the ILP formulation, which can affect the difficulty of the mathematical model even when an optimal solution is eventually obtained.}
		
		\item {The $|\Sigma|$ influence. 
			Across all six algorithms, $|\Sigma|$ is consistently the least influential feature. Its SHAP values form a small cluster that is slightly shifted towards the negative side. This suggests that larger alphabets tend to reduce the predicted objective value, since a larger alphabet generally implies fewer repeated symbols and, consequently, fewer potentially exploitable matches.\\
			Nevertheless, the influence of $|\Sigma|$ is substantially weaker than that of $n$, $|B|$, $\rho_{\mathcal{M},A}$, and $k$ for every considered algorithm.}
	\end{itemize}

	
	\begin{figure}[H]
		\centering
		\begin{subfigure}[b]{0.48\textwidth}
			\centering        \includegraphics[width=\textwidth]{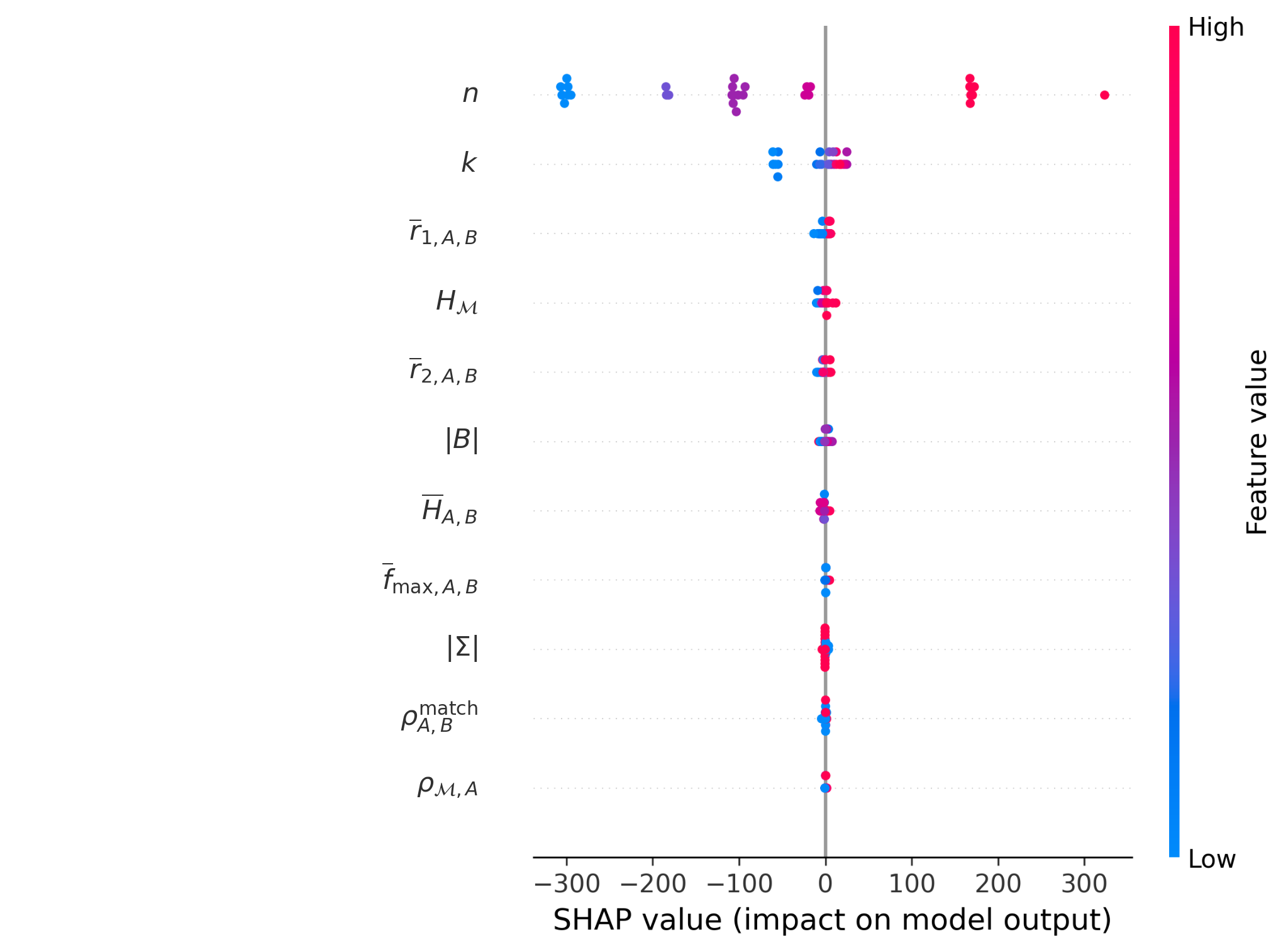}
			\caption{SHAP summary plot for \textsc{Adapt-Cmsa}}
		\end{subfigure}
		\hfill
		\begin{subfigure}[b]{0.48\textwidth}
			\centering
			\includegraphics[width=\textwidth]{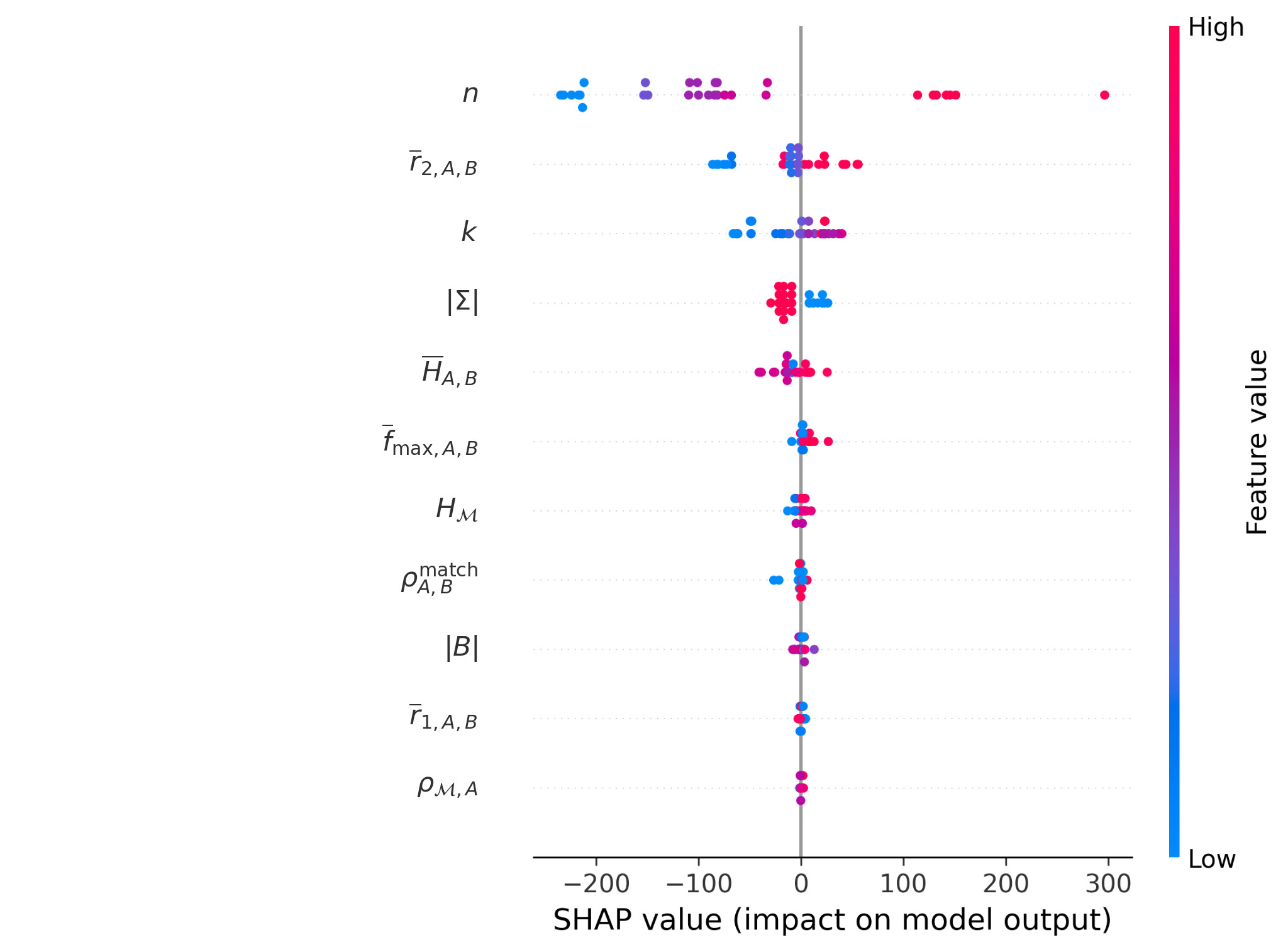}
			\caption{SHAP summary plot for \textsc{Approx}}    
		\end{subfigure}
		\hfill
		\begin{subfigure}[b]{0.48\textwidth}
			\centering
			\includegraphics[width=\textwidth]{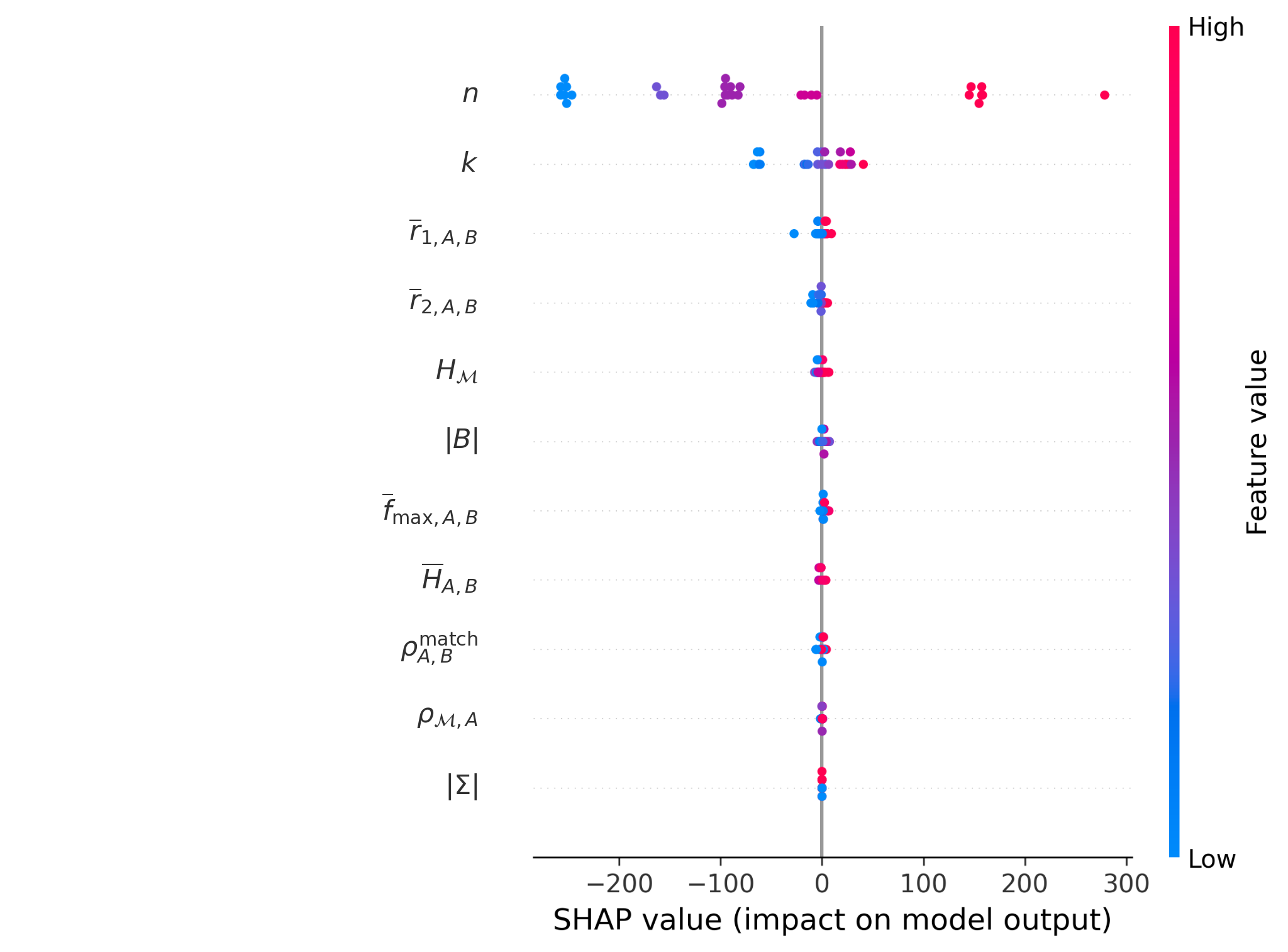}
			\caption{SHAP summary plot for \textsc{Rand-Sample}}    
		\end{subfigure}
		\hfill
		\begin{subfigure}[b]{0.48\textwidth}
			\centering
			\includegraphics[width=\textwidth]{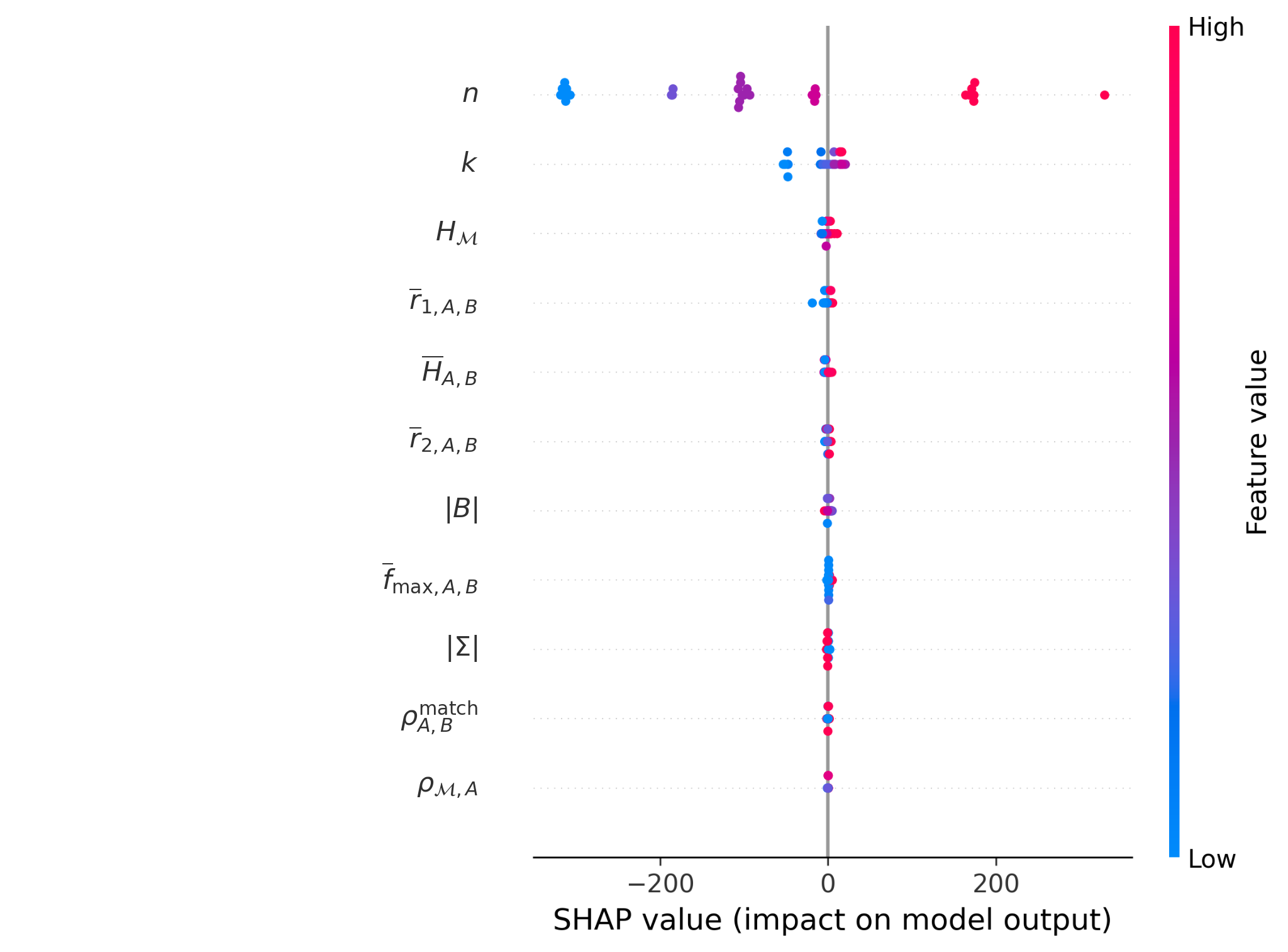}
			\caption{SHAP summary plot for \textsc{LS}2}    
		\end{subfigure}
		
		\begin{subfigure}[b]{0.48\textwidth}
			\centering
			\includegraphics[width=\textwidth]{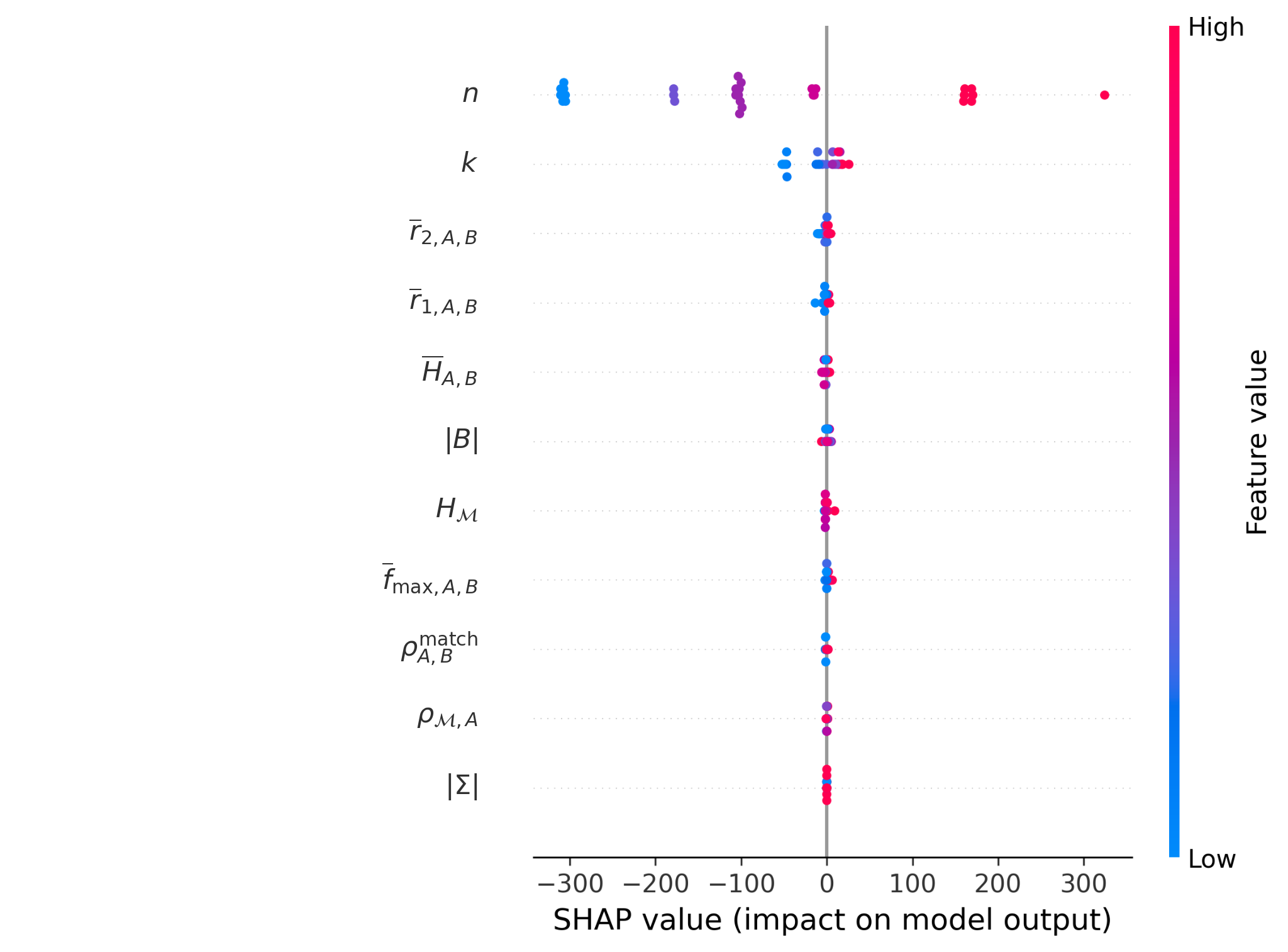}
			\caption{SHAP summary plot for \textsc{LS}4}
		\end{subfigure}
		\hfill
		\begin{subfigure}[b]{0.48\textwidth}
			\centering
			\includegraphics[width=\textwidth]{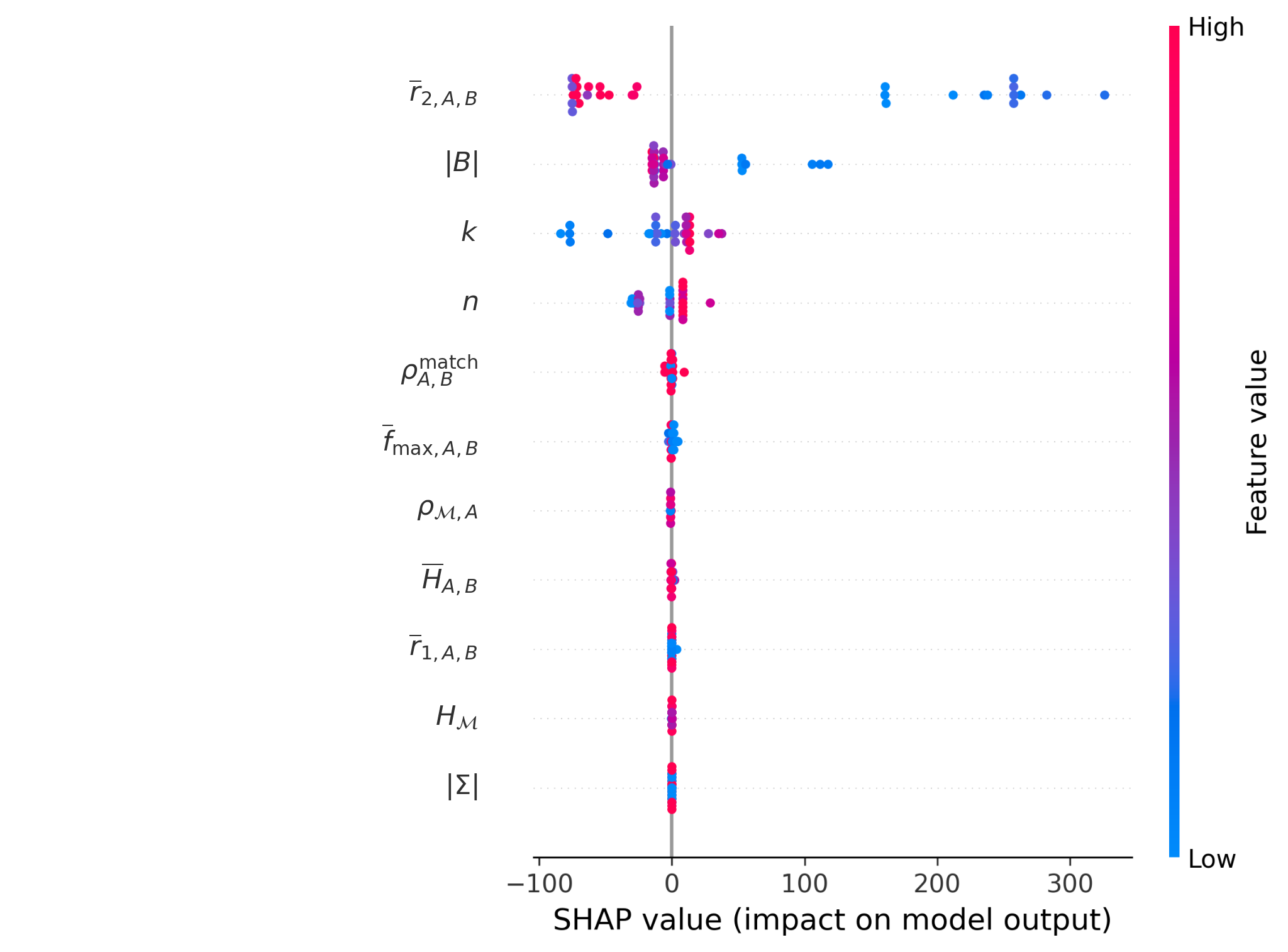}
			\caption{SHAP summary plot for \textsc{Ilp}}    
		\end{subfigure}    
		
		\caption{{SHAP summary plots showing the global feature importance and the direction of influence of different feature values (i.e., low in blue to high in red)  on the benchmark set~\textsc{Large} (the test data).} }
		\label{fig:large-summary-globalplots-lcfsp}
	\end{figure}

	{
		For the performance of the algorithms on the benchmark set \textsc{Large}, the following observations are made based on the plots provided in Figure~\ref{fig:large-summary-globalplots-lcfsp}. \\
		Compared with the \textsc{Small} benchmark, the ordering of the features changes substantially. Most notably, several statistical features that had only a negligible influence on the \textsc{Small} benchmark, including $\overline{r}_{2,A,B}$, $\overline{r}_{1,A,B}$, and $H_{\mathcal{M}}$, move into the highest positions for most algorithms. By contrast, $\rho_{\mathcal{M},A}$ and $\rho_{A,B}^{\mathrm{match}}$, which were consistently among the four most influential features for \textsc{Small}, move towards the bottom of the ranking for five of the six algorithms. \\
		This indicates a genuinely different operating regime. For large instances, the raw compatibility among $A$, $B$, and $\mathcal{M}$ no longer appears to be the principal limiting factor. Instead, characteristics related to the distributional structure of the sequences, particularly repetition patterns and entropy, become more influential.}
	
	{Algorithm-by-algorithm comparison reveals the following conclusions.}
	
	\begin{itemize}
		
		\item {\textsc{Rand-Sample}, \textsc{LS4}, \textsc{LS2}, and \textsc{Adapt-Cmsa}. \\
			These four algorithms are again almost indistinguishable with respect to their two most influential features. The sequence length $n$ dominates by a wide margin, with SHAP values ranging approximately from $-250$ to $+300$ and forming five clearly separated colour clusters. It is followed by $k$, with both features exhibiting the expected positive direction: longer sequences and a larger number of available multiset symbols generally increase the predicted objective value. This similarity is consistent with the shared solution construction mechanism discussed for the \textsc{Small} benchmark, namely randomized-greedy matching between $A$ and $\mathcal{M}$ followed by an LCS-style refinement involving $B$.\\
			The methods differ mainly in the features ranked between the third and sixth positions. For \textsc{Rand-Sample} and \textsc{Adapt-Cmsa}, $\overline{r}_{1,A,B}$ occupies the third position. In contrast, \textsc{LS2} assigns greater importance to $H_{\mathcal{M}}$, whereas \textsc{LS4} places $\overline{r}_{2,A,B}$ among the most influential features.\\
			These differences are relatively small and are plausible given the different local-refinement mechanisms employed by the algorithms. Nevertheless, the common pattern remains clear: $n$ and $k$ dominate, while $\rho_{\mathcal{M},A}$, $\rho_{A,B}^{\mathrm{match}}$, and $|\Sigma|$ become considerably less influential.}
		
		\item {\textsc{Approx}. 	The behaviour of \textsc{Approx} diverges more strongly from the other algorithms than it did on the \textsc{Small} benchmark. Although $n$ remains the most influential feature, the second-ranked feature is now $\overline{r}_{2,A,B}$ rather than $\rho_{\mathcal{M},A}$, representing a clear change from the pattern observed for \textsc{Small}.	The alphabet cardinality $|\Sigma|$ also rises to the fourth position and exhibits a clearly negative direction. Larger alphabets reduce the predicted solution quality more visibly for \textsc{Approx} than for the other algorithms. This supports the observation that, for larger instances, the behaviour of \textsc{Approx} changes and $|\Sigma|$ becomes a major factor.\\
			This behaviour is consistent with the algorithmic structure of \textsc{Approx}. Its \textsc{GreedyMatch} procedure performs a single left-to-right pass without backtracking. As the alphabet becomes larger, repeated symbols become less frequent and fewer alternative matches are available. Similarly, changes in the local repetition structure affect which symbols are available at each stage of the greedy pass. At the larger scale, these effects become more visible than the coarser overlap feature $\rho_{\mathcal{M},A}$.}
		
		\item {\textsc{ILP}. The \textsc{ILP} model exhibits the most pronounced departure from both the other algorithms and its own feature-importance profile on the \textsc{Small} benchmark. Its most influential feature is now $\overline{r}_{2,A,B}$, whose SHAP values have a notably wide and asymmetric spread, ranging approximately from $-80$ to $+325$. Low values of bigram repetition strongly decrease the predicted objective value, whereas several medium-range values produce large positive SHAP contributions.\\
			This feature is followed by $|B|$, $k$, and $n$, with $n$ occupying only the fourth position. This represents a striking reversal because $n$ is the dominant feature for every other algorithm.\\
			This pattern is consistent with the observed failure mode of the \textsc{ILP} approach on the \textsc{Large} benchmark. For many large instances with low repetition and small alphabet cardinality, the solver returns only the trivial empty solution, as discussed in Section~\ref{sec:experimental_section}. Consequently, the resulting performance surface does not necessarily reflect the true combinatorial difficulty of the instances. Instead, it may reflect structural characteristics that correlate with the solver occasionally finding a non-trivial feasible solution before reaching the time limit.\\
			The strong dependence on $\overline{r}_{2,A,B}$ may therefore indicate a solver-tractability effect rather than a direct solution-quality effect. Highly repetitive sequences can generate many symmetric or interchangeable matching possibilities. Such structure may affect the root relaxation and the subsequent branch-and-cut process, potentially allowing the solver to identify non-trivial feasible solutions more effectively within the prescribed time limit.}
		
	\end{itemize}

	\subsection{Local Explanations}
	\label{subsubsec:local-shap-explanations}
	
	\subsubsection{Small-to-medium-sized instance}

	\begin{figure}[H]
		\centering
		\begin{subfigure}[b]{0.49\textwidth}
			\centering        \includegraphics[width=\textwidth]{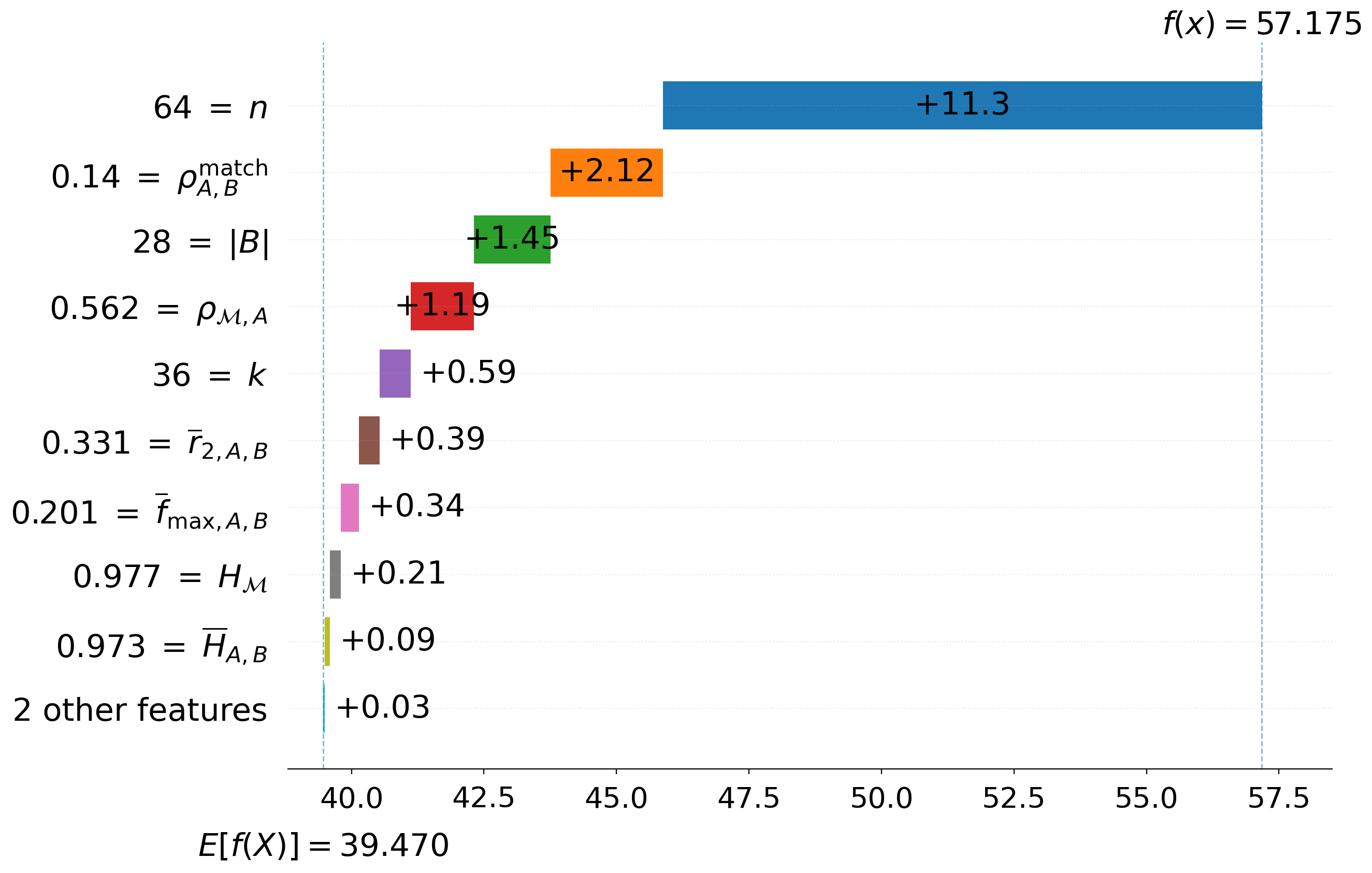} 
			\caption{SHAP waterfall plot for \textsc{Adapt-Cmsa}} \label{fig:waterfall-small-plot-adapt_cmsa}
		\end{subfigure}
		\hfill
		\begin{subfigure}[b]{0.49\textwidth}
			\centering
			\includegraphics[width=\textwidth]{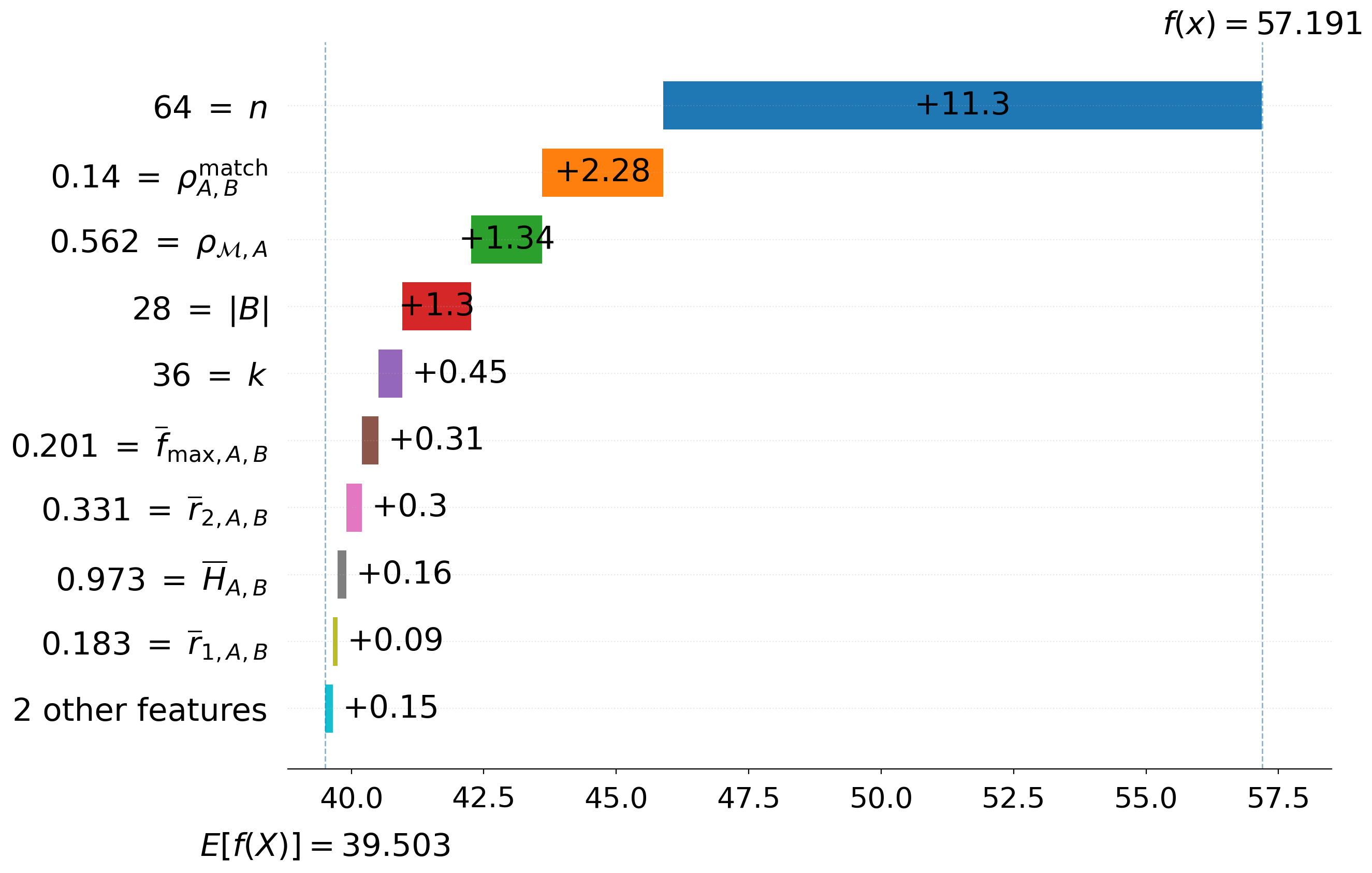}
			\caption{SHAP waterfall plot for \textsc{Ilp}} \label{fig:waterfall-small-plot-ilp}
		\end{subfigure}
		
		\caption{SHAP local waterfall plots show the local feature importance on the final model's prediction for an instance with $n=64$ and $|\Sigma|=8$ from dataset~\textsc{Small}. }     \label{fig:waterfall-small}
		
	\end{figure}
	
	{
		For a small-to-medium-sized instance sampled from the \textsc{Small} benchmark set, we conduct a local feature-importance analysis using a representative test instance from the group characterized by $n=64$ and $|\Sigma|=8$. More specifically, the selected instance has $k=36$ and $|B|=28$. The SHAP waterfall plots for \textsc{ILP} and \textsc{Adapt-Cmsa} are presented in Figure~\ref{fig:waterfall-small}.\\
		For both algorithms, the predicted solution values are almost identical:
		$f(\mathbf{x})=57.191$ for \textsc{Ilp} and
		$f(\mathbf{x})=57.175$ for \textsc{Adapt-Cmsa}.
		The predictions also start from almost the same baseline value,
		$\mathbb{E}[f(\mathbf{X})]\approx39.5$ for both methods.
		This agreement is expected because, on the \textsc{Small} benchmark set, both methods reliably reach the optimum. Consequently, the corresponding regression models learn essentially the same relationship between the instance characteristics and the achieved objective value.\\
		For both algorithms, the two most influential features are identical and clearly dominant. The sequence length $n=64$ provides by far the largest contribution, amounting to approximately $+11.3$ for both methods. This is consistent with the role of $n$ as a hard upper bound on the objective value and with its position as the most influential feature in the global SHAP summary plots.\\
		The matching density
		$\rho_{A,B}^{\mathrm{match}}=0.14$
		is the second-largest contributor for both algorithms. Its contribution is approximately $+2.28$ for \textsc{ILP} and $+2.12$ for \textsc{Adapt-Cmsa}. This indicates that a non-negligible proportion of position pairs between $A$ and $B$ represent compatible matches. Such pairs constitute directly exploitable information for both the $x$-variables of the \textsc{ILP} formulation and the $\texttt{Match}^s_{\mathcal{B}}$ component of the restricted \textsc{CMSA} subproblem.
		\\
		The two methods differ slightly in the ordering of $|B|$ and
		$\rho_{\mathcal{M},A}$. For \textsc{ILP},
		$\rho_{\mathcal{M},A}$ is ranked third, with a contribution of approximately $+1.34$, while $|B|$ is ranked fourth, with a contribution of approximately $+1.30$. The two contributions are therefore very similar, although the overlap feature is marginally more influential. For \textsc{Adapt-Cmsa}, the ordering is reversed. The feature $|B|$ is ranked third, contributing approximately $+1.45$, whereas
		$\rho_{\mathcal{M},A}$ is ranked fourth, with a contribution of approximately $+1.19$.\\
		This subtle distinction suggests that the \textsc{ILP} performance model is slightly more sensitive to the proportion of $A$ that can be explained by the multiset $\mathcal{M}$. In particular, constraint~(\ref{ilp:budget_per_symbols}) directly restricts the $y$-variables according to the multiplicities $|\mathcal{M}_{\sigma}|$. In contrast, the \textsc{Adapt-Cmsa} performance model assigns marginally greater importance to the length of the ordered target sequence $B$. A possible explanation is that the merge-and-adapt cycle repeatedly solves an LCS-type problem between the constructed sequence $s$ and the complete sequence $B$, making the method somewhat more responsive to $|B|$.\\
		Both waterfall plots show that $k=36$ has a moderate positive influence, contributing approximately $+0.45$ for \textsc{ILP} and $+0.59$ for \textsc{Adapt-Cmsa}. The remaining structural and entropy-related features,
		$\overline{f}_{\max,A,B}$,
		$\overline{r}_{2,A,B}$,
		$\overline{H}_{A,B}$,
		$\overline{r}_{1,A,B}$,
		and, for \textsc{Adapt-Cmsa}, $H_{\mathcal{M}}$,
		each provide only a small positive contribution, approximately between $+0.03$ and $+0.40$. This is consistent with their relatively low global importance on the \textsc{Small} benchmark set.\\ 
		Notably, every feature contributes positively in both waterfall plots, and no negative SHAP contributions are observed. This indicates that the characteristics of this particular instance, including $n=64$, a moderately high matching density, and a relatively high overlap between $A$ and $\mathcal{M}$, jointly reinforce a high predicted objective value. As a result, the final prediction is approximately $17.7$ units above the average predicted solution value of the corresponding dataset.}
	
	\subsubsection{Middle-to-large-sized instance}
	
	
	\begin{figure}[H]
		\centering
		\begin{subfigure}[b]{0.45\textwidth}
			\centering        \includegraphics[width=\textwidth]{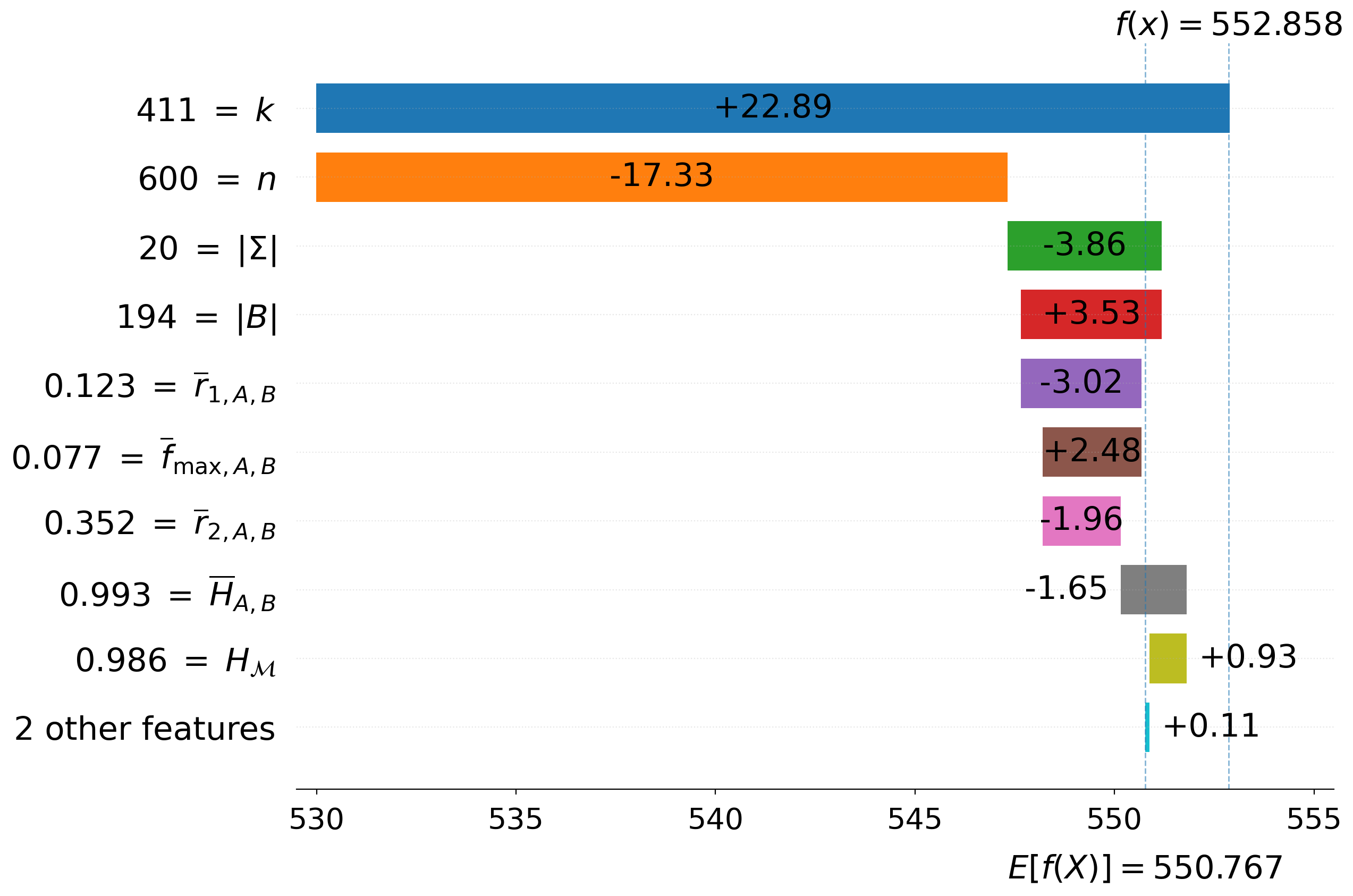} 
			\caption{SHAP waterfall plot for \textsc{Adapt-Cmsa}} \label{fig:waterfall-large-plot-adapt_cmsa}
		\end{subfigure}
		\hfill
		\begin{subfigure}[b]{0.45\textwidth}
			\centering
			\includegraphics[width=\textwidth]{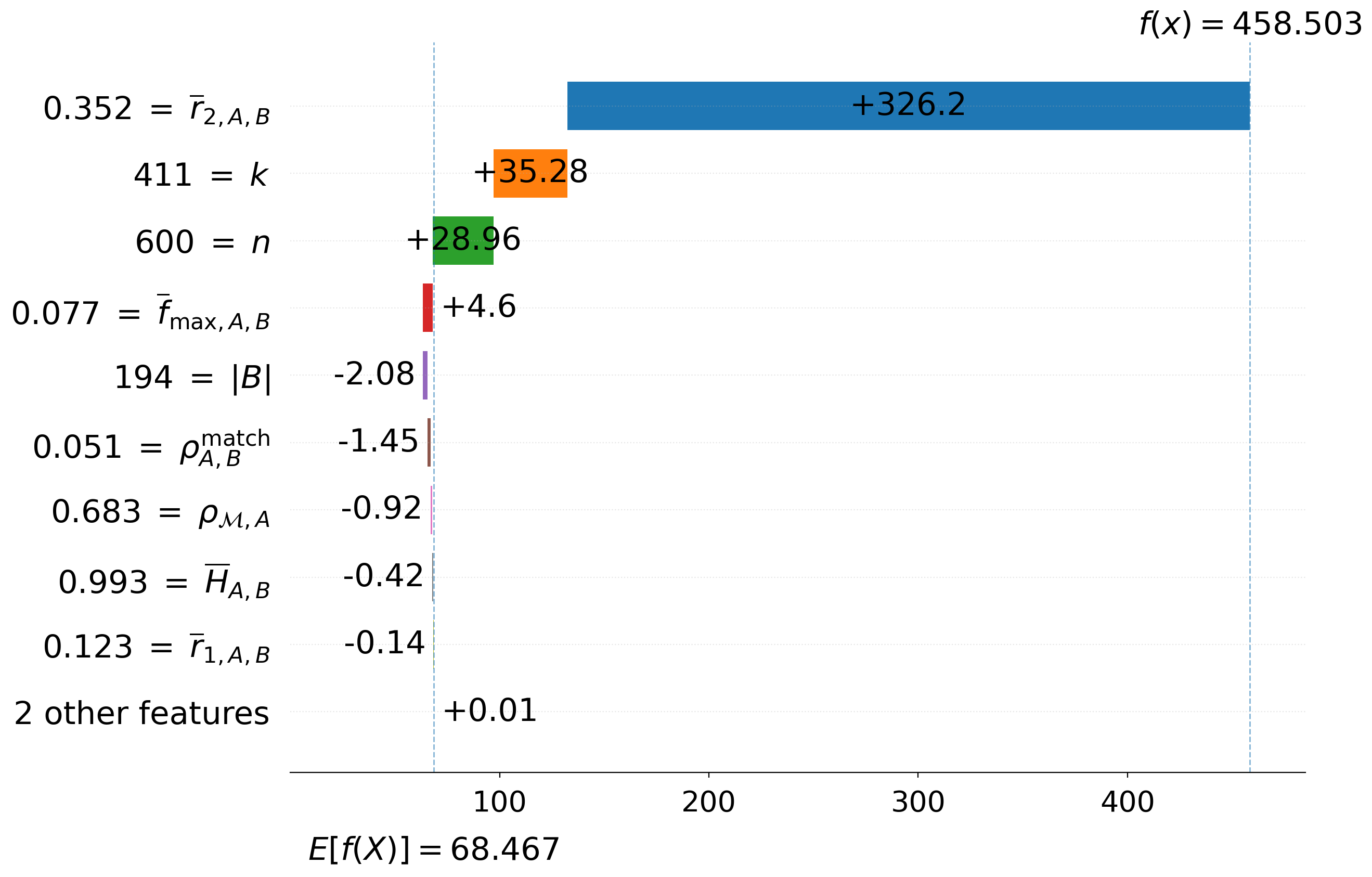}
			\caption{SHAP waterfall plot for \textsc{Ilp}} \label{fig:waterfall-large-plot-ilp}
		\end{subfigure}
		
		\caption{SHAP local waterfall plots show the local feature importance on the final model's prediction for an instance with $n=600$ and $|\Sigma|=20$ from dataset~\textsc{Large}. }     \label{fig:waterfall-large}
		
	\end{figure}
	{
		For a middle-to-large-sized instance sampled from the \textsc{Large} benchmark set, we conduct a local feature-importance analysis using a representative test instance from the group characterized by $n=600$ and $|\Sigma|=20$. The selected instance has $k=411$ and $|B|=194$. The SHAP waterfall plots for \textsc{ILP} and \textsc{Adapt-Cmsa} are presented in Figure~\ref{fig:waterfall-large}.\\
		Unlike the representative instance from the \textsc{Small} benchmark set, for which \textsc{ILP} and \textsc{Adapt-Cmsa} agreed almost feature by feature, this \textsc{Large} instance reveals a substantial difference in the factors driving the two model predictions. The differences concern both the magnitudes of the SHAP contributions and the directions in which individual features affect the predictions.}
	
	\paragraph{\textsc{ILP}}  {For \textsc{ILP}, the baseline value is
		$68.467$,
		which is comparatively low. This reflects the fact that, across the \textsc{Large} test set, solving the \textsc{ILP} model frequently results in trivial or nearly trivial solutions, as discussed earlier in the paper.\\
		The mean repeated-bigram ratio,
		$\overline{r}_{2,A,B}=0.352$,
		is overwhelmingly the most influential feature, with a SHAP contribution of approximately $+326.2$. This contribution exceeds those of all other features by roughly one order of magnitude and explains almost the entire increase from the baseline value of approximately $68.5$ to the final prediction $458.503$. The multiset size $k$ and sequence length $n$ provide additional positive contributions of approximately $+35.28$ and $+28.96$, respectively. Although these effects are substantial in absolute terms, they remain modest in comparison with the dominant repeated-bigram contribution. The mean maximum-frequency ratio
		$\overline{f}_{\max,A,B}$
		provides a smaller additional positive contribution of approximately $+4.60$.\\
		The remaining features, including $|B|$,
		$\rho_{A,B}^{\mathrm{match}}$,
		$\rho_{\mathcal{M},A}$,
		$\overline{H}_{A,B}$,
		and $\overline{r}_{1,A,B}$,
		all have small negative contributions, generally ranging from approximately $-0.1$ to $-2.0$. Relative to the dominant bigram effect, these contributions are negligible.\\
		This local explanation strongly supports the global SHAP findings for \textsc{ILP} on the \textsc{Large} benchmark set, where
		$\overline{r}_{2,A,B}$
		was identified as the most influential feature. For this particular instance, a moderately high repeated-bigram ratio is sufficient to increase the predicted solution value beyond $450$.\\
		This behaviour is consistent with the interpretation that repetitive structure may improve the tractability of the corresponding \textsc{ILP} formulation. Repetition can introduce symmetric or interchangeable matching possibilities and alter the structure of the conflict graph. These effects may simplify the root relaxation or the subsequent branch-and-cut process sufficiently for the solver to escape the empty-solution failure mode that otherwise dominates its baseline behaviour.}
	
	\paragraph{\textsc{Adapt-Cmsa}} { For \textsc{Adapt-Cmsa}, the baseline value is
		$ 550.767$,
		which is already very high. This is consistent with the fact that \textsc{Adapt-Cmsa} reliably produces high-quality solutions across the \textsc{Large} test set. The multiset size $k=411$ is the single largest positive contributor, with a SHAP value of approximately $+22.89$. This direction is expected because a larger multiset provides more symbols that can potentially be used to explain positions of $A$.\\
		In contrast, the sequence length $n=600$ contributes negatively, with a SHAP value of approximately $-17.33$. This sign is opposite to the general pattern observed in the global SHAP summary plot for \textsc{Adapt-Cmsa}, where $n$ is usually the strongest positive driver. This represents a genuine local deviation from the global behaviour. For this particular combination of feature values, $n=600$ lies below the values that the regression model associates with the highest-quality outcomes in this region of the feature space. Consequently, it reduces the prediction relative to the already high baseline instead of increasing it, as it typically does for other instances.\\
		The alphabet cardinality $|\Sigma|=20$ and the mean adjacent-repetition ratio
		$\overline{r}_{1,A,B}$
		also reduce the predicted value, with contributions of approximately $-3.86$ and $-3.02$, respectively. By contrast, $|B|$ contributes approximately $+3.53$,
		$\overline{f}_{\max,A,B}$ contributes approximately $+2.48$, and
		$H_{\mathcal{M}}$ contributes approximately $+0.93$. The repeated-bigram feature
		$\overline{r}_{2,A,B}$
		has only a small negative contribution of approximately $-1.96$. This is in sharp contrast to its role in the \textsc{ILP} waterfall plot for the same instance, where it is the overwhelmingly dominant positive feature.\\
		The total effect of all feature contributions is comparatively small. The prediction increases from the baseline value
		$ 550.767$
		to $ 552.858$,
		representing an increase of only approximately $2.1$ units. This limited deviation from the baseline is reasonable because the \textsc{Adapt-Cmsa} predictions are concentrated close to near-optimal values across the \textsc{Large} benchmark set. Consequently, there is relatively little room for any single feature to produce a large change in the predicted solution value.}\\
	
	{The findings of the SHAP analysis could guide future algorithm development, parameter adaptation, and instance-specific search strategies. First, instances with similar SHAP profiles and comparable solution-quality behavior could be grouped into common categories for joint parameter tuning, rather than being treated together with structurally different instances, such as small-sized and medium-sized instances. Such grouping could support more effective parameter selection and provide a better balance between intensification and diversification.\\
		Second, influential features identified by the explainability analysis but not explicitly used by the current algorithms could be incorporated into more advanced solution-construction and local-search mechanisms. Promising candidates include the number of matching bigrams or trigrams, the total number of feasible matchings between the input sequences, and relative matching ratios involving $\mathcal{M}$ and $A$. These features could motivate neighborhood structures that manipulate groups of consecutive symbols rather than individual symbols. They could also be used to select among alternative construction, modification, or local-search strategies according to the structural properties of a given instance. \\
		This emerging research direction, in which explainable artificial intelligence is used not only to interpret but also to improve optimization methods, has already attracted considerable attention; see, for example,~\cite{kim2025enhancing,ha2019improvement}. As a concrete example, the SHAP profiles may also support instance-specific parameter adaptation during the search; many features could be used to determine the initial restricted-model size, the number of generated solutions, the component aging threshold, or the proportion of time budget assigned to exact subproblem solving. In this way, the explainability analysis could be transformed into a decision mechanism that selects or adapts search settings according to the structural characteristics of each instance.}

	\section{Conclusions and Future Work}\label{sec:conclusions}
	
	This paper addresses the Longest Filled Common Subsequence (LFCS) problem, a combinatorial optimization problem with practical relevance in bioinformatics, particularly for identifying gene mutations and reconstructing genomic data. To effectively solve this problem, we use a variant of the robust metaheuristic framework known as Construct, Merge, Solve, Adapt (CMSA). Our proposed variant incorporates several tailored components. In the construction phase, we adapted a randomized algorithm from the literature to generate diverse solutions within a parameterized neighborhood of the current best solution. The generated solution components are then merged efficiently to build a promising subproblem, which is solved using an Integer Linear Programming (ILP) technique. At subsequent iterations, new subproblems are dynamically generated based on the quality of the obtained solution, automatically adjusting the number of solutions to be constructed in each iteration as well as the value of the parameter that controls the level of diversification between the solutions. To evaluate the scalability and robustness of our approach relative to five existing methods from the literature, we introduced an additional synthetic dataset comprising 120 large-scale problem instances. These instances are approximately an order of magnitude larger than those previously studied, allowing for a more thorough and realistic assessment of algorithmic performance under practical conditions. A comprehensive experimental evaluation demonstrates the effectiveness of our CMSA-based approach, {which achieved the best average solution quality across all groups of the \textsc{Large} benchmark set while matching the known optimal solution quality for almost all instances from the \textsc{Small} benchmark set}. 
	{Specifically, among the 1,510 instances for which the \textsc{Ilp} solver proved optimality, \textsc{Adapt-Cmsa} produced solutions of the same quality for 1,486 instances---substantially more than the second-best heuristic approach.} In addition to the synthetic problem instances, we generated a set of real-world problem instances. These are used in an audio-fingerprinting case study, demonstrating the ability of the LFCSP to identify the correct song in a noisy audio-query scenario. {Beyond their immediate use in this study, we expect these benchmark datasets to have lasting value for the community: the \textsc{Large} synthetic set enables scalability comparisons roughly an order of magnitude beyond previously available LFCS benchmarks, while the audio-derived \textsc{Song} instances -- extended in \ref{app:jamendo} with 19 instances drawn from a 1000-track real-world corpus -- connect the LFCS problem to a concrete, reproducible downstream application. All benchmark sets are released alongside our source code, which is intended to serve as a standardized reference point for future algorithmic comparisons on the LFCS problem.} To further investigate the relationship between instance features and algorithmic performance across all six approaches, we apply an explainable AI technique, namely Shapley Additive Explanations (SHAP). {  This analysis identifies key combinations of features that significantly influence algorithm performance. For our CMSA-based method, the most influential feature for middle-to-large instances is the length of the first input sequence ($n$), followed by the cardinality of the multiset $\mathcal{M}$ ($k$). For small-to-middle-sized instances $|B|$ is also important, together with the number of matchings between input sequences. This observation suggests that effectively identifying a substantial number of relevant matchings is a critical factor in producing high-quality solutions.}
	
	Future work could investigate the applicability and robustness of our approach in practical gene-data reconstruction scenarios. Furthermore, several variants of CMSA from the literature offer substantial potential for further performance improvements. In particular, the reinforcement learning-based CMSA method proposed by Reixach et al.~\cite{reixach2024improve} avoids traditional greedy decision-making during solution construction, potentially improving solution diversity and adaptability. {Another avenue involves strengthening the current ILP formulation through valid inequalities, cutting planes, or tighter bounds, and by developing an advanced branch-and-cut algorithm~\cite{mitchell2002branch}. Such enhancements could improve the LP relaxation and accelerate the solution of the subproblems within \textsc{Adapt-Cmsa}.} Moreover, integrating insights from various machine learning models tailored to the problem domain presents an exciting research direction~\cite{reixach2024neural}. For example, a trained model could inform better symbol deletion strategies by identifying promising symbols and corresponding positions for rapid omission when constructing high-quality solutions. {Another promising direction is to use the output from explainable AI to guide more efficient variants of algorithms}.

	\section*{Acknowledgments} 
	
	This article is based upon work from COST Action ``Randomised Optimisation Algorithms Research Network'' (ROAR-NET), CA22137, supported by
	COST (European Cooperation in Science and Technology). M. D. is partially supported by the project entitled ``Utilizing deep learning to boost the performance of classical optimization techniques'' funded by the Ministry of Scientific and technological development and Higher Education of Republic of Srpska (project no.\ 1259114). A. N. acknowledges the support of the Slovenian Research and Innovation Agency through program grant P2-0098, and project grants No.J2-4460 and No. GC-0001, and young researcher grant No. PR-12897. Her work is also funded by the European Union under Grant Agreement 101187010 (HE ERA Chair AutoLearn-SI). Additionally, this	 publication is co-funded by the European Union’s Horizon Europe research and innovation program under the Marie Sklodowska-Curie COFUND Postdoctoral Programme (grant agreement No.~101081355 -- SMASH), and by the Republic of Slovenia and the European Union through the European Regional Development Fund. Views and opinions expressed are those of the authors only and do not necessarily reflect those of the European Union or the European Research Executive Agency (REA). Neither the European Union nor the REA can be held responsible for them.
	
	\bibliographystyle{elsarticle-num} 
	\bibliography{references} 

@inproceedings{hershey2017cnn,
	title={CNN architectures for large-scale audio classification},
	author={Hershey, Shawn and Chaudhuri, Sourish and Ellis, Daniel PW and Gemmeke, Jort F and Jansen, Aren and Moore, R Channing and Plakal, Manoj and Platt, Devin and Saurous, Rif A and Seybold, Bryan and others},
	booktitle={2017 {IEEE} international conference on acoustics, speech and signal processing (icassp)},
	pages={131--135},
	year={2017},
	organization={IEEE}
}

@inproceedings{cramer2019look,
	title={Look, listen, and learn more: Design choices for deep audio embeddings},
	author={Cramer, Aurora Linh and Wu, Ho-Hsiang and Salamon, Justin and Bello, Juan Pablo},
	booktitle={ICASSP 2019-2019 IEEE International Conference on Acoustics, Speech and Signal Processing (ICASSP)},
	pages={3852--3856},
	year={2019},
	organization={IEEE}
}

@inproceedings{wang2003shazam,
	title={An industrial strength audio search algorithm.},
	author={Wang, Avery and others},
	booktitle={Ismir},
	volume={2003},
	pages={7--13},
	year={2003},
	organization={Washington, DC}
}

@inproceedings{joren2014panako,
	title={Panako-A scalable acoustic fingerprinting system handling time-scale and pitch modification},
	author={Joren, Six and Leman, Marc},
	booktitle={Proc. ISMIR},
	pages={259--264},
	year={2014}
}

@inproceedings{ellis20142014,
	title={The 2014 labrosa audio fingerprint system},
	author={Ellis, Daniel},
	booktitle={ISMIR},
	year={2014}
}

@inproceedings{bogdanov2019mtg,
	title={The MTG-Jamendo dataset for automatic music tagging},
	author={Bogdanov, Dmitry and Won, Minz and Tovstogan, Philip and Porter, Alastair and Serra, Xavier},
	booktitle={Machine Learning for Music Discovery Workshop, International Conference on Machine Learning (ICML 2019)},
	year={2019}
}

@inproceedings{castelli2017longest,
	title={The longest filled common subsequence problem},
	author={Castelli, Mauro and Dondi, Riccardo and Mauri, Giancarlo and Zoppis, Italo},
	booktitle={28th Annual Symposium on Combinatorial Pattern Matching (CPM 2017)},
	year={2017},
	organization={Schloss-Dagstuhl-Leibniz Zentrum f{\"u}r Informatik}
}

@inproceedings{mincu2018heuristic,
	title={Heuristic algorithms for the longest filled common subsequence problem},
	author={Mincu, Radu Stefan and Popa, Alexandru},
	booktitle={The 20th International Symposium on Symbolic and Numeric Algorithms for Scientific Computing (SYNASC)},
	pages={449--453},
	year={2018},
	organization={IEEE}
}

@article{akbay2022self,
	title={A self-adaptive variant of CMSA: application to the minimum positive influence dominating set problem},
	author={Akbay, Mehmet An{\i}l and L{\'o}pez Serrano, Albert and Blum, Christian},
	journal={International Journal of Computational Intelligence Systems},
	volume={15},
	number={1},
	pages={44},
	year={2022},
	publisher={Springer}
}

@article{BluBleLop2009,
	author    = {Christian Blum and Maria J. Blesa and Manuel L{\'{o}}pez{-}Ib{\'{a}}{\~{n}}ez},
	title     = {Beam search for the longest common subsequence problem},
	journal   = {Computers {\&} Operations Research},
	volume    = {36},
	number    = {12},
	pages     = {3178--3186},
	year      = {2009},
}

@Article{ijms160920748,
	AUTHOR = {Minkiewicz, Piotr and Darewicz, Małgorzata and Iwaniak, Anna and Sokołowska, Jolanta and Starowicz, Piotr and Bucholska, Justyna and Hrynkiewicz, Monika},
	TITLE = {Common Amino Acid Subsequences in a Universal  Proteome—Relevance for Food Science},
	JOURNAL = {International Journal of Molecular Sciences},
	VOLUME = {16},
	YEAR = {2015},
	NUMBER = {9},
	PAGES = {20748--20773},
}

@Article{ijgi6010001,
	AUTHOR = {Xie, Xingzhe and Liao, Wenzhi and Aghajan, Hamid and Veelaert, Peter and Philips, Wilfried},
	TITLE = {Detecting Road Intersections from {GPS} Traces Using Longest Common Subsequence Algorithm},
	JOURNAL = {ISPRS International Journal of Geo-Information},
	VOLUME = {6},
	YEAR = {2017},
	NUMBER = {1},
	ARTICLE-NUMBER = {1},
}

@inproceedings{bergroth2000survey,
	title={A survey of longest common subsequence algorithms},
	author={Bergroth, Lasse and Hakonen, Harri and Raita, Timo},
	booktitle={In Proceedings of SPIRE 2000 -- The 7th International Symposium on String Processing and Information Retrieval},
	pages={39--48},
	year={2000},
	organization={IEEE}
}

@article{irace,
	author = { Manuel L{\'o}pez-Ib{\'a}{\~n}ez  and  J{\'e}r{\'e}mie Dubois-Lacoste  and  Leslie {P{\'e}rez C{\'a}ceres}  and  Thomas St{\"u}tzle  and  Mauro Birattari },
	title = {The irace package: Iterated Racing for Automatic
	Algorithm Configuration},
	journal = {Operations Research Perspectives},
	year = 2016,
	supplement = {http://iridia.ulb.ac.be/supp/IridiaSupp2016-003/},
	doi = {10.1016/j.orp.2016.09.002},
	volume = 3,
	pages = {43--58}
}

@article{Maier78,
	title={The Complexity of Some Problems on Subsequences and Supersequences},
	author={David Maier},
	journal={Journal of the ACM},
	pages={322--336},
	volume={25},
	number=2,
	year=1978,
}

@BOOK{Gus97:sequence-algorithms,
	author = {D.~Gusfield},
	year = {1997},
	title = {Algorithms on Strings, Trees, and Sequences},
	series = {Computer Science and Computational Biology},
	publisher = {Cambridge University Press}
}

@BOOK{Sto88:datacompr,
	author = {J.~Storer},
	year = {1988},
	title = {Data Compression: Methods and Theory},
	publisher = {Computer Science Press},
	address = {MD, USA}
}

@Article{Beal2016,
	author="Beal, Richard
	and Afrin, Tazin
	and Farheen, Aliya
	and Adjeroh, Donald",
	title="A new algorithm for the {LCS} problem with application in compressing genome resequencing data",
	journal="BMC Genomics",
	year="2016",
	volume="17",
	number="4",
	pages="544",
	issn="1471-2164",
	doi="10.1186/s12864-016-2793-0",
	url="https://doi.org/10.1186/s12864-016-2793-0"
}

@article{DJUKANOVIC2020106499,
	title = {Finding Longest Common Subsequences: New anytime {A}* search results},
	journal = {Applied Soft Computing},
	volume = {95},
	pages = {106499},
	year = {2020},
	issn = {1568--4946},
	doi = {https://doi.org/10.1016/j.asoc.2020.106499},
	author = {Marko Djukanovic and Guenther R. Raidl and Christian Blum}
}

@InProceedings{DjukanovicRaidlBlum19lod,
	title={A Beam Search for the Longest Common
	Subsequence Problem Guided by a Novel
	Approximate Expected Length Calculation},
	author       = {Marko Djukanovic and G\"{u}nther Raidl and Christian Blum},
	booktitle = {Proceedings of {LOD 2019} -- The 5th International Conference on Machine Learning, Optimization, and Data Science},
	year={2019},
	publisher={Springer},
	series = {LNCS},
	note={to appear}
}

@article{kruskal1983overview,
	title={An overview of sequence comparison: Time warps, string edits, and macromolecules},
	author={Kruskal, Joseph B},
	journal={SIAM review},
	volume={25},
	number={2},
	pages={201--237},
	year={1983},
	publisher={SIAM}
}

@article{nikolic2021solving,
	title={Solving the longest common subsequence problem concerning non-uniform distributions of letters in input strings},
	author={Nikolic, Bojan and Kartelj, Aleksandar and Djukanovic, Marko and Grbic, Milana and Blum, Christian and Raidl, G{\"u}nther},
	journal={Mathematics},
	volume={9},
	number={13},
	pages={1515},
	year={2021},
	publisher={MDPI}
}

@article{lin2002longest,
	title={The longest common subsequence problem for sequences with nested arc annotations},
	author={Lin, Guohui and Chen, Zhi-Zhong and Jiang, Tao and Wen, Jianjun},
	journal={Journal of Computer and System Sciences},
	volume={65},
	number={3},
	pages={465--480},
	year={2002},
	publisher={Elsevier}
}

@article{adi2010repetition,
	title={Repetition-free longest common subsequence},
	author={Adi, Said S and Braga, Mar{\'\i}lia DV and Fernandes, Cristina G and Ferreira, Carlos E and Martinez, F{\'a}bio Viduani and Sagot, Marie-France and Stefanes, Marco A and Tjandraatmadja, Christian and Wakabayashi, Yoshiko},
	journal={Discrete Applied Mathematics},
	volume={158},
	number={12},
	pages={1315--1324},
	year={2010},
	publisher={Elsevier}
}

@article{tsai2003constrained,
	title={The constrained longest common subsequence problem},
	author={Tsai, Yin-Te},
	journal={Information Processing Letters},
	volume={88},
	number={4},
	pages={173--176},
	year={2003},
	publisher={Elsevier}
}

@article{blum2016construct,
	title={Construct, merge, solve \& adapt a new general algorithm for combinatorial optimization},
	author={Blum, Christian and Pinacho, Pedro and L{\'o}pez-Ib{\'a}{\~n}ez, Manuel and Lozano, Jos{\'e} A},
	journal={Computers \& Operations Research},
	volume={68},
	pages={75--88},
	year={2016},
	publisher={Elsevier}
}

@incollection{blum2024self,
	title={Self-adaptive CMSA},
	author={Blum, Christian},
	booktitle={Construct, Merge, Solve \& Adapt: A Hybrid Metaheuristic for Combinatorial Optimization},
	pages={41--70},
	year={2024},
	publisher={Springer}
}

@inproceedings{blum2016constructMcsp,
	title={Construct, merge, solve and adapt: application to unbalanced minimum common string partition},
	author={Blum, Christian},
	booktitle={Hybrid Metaheuristics: 10th International Workshop, HM 2016, Plymouth, UK, June 8-10, 2016, Proceedings 10},
	pages={17--31},
	year={2016},
	organization={Springer}
}

@article{djukanovic2023self,
	title={Self-adaptive {CMSA} for solving the multidimensional multi-way number partitioning problem},
	author={Djukanovi{\'c}, Marko and Kartelj, Aleksandar and Blum, Christian},
	journal={Expert Systems with Applications},
	volume={232},
	pages={120762},
	year={2023},
	publisher={Elsevier}
}

@article{akbay2024cmsa,
	title={{CMSA} based on set covering models for packing and routing problems},
	author={Akbay, Mehmet An{\i}l and Blum, Christian and Kalayci, Can Berk},
	journal={Annals of Operations Research},
	volume={343},
	number={1},
	pages={1--38},
	year={2024},
	publisher={Springer}
}

@article{woolson2005wilcoxon,
	title={Wilcoxon signed-rank test},
	author={Woolson, Robert F},
	journal={Encyclopedia of biostatistics},
	volume={8},
	year={2005},
	publisher={Wiley Online Library}
}

@article{nohara2022explanation,
	title={Explanation of machine learning models using shapley additive explanation and application for real data in hospital},
	author={Nohara, Yasunobu and Matsumoto, Koutarou and Soejima, Hidehisa and Nakashima, Naoki},
	journal={Computer Methods and Programs in Biomedicine},
	volume={214},
	pages={106584},
	year={2022},
	publisher={Elsevier}
}

@article{avanijaa2021prediction,
	title={Prediction of house price using xgboost regression algorithm},
	author={Avanijaa, Jangaraj and others},
	journal={Turkish Journal of Computer and Mathematics Education (TURCOMAT)},
	volume={12},
	number={2},
	pages={2151--2155},
	year={2021}
}

@article{reixach2024improve,
	title={How to improve “construct, merge, solve and adapt"? Use reinforcement learning!},
	author={Reixach, Jaume and Blum, Christian},
	journal={Annals of Operations Research},
	pages={1--32},
	year={2024},
	publisher={Springer}
}

@article{munoz2010scaffold,
	title={Scaffold filling, contig fusion and comparative gene order inference},
	author={Mu{\~n}oz, Adriana and Zheng, Chunfang and Zhu, Qian and Albert, Victor A and Rounsley, Steve and Sankoff, David},
	journal={BMC bioinformatics},
	volume={11},
	pages={1--15},
	year={2010},
	publisher={Springer}
}

@article{bulteau2015fixed,
	title={Fixed-parameter algorithms for scaffold filling},
	author={Bulteau, Laurent and Carrieri, Anna Paola and Dondi, Riccardo},
	journal={Theoretical Computer Science},
	volume={568},
	pages={72--83},
	year={2015},
	publisher={Elsevier}
}

@inproceedings{reixach2024neural,
	title={A Neural Network Based Guidance for a {BRKGA}: An Application to the Longest Common Square Subsequence Problem},
	author={Reixach, Jaume and Blum, Christian and Djukanovi{\'c}, Marko and Raidl, G{\"u}nther R},
	booktitle={European Conference on Evolutionary Computation in Combinatorial Optimization (Part of EvoStar)},
	pages={1--15},
	year={2024},
	organization={Springer}
}

@article{lopez2016irace,
	title={The irace package: Iterated racing for automatic algorithm configuration},
	author={L{\'o}pez-Ib{\'a}{\~n}ez, Manuel and Dubois-Lacoste, J{\'e}r{\'e}mie and C{\'a}ceres, Leslie P{\'e}rez and Birattari, Mauro and St{\"u}tzle, Thomas},
	journal={Operations Research Perspectives},
	volume={3},
	pages={43--58},
	year={2016},
	publisher={Elsevier}
}

@inproceedings{holzinger2020explainable,
	title={Explainable {AI} methods-a brief overview},
	author={Holzinger, Andreas and Saranti, Anna and Molnar, Christoph and Biecek, Przemyslaw and Samek, Wojciech},
	booktitle={International workshop on extending explainable AI beyond deep models and classifiers},
	pages={13--38},
	year={2020},
	organization={Springer}
}

@article{dai2014autonomous,
	title={Autonomous document cleaning—a generative approach to reconstruct strongly corrupted scanned texts},
	author={Dai, Zhenwen and Luecke, Joerg},
	journal={IEEE transactions on pattern analysis and machine intelligence},
	volume={36},
	number={10},
	pages={1950--1962},
	year={2014},
	publisher={IEEE}
}

@article{ferrer2021cmsa,
	title={{CMSA} algorithm for solving the prioritized pairwise test data generation problem in software product lines},
	author={Ferrer, Javier and Chicano, Francisco and Ortega-Toro, Jos{\'e} Antonio},
	journal={Journal of Heuristics},
	volume={27},
	number={1},
	pages={229--249},
	year={2021},
	publisher={Springer}
}

@article{dupin2021matheuristics,
	title={Matheuristics to optimize refueling and maintenance planning of nuclear power plants},
	author={Dupin, Nicolas and Talbi, El-Ghazali},
	journal={Journal of Heuristics},
	volume={27},
	number={1},
	pages={63--105},
	year={2021},
	publisher={Springer}
}

@incollection{blum2025hybrid,
	title={The Hybrid Metaheuristic CMSA},
	author={Blum, Christian and Reixach, Jaume},
	booktitle={Handbook of Heuristics},
	pages={1--24},
	year={2025},
	publisher={Springer}
}

@article{ali2023shannon,
	title={Shannon entropy in artificial intelligence and its applications based on information theory},
	author={Ali, Aqib and Anam, Sania and Ahmed, Muhammad Munawar},
	journal={J. Appl. Emerg. Sci},
	volume={13},
	number={1},
	pages={9--17},
	year={2023}
}

@article{mitchell2002branch,
	title={Branch-and-cut algorithms for combinatorial optimization problems},
	author={Mitchell, John E},
	journal={Handbook of applied optimization},
	volume={1},
	number={1},
	pages={65--77},
	year={2002},
	publisher={Oxford, UK}
}

@article{kim2025enhancing,
	title={Enhancing genetic algorithm with explainable artificial intelligence for last-mile routing},
	author={Kim, Yonggab and Khir, Reem and Lee, Seokcheon},
	journal={IEEE Transactions on Evolutionary Computation},
	year={2025},
	publisher={IEEE}
}

@inproceedings{ha2019improvement,
	title={Improvement in deep networks for optimization using explainable artificial intelligence},
	author={ha Lee, Jin and hee Shin, Ik and gu Jeong, Sang and Lee, Seung-Ik and Zaheer, Muhamamad Zaigham and Seo, Beom-Su},
	booktitle={2019 International Conference on Information and Communication Technology Convergence (ICTC)},
	pages={525--530},
	year={2019},
	organization={IEEE}
}

	\appendix
	\section{Extended Validation on a Larger, Real-World Audio Corpus}\label{app:jamendo}

	{To extend the audio fingerprinting case study presented in Section~\ref{sec:experimental_section} toward a more realistic use case, we repeated the experiment at a considerably larger and less curated scale, using 1000 full-length tracks---500 rock songs and 500 classical pieces, 3--4 minutes each---drawn from the Creative-Commons-licensed MTG-Jamendo dataset~\cite{bogdanov2019mtg}, applying the same degradation model used in Section~\ref{sec:experimental_section}. From this corpus, we randomly selected 20 query tracks (10 rock, 10 classical) and degraded each one at 16 increasing degradation levels, yielding $20 \times 16 = 320$ degraded query instances in total. Of these, 89 reached a degradation level at which Audfprint's own confidence (raw shared-hash count) fell below 10, and, of those 89, the 19 query/level instances listed in Table~\ref{tab:jamendo-instances} are the ones for which the true track nonetheless remained within its top-16 candidate shortlist; these constitute the hard instances used in the remainder of this analysis.}

	\begin{table}[ht]
		\centering
		\begin{tabular}{|l|l|c|c|c|}
			\hline
			\textbf{Query track} & \textbf{Genre} & \textbf{Rem} & \textbf{Audfprint rank} & \textbf{LFCSP rank} \\
			\hline \hline
			0000242 & rock & 0.75 & 8  & 2 \\
			0000242 & rock & 0.85 & 3  & 2 \\
			1265963 & rock & 0.90 & 4  & 7 \\
			1352255 & rock & 0.90 & 12 & 1 \\
			1370673 & rock & 0.75 & 3  & 4 \\
			1370673 & rock & 0.80 & 1  & 6 \\
			1389561 & rock & 0.70 & 13 & 1 \\
			\hline
			0000387 & classical & 0.65 & 10 & 2 \\
			0949644 & classical & 0.60 & 6  & 6 \\
			0965319 & classical & 0.40 & 3  & 5 \\
			0965319 & classical & 0.50 & 3  & 5 \\
			1012003 & classical & 0.40 & 8  & 6 \\
			1012003 & classical & 0.55 & 13 & 6 \\
			1092337 & classical & 0.85 & 1  & 2 \\
			1114849 & classical & 0.70 & 9  & 1 \\
			1114849 & classical & 0.90 & 7  & 1 \\
			1129581 & classical & 0.65 & 10 & 1 \\
			1129581 & classical & 0.75 & 5  & 2 \\
			1129581 & classical & 0.90 & 5  & 5 \\
			\hline
		\end{tabular}
		\caption{The 19 hard query/level instances retained after filtering, with the rank (out of 16) of the true candidate before (Audfprint) and after re-ranking with \textsc{Ls2} (representative of the LCS-exact heuristics; see Table~\ref{tab:jamendo-approaches} for the full per-approach breakdown).}
		\label{tab:jamendo-instances}
	\end{table}

	{Table~\ref{tab:jamendo-approaches} reports these results for each of the six approaches from Section~\ref{sec:experimental_section}, together with their mean solving time and a paired Wilcoxon signed-rank test against Audfprint's own ranking. The rank improvement is statistically significant for \textsc{Rand-Sample}, \textsc{Ls2}, \textsc{Ls4}, and \textsc{Adapt-Cmsa} ($p<0.05$), but not for \textsc{Approx}, which performs slightly worse than Audfprint's own ranking on average; \textsc{Ilp} produced a usable rank for only 1 of the 19 instances and is therefore not tested. Notably, the lightweight \textsc{Ls2} heuristic matches \textsc{Adapt-Cmsa}'s rank improvement in roughly 2~seconds per candidate, two orders of magnitude faster than \textsc{Adapt-Cmsa}'s 600-second budget -- a relevant property for a re-ranking layer intended to run alongside an existing fingerprinting system.}

	\begin{table}[ht]
		\centering
		\footnotesize
		\begin{tabular}{|l|c|c|c|c|c|}
			\hline
			\textbf{Approach} & \textbf{Mean} & \textbf{Median} & \textbf{Rank~1} & \textbf{t}[$s$] & \textbf{Wilcoxon} \\
			 & \textbf{rank} & \textbf{rank} & \textbf{(or tied)} & & \textbf{$p$} \\
			\hline \hline
			Audfprint (baseline) & 6.53 & 6 & 2/19 & -- & -- \\
			\textsc{Rand-Sample}  & 3.47 & 2 & 5/19 & 1.26 & 0.037 \\
			\textsc{Approx}       & 7.26 & 9 & 4/19 & $<0.001$ & 0.559 \\
			\textsc{Ls2}          & 3.42 & 2 & 5/19 & 2.12 & 0.029 \\
			\textsc{Ls4}          & 3.42 & 2 & 5/19 & 5.65 & 0.029 \\
			\textsc{Adapt-Cmsa}   & 3.47 & 2 & 5/19 & 602.96 & 0.037 \\
			\textsc{Ilp}          & n/a$^{a}$ & n/a$^{a}$ & 0/19 & 275.1$^{b}$ & n/a \\
			\hline
		\end{tabular}
		\normalsize
		\caption{Rank of the true candidate within the same 16-item shortlist (Audfprint's own ranking versus each approach's re-ranking), and mean solving time over the $19\times16=304$ candidate sub-instances, each solved under the 600~CPU-second limit used throughout this paper. $^{a}$\textsc{Ilp} produced a usable rank for only 1 of the 19 instances. $^{b}$Based only on the 67/304 (22\%) sub-instances \textsc{Ilp} solved to completion within the time limit.}
		\label{tab:jamendo-approaches}
	\end{table}

	{In the two cases where Audfprint's own top-1 pick was already correct, LFCSP demoted it, indicating a re-ranking aid rather than a strict improvement. We repeated a smaller version of this evaluation using Panako~\cite{joren2014panako} as the underlying fingerprinting system and observed a comparable prefiltering effect; we report the full analysis for Audfprint only, as it integrates more directly with our existing Python pipeline. These results support LFCSP as a re-ranking layer on top of existing fingerprinting systems for hard, heavily degraded queries, rather than a stand-alone replacement, on a corpus still far smaller than production scale (millions of tracks). The 19 instances are released alongside the \textsc{Small} and \textsc{Large} benchmark sets.}

\section{Evolution of the restricted ILP subproblem on benchmark set \textsc{SONG}}\label{app:evolution_subproblem_songs}

\begin{itemize}
	\item {For the \texttt{DaftPunk} instance from the \textsc{Song} benchmark, the restricted subproblem initially contains only a subset of the available $\textbf{y}$-variables, but their number rapidly increases to $\approx$60\% (i.e., to 204 out of 322 variables) and remains at that level throughout almost the entire execution. The number of $\textbf{x}$-variables exhibits a more progressive increase, growing from a few percentage to approximately 15\% (700) of all variables. After the initial growth phase, the number of $\textbf{x}$-variables continues to slightly fluctuate within a relatively narrow range.}  
\end{itemize}

\begin{figure}[H]
	
\begin{subfigure}[t]{0.485\textwidth}
	\centering
	\includegraphics[width=\linewidth]{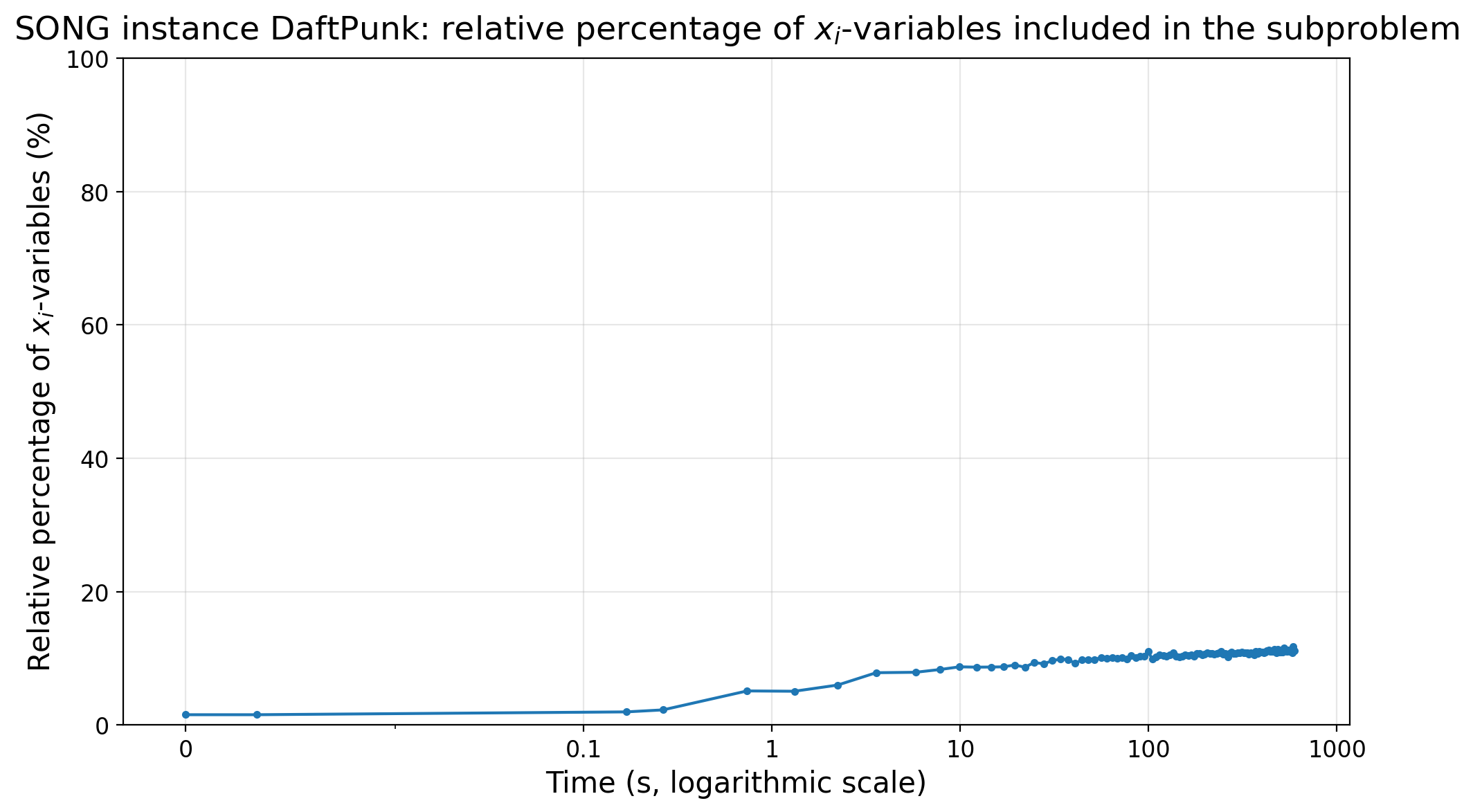}
	\caption{Number of  $\textbf{x}$-variables for the \texttt{DaftPunk} instance from \textsc{Song}.  The number of $\textbf{x}$-variables in the ILP model of this instance equals 6367.}
	\label{fig:song-x-vars}
\end{subfigure}
\hfill
\begin{subfigure}[t]{0.485\textwidth}
	\centering
	\includegraphics[width=\linewidth]{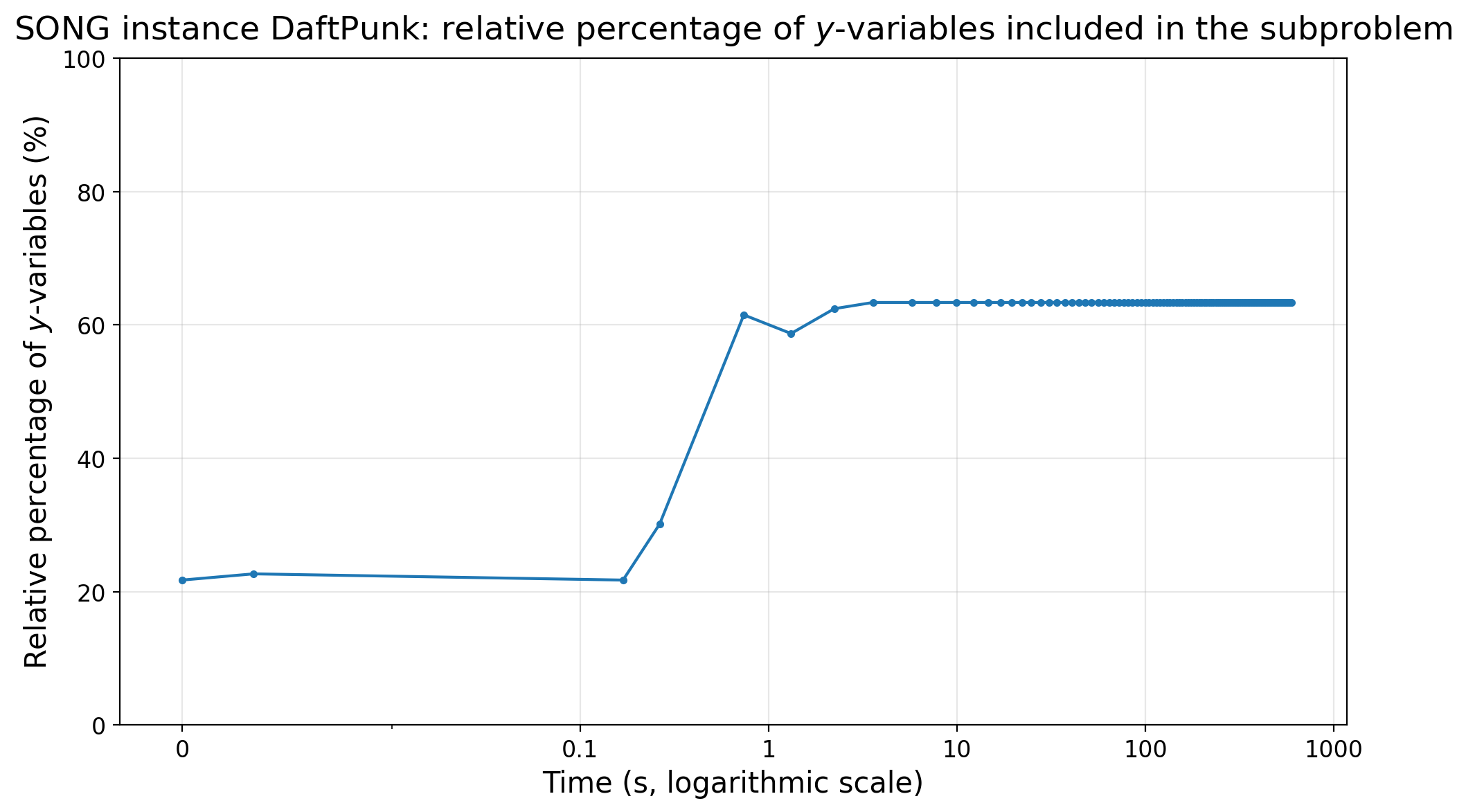}  
	\caption{Number of $\textbf{y}$-variables for the \texttt{DaftPunk} instance from \textsc{Song}. The number of $\textbf{y}$-variables in the complete ILP model equals 322.}
	\label{fig:song-y-vars}
\end{subfigure}

\caption{Evolution of the restricted ILP subproblem size during the execution of \textsc{Adapt-Cmsa}.} \label{fig:subproblem-size-evolution-songs}
\end{figure}

\end{document}